# Derivation of a multilayer approach to model suspended sediment transport: application to hyperpycnal and hypopycnal plumes

Morales de Luna, T. Fernández-Nieto, E.D. Castro Díaz, M.J.

May 11, 2017

**Abstract**

We propose a multi-layer approach to simulate hyperpycnal and hypopycnal plumes in flows with free surface. The model allows to compute the vertical profile of the horizontal and the vertical components of the velocity of the fluid flow. The model can describe as well the vertical profile of the sediment concentration and the velocity components of each one of the sediment species that form the turbidity current. To do so, it takes into account the settling velocity of the particles and their interaction with the fluid. This allows to better describe the phenomena than a single layer approach. It is in better agreement with the physics of the problem and gives promising results. The numerical simulation is carried out by rewriting the multi-layer approach in a compact formulation, which corresponds to a system with non-conservative products, and using path-conservative numerical scheme. Numerical results are presented in order to show the potential of the model.

## 1 Introduction

When a river that carries sediment in suspension enters into lake or the ocean, it can form a plume that advects the sediment from the river mouth. Based on the difference of density [5, 30], these particle-bearing flows are said to be 'hypopycnal' (or an 'overflow') if the combined density of the sediment and interstitial fluid is lower than that of the ambient. If the combined density is higher than that of the ambient, it is said to be 'hyperpycnal' (or an 'underflow'). Hyperpycnal plumes are a class of sediment-laden gravity current commonly referred to as turbidity currents [23, 27, 30].

Turbidity currents occur in many different circumstances in nature, for example, at the outflow of rivers into the ocean [23, 22], where they may be generated by storm waves impacting the coast [29], in regions of submarine landslides resulting for example from tectonic activity [18] and where tidal activity acts on steep slopes [31]. They are responsible for the transportation of sediment on a global scale, defining the main mechanism that allows sediment to be transported to the deeper ocean [19, 26, 13, 15]. Because of their impacts in global sediment transport, their role in erosion and deposition over continental slopes and submarine canyons, and the effects on marine constructions and infrastructures near river mouths and continental shelves, understanding the evolution of turbidity currents is of great importance.

Only limited observational records exist for the occurrence and flow of turbidity currents. This is due to the difficulty in predicting the time and frequency of turbidity currents as well as the destructive nature of such sediment-laden flows. As a result, most of our knowledge about these flows is derived from small scale laboratory experiments like the ones described in [20, 13, 1, 17, 16]. Given the lack of observations, numerical modeling is an excellent tool to gain an increased understanding of the evolution of turbidity currents.

Some layer-averaged models have been previously developed on the basis of small-scale tank experiments of particle-driven density currents in [8, 26, 6, 17, 17, 14]. Although this layer-averaged approach gives a fast and valuable information, it has the disadvantage that the vertical distribution of the sediment in suspension is lost.

A recent technique based on a multilayer approach [2, 3, 11] has shown to be specially useful to generalize shallow water type models in order to keep track of the vertical components of the averaged variables in the classical shallow water equations. Thus, instead of averaging between the bottom and the free surface, as in shallow water system, a partition of the water height is considered. The system is then integrated inside of each layer defined by the partition, combined with a set of kinematic conditions at each interface, defined in terms on the mass transference term.

In [11] another multilayer approach is proposed for the case of the Navier-Stokes equations with constant density and viscosity. In that work, a multilayer model is obtained using a vertical discontinuous Galerkin approach for which the vertical velocity is supposed to be piecewise linear and the horizontal velocity is supposed to be piecewise constant. The mass and momentum transfer terms among the layers are obtained from the jump conditions of the conserved principles. In the numerical tests presented in [11] authors show that this discontinuous piecewise linear profile of the vertical profile allows to approximate properly the vertical velocity of Navier-Stokes equations. The key point is to compute the jump of the vertical velocity at each interface in terms of the jump condition associated to the definition of the mass transference.

In [12] authors propose an application of the multilayer approach introduced in [2, 3] to study polydisperse sedimentation. Here the technique introduced in [11] is generalized to derive a model for turbidity currents. As in [11], we also obtain a piecewise linear profile for the vertical velocity profile of the fluid and each sediment species involved in the turbidity current. As it will be shown, this model is really promising and allows to simulate hyperpycnal as well as hypopycnal plumes.

The paper is organized as follows: in Section 2 we describe the governing equations for the phenomena. These equations are the starting point for Section 3 where a multilayer approach is used to develop a new set of equations. In Section 4 the assumption of hydrostatic pressure is used and a particular weak solution of the system is defined. The system needs some closure relations that are introduced in Section 5 and a compact formulation of the system is shown. Then, in Section 6 a numerical scheme is proposed and some numerical tests are shown in Section 7. The paper contains some Appendixes with the technical details of the different sections.

## 2 Governing equations

Let us consider $N \in \mathbb{N}^*$ species of spherical solid particles dispersed in a viscous fluid. For each solid species $j$, $j = 1, \dots, N$, we denote by $\phi_j$, $\rho_j$, $d_j$, and $\vec{\boldsymbol{v}}_j = (\vec{\boldsymbol{u}}_j, w_j)$, $j = 1, \dots, N$, its volumetric concentration, density, particle diameter, and phase velocity, respectively. $\vec{\boldsymbol{u}}_j \in \mathbb{R}^{d-1}$ $(d = 2, 3)$ represents the horizontal component of the velocity and $w_j \in \mathbb{R}$ the vertical one. The same notation is used for the fluid indexed by $j = 0$. We assume that effects of sediment compressibility can be neglected. The model is based on the continuity and linear momentum balance equations for the $N$ solid species and the fluid. The continuity equations are given by

$$\partial_t \phi_j + \nabla \cdot (\phi_j \vec{\boldsymbol{v}}_j) = 0, \quad j = 0, \dots, N. \tag{2.1}$$

Taking into account that $\phi = 1 - \phi_0$, where $\phi := \phi_1 + \cdots + \phi_N$ denotes the total solids concentration, we see by summing all equations in (2.1) that the volume average velocity of the mixture

$$\vec{\boldsymbol{v}} := (\vec{\boldsymbol{u}}, w) := \phi_0 \vec{\boldsymbol{v}}_0 + \phi_1 \vec{\boldsymbol{v}}_1 + \cdots + \phi_N \vec{\boldsymbol{v}}_N = (1 - \phi)\vec{\boldsymbol{v}}_0 + \phi_1 \vec{\boldsymbol{v}}_1 + \cdots + \phi_N \vec{\boldsymbol{v}}_N$$

satisfies the simple mass balance of the mixture

$$\nabla \cdot \vec{\boldsymbol{v}} = 0. \tag{2.2}$$

Here, we shall assume that the velocity of each sediment species is equal to the volume average velocity of the mixture plus a vertical downwards component due to the sedimentation of the particles, that is

$$\vec{\boldsymbol{v}}_j = \vec{\boldsymbol{v}} + \delta w_j \vec{\boldsymbol{k}}, \quad j = 1, \dots, N, \tag{2.3}$$

where $\vec{\boldsymbol{k}}$ is the upward-pointing unit vector and $\delta w_j$ is a negative velocity related to the settling velocity of the $j$ species which will be discussed in Section 2.1.

Remark that this implies

$$\phi_0 w_0 = w - \sum_{j=1}^{N} \phi_j w_j = w - \sum_{j=1}^{N} \phi_j (w + \delta w_j) = \phi_0 w - \sum_{j=1}^{N} \delta w_j \phi_j, \tag{2.4}$$

and we shall denote

$$\delta w_0 = -\frac{1}{\phi_0} \sum_{j=1}^{N} \delta w_j \phi_j, \tag{2.5}$$

and we have

$$\vec{\boldsymbol{v}}_j = \vec{\boldsymbol{v}} + \delta w_j \vec{\boldsymbol{k}}, \quad j = 0, 1, \dots, N, \tag{2.6}$$

The model also involves the sum up of the linear momentum balance equations for the solid phases:

$$\rho_j \partial_t (\phi_j \vec{\boldsymbol{v}}_j) + \nabla \cdot (\rho_j \phi_j \vec{\boldsymbol{v}}_j \otimes \vec{\boldsymbol{v}}_j) = -\rho_j \phi_j g \vec{\boldsymbol{k}} + \nabla \cdot \boldsymbol{\Sigma}_j. \tag{2.7}$$

Here g is the gravity acceleration and the stress tensor is

$$\boldsymbol{\Sigma}_j = -\phi_j p \boldsymbol{I} + \boldsymbol{T}_j. \tag{2.8}$$

In what follows, let us denote by $\Phi = \{\phi_j\}_{j=0}^{N}$ the set of concentrations corresponding to each species and by $\rho(\Phi)$ the averaged density of the mixing,

$$\rho(\Phi) = \sum_{j=0}^{N} \rho_j \phi_j.$$

Let us also denote by $\boldsymbol{\Sigma} = \boldsymbol{\Sigma}_0 + \boldsymbol{\Sigma}_1 + \ldots + \boldsymbol{\Sigma}_N = -p\boldsymbol{I} + \boldsymbol{T}$, with

$$\boldsymbol{T} = \sum_{j=0}^{N} \boldsymbol{T}_j = \left( \begin{array}{c|c} T_H & T_{xz} \\ \hline T'_{xz} & T_{zz} \end{array} \right). \tag{2.9}$$

The symbol $\boldsymbol{I}$ stands for the identity tensor. Then, by summing up, from $j = 0$ to $j = N$ the equations in (2.7) we obtain

$$\partial_t \left( \sum_{j=0}^{N} \rho_j \phi_j \vec{\boldsymbol{v}}_j \right) + \nabla \cdot \left( \sum_{j=0}^{n} \rho_j \phi_j \vec{\boldsymbol{v}}_j \otimes \vec{\boldsymbol{v}}_j \right) = -\rho(\Phi) g \vec{\boldsymbol{k}} + \nabla \cdot \boldsymbol{\Sigma}. \tag{2.10}$$

And by using (2.3) we can write each one of the equations of previous system as follows:

$$\partial_t (\rho(\Phi) \vec{\boldsymbol{u}}) + \nabla_x \cdot (\rho(\rho) \vec{\boldsymbol{u}} \otimes \vec{\boldsymbol{u}}) + \partial_z \left( \left( \rho(\Phi) w + \sum_{j=0}^{N} \rho_j \delta w_j \phi_j \right) \vec{\boldsymbol{u}} \right) = \nabla_x \cdot (-p\boldsymbol{I} + T_H) + \partial_z (T_{xz}), \tag{2.11}$$

$$\partial_t \left( \rho(\Phi) w + \sum_{j=0}^{N} \rho_j \delta w_j \phi_j \right) + \nabla_x \cdot \left( \left( \rho(\Phi) w + \sum_{j=0}^{N} \rho_j \delta w_j \phi_j \right) \vec{\boldsymbol{u}} \right)$$
$$+ \partial_z \left( \left( \rho(\Phi) w + \sum_{j=0}^{N} \rho_j \delta w_j \phi_j \right)^2 \right) = \nabla_x \cdot (T_{xz}) + \partial_z (-p + T_{zz}). \tag{2.12}$$

The generic space variable is $(x, z) \in \mathbb{R}^d$ such that the horizontal variable corresponds to $x = (x_1, ..., x_{d-1}) \in \mathbb{R}^{d-1}$.

## 2.1 Settling velocity

As we said before, we assume that the velocity of sediment species differ from the volume average of the mixture by a vertical downwards component due to the sedimentation of the particle. This vertical component is related to the so-called settling velocity ($-w_{s_j} \leq 0$). Settling velocity, also known as fall velocity or terminal velocity of a sediment particle is defined as the rate at which the sediment settles in still fluid. The factors influencing the settling velocity include the particle properties (i.e. size, shape, structure) as well as the viscosity and density of the fluid.
For monodisperse particles settling at infinite dilution, discrete particles will settle within still, homogeneous fluid conditions at a terminal fall velocity determined by application of Stokes' law, if the particle Reynolds number $Re_p << 1$ or by one of various empirically-derived formulae if $Re_p > 1$ (see [9]). Nevertheless, when the concentration of sediment particles increases, particles cease to behave independently. Instead, their motions are correlated through hydrodynamic and particle-particle interactions, often resulting in settling rates that are lower than that for individual, isolated particles, i.e. hindered settling.
Hindered settling is often accounted for by estimating an actual settling velocity, $\delta w_j$, given by

$$\delta w_j = -w_{s_j} \chi(\Phi). \tag{2.13}$$

$\chi(\phi)$ is the so-called hindered settling factor, which may be described following [28] by:

$$\chi(\Phi) = \begin{cases} \left(1 - \sum_{j=1}^{N} \phi_j\right)^n, & \text{for } \sum_{j=1}^{N} \phi_j < \phi_{max}, \\ 0, & \text{otherwise}, \end{cases} \tag{2.14}$$

where $n$ is a parameter that depends on particle Reynolds number (typically between 2.5 and 5.5) and $\phi_{max}$ is a maximal solids concentration.
Nevertheless, other methods may be proposed for predicting the hindered settling conditions. We refer the reader to [9, 4] and the references therein for this purpose.

# 3 A multilayer approach

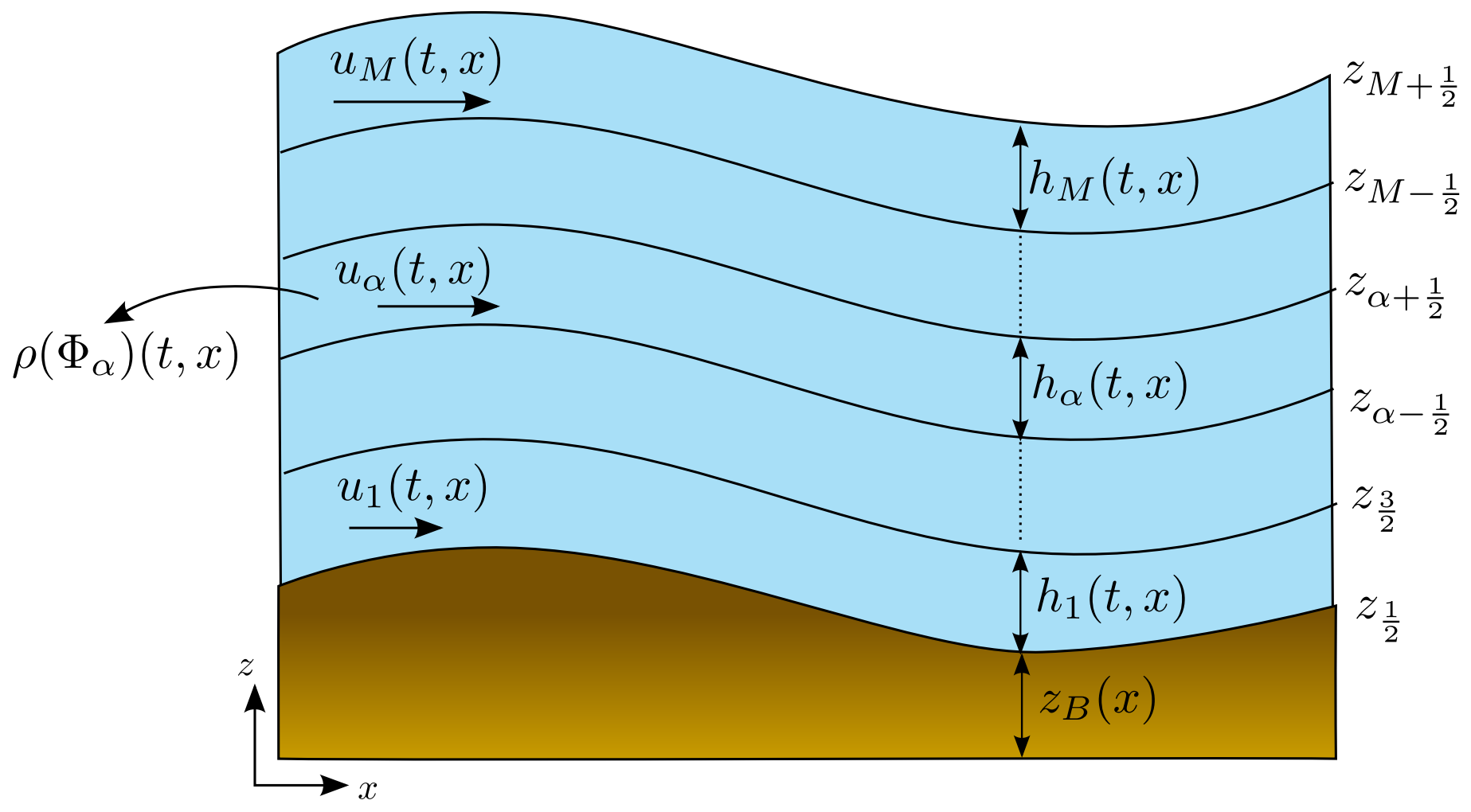

Figure 1: Sketch of the multilayer approach

We shall consider a $d$-dimensional space ($d = 2, 3$). For a given positive constant real number $T$, and each time $t \in [0, T]$ we denote by $\Omega_F(t)$, the fluid domain and by $I_F(t)$, its projection onto the horizontal plane. In order to introduce a multilayer system, the fluid domain is divided along the vertical direction into $M \in \mathbb{N}^*$ pre-set layers of thickness $h_\alpha(t, x)$ with $M + 1$ interfaces $\Gamma_{\alpha+\frac{1}{2}}(t)$ of equations $z = z_{\alpha+\frac{1}{2}}(t, x)$ for $\alpha = 0, 1, ..., M$ and $x \in I_F(t)$ (see Figure 1). We assume that the interfaces $\Gamma_{\alpha+\frac{1}{2}}(t)$ are smooth, concretely at least of class $\mathcal{C}^1$ in time and space. We shall denote by $z_B = z_{\frac{1}{2}}$ and $z_S = z_{M+\frac{1}{2}}$ the equations of the bottom and the free surface interfaces $\Gamma_B(t)$ and $\Gamma_S(t)$, respectively. We have $h_\alpha = z_{\alpha+\frac{1}{2}} - z_{\alpha-\frac{1}{2}}$ and $z_{\alpha+\frac{1}{2}} = z_B + \sum_{\beta=1}^{\alpha} h_\beta$ for $\alpha = 1, ..., M$. Then the height of the fluid is given by $h = z_S - z_B = \sum_{\alpha=1}^{M} h_\alpha$.
Actually we have $\partial\Omega_F(t) = \Gamma_B(t) \cup \Gamma_S(t) \cup \Theta(t)$, where $\Theta(t)$ is the inflow/outflow boundary which we assume here to be vertical. The fluid domain is split as $\overline{\Omega_F(t)} = \cup_{\alpha=1}^{M} \overline{\Omega_\alpha(t)}$ with the setting

$$\begin{aligned}
&\Omega_\alpha(t) = \Big\{(x, z);\ x \in I_F(t) \text{ and } z_{\alpha-\frac{1}{2}} < z < z_{\alpha+\frac{1}{2}}\Big\},\\
&\partial\Omega_\alpha(t) = \Gamma_{\alpha-\frac{1}{2}}(t) \cup \Gamma_{\alpha+\frac{1}{2}}(t) \cup \Theta_\alpha(t), \text{ with}\\
&\Theta_\alpha(t) = \Big\{(x, z);\ x \in \partial I_F(t) \text{ and } z_{\alpha-\frac{1}{2}} < z < z_{\alpha+\frac{1}{2}}\Big\}.
\end{aligned} \tag{3.1}$$

Hence the inflow/outflow boundary is split as $\overline{\Theta(t)} = \cup_{\alpha=1}^{M} \overline{\Theta_\alpha(t)}$.

Moreover, let us introduce the following notation:

(i) For two tensors $\boldsymbol{a}$ and $\boldsymbol{b}$ of sizes $(n, m)$ and $(n, p)$ respectively, we shall denote by $(\boldsymbol{a};\ \boldsymbol{b})$ the tensor of size $(n, m + p)$ which is the concatenation of $\boldsymbol{a}$ and $\boldsymbol{b}$ in this order.

(ii) Let consider the differential operator $\nabla = (\partial_{x_1}, ..., \partial_{x_{d-1}}, \partial_z)$, then we set $\overline{\nabla} := (\partial_t;\ \nabla) = (\partial_t, \partial_{x_1}, ..., \partial_{x_{d-1}}, \partial_z)$ and $\nabla_x := (\partial_{x_1}, ..., \partial_{x_{d-1}})$.

(iii) For $\alpha = 0, 1, ..., N$ and for a function $f$, we set

$$f^-_{\alpha+\frac{1}{2}} := (f_{|\Omega_\alpha(t)})_{|\Gamma_{\alpha+\frac{1}{2}}(t)} \text{ and } f^+_{\alpha+\frac{1}{2}} := (f_{|\Omega_{\alpha+1}(t)})_{|\Gamma_{\alpha+\frac{1}{2}}(t)}.$$

When the function $f$ is continuous, we will simply set

$$f_{\alpha+\frac{1}{2}} := f_{|\Gamma_{\alpha+\frac{1}{2}}(t)}.$$

We shall also use the notation

$$< f >_{\alpha+\frac{1}{2}} = \frac{f^+_{\alpha+\frac{1}{2}} + f^-_{\alpha+\frac{1}{2}}}{2}.$$

(iv) We will denote by $\vec{\boldsymbol{\eta}}_{\alpha+\frac{1}{2}}$ the space unit normal vector to the interface $\Gamma_\alpha(t)$ outward to the layer $\Omega_\alpha(t)$ for a given time $t$ and for $\alpha = 0, ..., M$. It is defined by

$$\vec{\boldsymbol{\eta}}_{\alpha+1/2} = \frac{\Big(\nabla_x z_{\alpha+\frac{1}{2}}, -1\Big)'}{\sqrt{1 + \Big|\nabla_x z_{\alpha+\frac{1}{2}}\Big|^2}}. \tag{3.2}$$

(v) We will denote by $\vec{\boldsymbol{n}}_{T,\alpha+\frac{1}{2}}$ the (space-time) unit normal vector to the interface $\Gamma_\alpha(t)$ outward to the layer $\Omega_\alpha(t)$ and for $\alpha = 0, ..., N$,

$$\vec{\boldsymbol{n}}_{T,\alpha+1/2} = \frac{\Big(\partial_t z_{\alpha+\frac{1}{2}}, \nabla_x z_{\alpha+\frac{1}{2}}, -1\Big)'}{\sqrt{1 + \Big|\nabla_x z_{\alpha+\frac{1}{2}}\Big|^2 + (\partial_t z_{\alpha-\frac{1}{2}})^2}}.$$

*Remark* 3.1. *Adding the time variable as one more dimension, the corresponding domain $\Omega_T$ is actually given by*

$$\begin{aligned}
&\Omega_T = \Big\{(t,x,z);\ t\in ]0,T] \ \textit{and}\ (x,z)\in\Omega_F(t)\Big\},\ \textit{with}\\
&\partial\Omega_T = \Lambda_T\cup\Lambda_1\cup\Lambda_2,\ \textit{where}\\
&\Lambda_T = \Big\{(t,x,z);\ t\in ]0,T] \ \textit{and}\ (x,z)\in\partial\Omega_F(t)\Big\},\\
&\Lambda_1 = \Big\{(0,x,z);\ (x,z)\in\Omega_F(0)\Big\},\\
&\Lambda_2 = \Big\{(T,x,z);\ (x,z)\in\Omega_F(T)\Big\}.
\end{aligned}$$

*Since we integrate over $\Omega_F(t)$, we retain here the boundary $\Lambda_T$ for the computations even if it means cancelling the tests functions over the boundaries $\Lambda_1$ and $\Lambda_2$.*

## 3.1 Weak solution with discontinuities

Let us recall the conditions to be verified by a piecewise smooth weak solution $(\vec{\boldsymbol{v}}_j, \phi_j, p)$ of (2.1)-(2.10). More precisely, let us suppose that the velocity $\vec{\boldsymbol{v}}_j$, the pressure $p$ and the volume fractions $\phi_j$ are smooth in each $\Omega_\alpha(t)$, but possibly discontinuous across the predetermined hypersurfaces $\Gamma_{\alpha+\frac{1}{2}}(t)$ for $\alpha = 1, ..., M-1$. Then the triplet $(\vec{\boldsymbol{v}}_j, \phi_j, p)$ for $j = 0, 1, \dots, N$ is a weak solution of (2.1)-(2.10) if the following conditions hold:

(i) $(\vec{\boldsymbol{v}}_j, \phi_j, p)$ is a standard weak solution of (2.1)-(2.10) in each layer $\Omega_\alpha(t)$.

(ii) $(\vec{\boldsymbol{v}}_j, \phi_j, p)$ satisfies the normal flux jump conditions at $\Gamma_{\alpha+\frac{1}{2}}(t)$, for $\alpha = 0, \dots, M$:

- For the mass conservation law,

$$[(\phi_j;\ \phi_j\vec{\boldsymbol{v}}_j)]_{|\Gamma_{\alpha+\frac{1}{2}}(t)} \cdot\ \vec{\boldsymbol{n}}_{T,\alpha+1/2} = 0, \tag{3.3}$$

- For the momentum conservation law (2.10),

$$\left[\left(\sum_{j=0}^{N}\rho_j\phi_j\vec{\boldsymbol{v}}_j;\ \sum_{j=0}^{N}\rho_j\phi_j\vec{\boldsymbol{v}}_j\otimes\vec{\boldsymbol{v}}_j - \boldsymbol{\Sigma}\right)\right]_{|\Gamma_{\alpha+\frac{1}{2}}(t)} \cdot\ \vec{\boldsymbol{n}}_{T,\alpha+1/2} = 0, \tag{3.4}$$

where $[(a;\ b)]_{|\Gamma_{\alpha+\frac{1}{2}}(t)}$ denotes the the jump of the pair $(a;\ b)$ across $\Gamma_{\alpha+\frac{1}{2}}(t)$,

$$[(a;\ b)]_{|\Gamma_{\alpha+\frac{1}{2}}(t)} = \Big((a;\ b)_{|\Omega_{\alpha+1}(t)} - (a;\ b)_{|\Omega_\alpha(t)}\Big)_{|\Gamma_{\alpha+\frac{1}{2}}(t)}$$

In order to develop the multilayer model, we adapt the preceding conditions to a particular class of triplets velocity-concentration-pressure: we assume the layers thicknesses small enough to neglect the dependence of the horizontal velocities, the concentrations and the pressure on the vertical variable inside each layer. Moreover, we assume that the vertical velocity is piecewise linear in $z$, and possibly discontinuous. Concretely, we set

$$(\vec{\boldsymbol{v}}_j)_{|\Omega_\alpha(t)} := \vec{\boldsymbol{v}}_{j,\alpha} := (\vec{\boldsymbol{u}}_\alpha, w_{j,\alpha})', \quad \phi_{j,\alpha} := (\phi_j)_{|\Omega_\alpha(t)}, \quad p_\alpha := p_{|\Omega_\alpha(t)}, \text{ for } j = 0, 1, \dots, N,$$

where

$$\vec{\boldsymbol{u}}_\alpha, \quad w_{j,\alpha} = w_\alpha + \delta w_{j,\alpha} \quad \text{and} \quad \phi_{j,\alpha},$$

respectively stand for the horizontal and vertical velocities and volumetric concentration of the species $j$ on layer $\alpha$. Let us also denote the averaged velocity at each lager by $\vec{\boldsymbol{v}}_\alpha$,

$$\vec{\boldsymbol{v}}_\alpha = \sum_{j=0}^{N} \phi_{j,\alpha}\vec{\boldsymbol{v}}_{j,\alpha} = (\vec{\boldsymbol{u}}_\alpha, w_\alpha),$$

and to assume that

$$\partial_z \vec{\boldsymbol{u}}_\alpha = 0, \quad \partial_z \phi_{j,\alpha} = 0, \quad \partial_z w_{j,\alpha} = d_{j,\alpha}(t,x), \quad \partial_z p_\alpha(t,x) = e_\alpha(t,x) \tag{3.5}$$

for some smooth functions $d_{j,\alpha}(t,x)$ and $e_\alpha(t,x)$. That is, we suppose that:

- the horizontal velocity $\vec{\boldsymbol{u}}_\alpha$ and the concentration of each of the species $\phi_{j,\alpha}$ do not depends on $z$ inside each layer,
- $w_{j,\alpha}$ and $p_\alpha$ are linear in $z$ inside each layer.

There is no hope for such a particular set $\left((\vec{\boldsymbol{u}}_\alpha, w_{j,\alpha})', \phi_{j,\alpha}, p_\alpha\right)$ to be a solution of the complete equations in the layer $\Omega_\alpha(t)$. Instead, we shall consider a reduced weak formulation with particular test functions, that we describe in Section 4.
Let us also denote $\Phi_\alpha = \{\phi_{j,\alpha}\}_{j=0}^N$ and

$$\rho(\Phi_\alpha) := \sum_{j=0}^{N} \rho_j \phi_{j,\alpha}.$$

#### 3.1.1 Mass conservation jump conditions

Remark that

$$\vec{\boldsymbol{u}}^+_{\alpha-\frac{1}{2}}(t,x) \;=\; \vec{\boldsymbol{u}}^-_{\alpha+\frac{1}{2}}(t,x) \;=\; \vec{\boldsymbol{u}}_\alpha(t,x), \quad \text{and} \quad \phi^+_{j,\alpha-\frac{1}{2}}(t,x) = \phi^-_{j,\alpha+\frac{1}{2}}(t,x) = \phi_{j,\alpha}(t,x). \tag{3.6}$$

Then mass conservation jump conditions are satisfied provided that

$$G_{j,\alpha+\frac{1}{2}} := G^-_{j,\alpha+\frac{1}{2}} = G^+_{j,\alpha+\frac{1}{2}}, \tag{3.7}$$

where

$$\left\{\begin{array}{l} G^+_{j,\alpha+\frac{1}{2}} = \phi_{j,\alpha+1}\left(\partial_t z_{\alpha+\frac{1}{2}} + \vec{\boldsymbol{u}}_{\alpha+1}\cdot\nabla_x z_{\alpha+\frac{1}{2}} - w^+_{j,\alpha+\frac{1}{2}}\right), \\ \\ G^-_{j,\alpha+\frac{1}{2}} = \phi_{j,\alpha}\left(\partial_t z_{\alpha+\frac{1}{2}} + \vec{\boldsymbol{u}}_{\alpha}\cdot\nabla_x z_{\alpha+\frac{1}{2}} - w^-_{j,\alpha+\frac{1}{2}}\right) \end{array}\right. \tag{3.8}$$

for $j = 0, 1, \ldots, N$.
Remark that $G_{j,\alpha+\frac{1}{2}}$ is the normal mass flux for the species $j$ at the interface $\Gamma_{\alpha+\frac{1}{2}}(t)$.
Taking into account the structure of the vertical velocity (2.3), let us denote

$$w^\pm_{j,\alpha+\frac{1}{2}} = w^\pm_{\alpha+\frac{1}{2}} + \delta w^\pm_{j,\alpha+\frac{1}{2}}. \tag{3.9}$$

where $\delta w^\pm_{j,\alpha+\frac{1}{2}}$ must verify that

$$\sum_{j=0}^{N} \phi_{j,\alpha+1}\delta w^+_{j,\alpha+\frac{1}{2}} = \sum_{j=0}^{N} \phi_{j,\alpha}\delta w^-_{j,\alpha+\frac{1}{2}} = 0.$$

It is now clear, adding up in $j$, that

$$G_{\alpha+\frac{1}{2}} := G^-_{\alpha+\frac{1}{2}} = G^+_{\alpha+\frac{1}{2}}, \quad \text{where} \quad G_{\alpha+\frac{1}{2}} = \sum_{j=0}^{N} G_{j,\alpha+\frac{1}{2}} \tag{3.10}$$

and

$$\begin{cases} G^+_{\alpha+\frac{1}{2}} = \partial_t z_{\alpha+\frac{1}{2}} + \vec{\boldsymbol{u}}_{\alpha+1} \cdot \nabla_x z_{\alpha+\frac{1}{2}} - w^+_{\alpha+\frac{1}{2}}, \\ \\ G^-_{\alpha+\frac{1}{2}} = \partial_t z_{\alpha+\frac{1}{2}} + \vec{\boldsymbol{u}}_{\alpha} \cdot \nabla_x z_{\alpha+\frac{1}{2}} - w^-_{\alpha+\frac{1}{2}}, \end{cases} \tag{3.11}$$

which corresponds to the jump condition for the equation $\nabla \cdot \vec{\boldsymbol{v}} = 0$.
Then, from (3.8), (3.9) and (3.11) we have that

$$G^+_{j,\alpha+\frac{1}{2}} = \phi_{j,\alpha+1}(G_{\alpha+\frac{1}{2}} - \delta w^+_{j,\alpha+\frac{1}{2}}), \quad G^-_{j,\alpha+\frac{1}{2}} = \phi_{j,\alpha}(G_{\alpha+\frac{1}{2}} - \delta w^-_{j,\alpha+\frac{1}{2}}). \tag{3.12}$$

This gives

$$\begin{aligned} G_{j,\alpha+\frac{1}{2}} &= < \phi_j(G - \delta w_j) >_{\alpha+\frac{1}{2}} \\ &= \frac{\phi_{j,\alpha+1} + \phi_{j,\alpha}}{2} G_{\alpha+\frac{1}{2}} - \frac{\phi_{j,\alpha+1}\delta w^+_{j,\alpha+\frac{1}{2}} + \phi_{j,\alpha}\delta w^-_{j,\alpha+\frac{1}{2}}}{2}. \end{aligned} \tag{3.13}$$

As a consequence,

$$\begin{aligned} \sum_{j=0}^{N} \rho_j G_{j,\alpha+\frac{1}{2}} &= < \rho(\Phi) G + \sum_{j=0}^{N} \rho_j \delta w_j >_{\alpha+\frac{1}{2}} \\ &= \frac{\rho(\Phi_{\alpha+1}) + \rho(\Phi_\alpha)}{2} G_{\alpha+\frac{1}{2}} - \sum_{j=0}^{N} \rho_j \frac{\phi_{j,\alpha+1}\delta w^+_{j,\alpha+\frac{1}{2}} + \phi_{j,\alpha}\delta w^-_{j,\alpha+\frac{1}{2}}}{2}. \end{aligned} \tag{3.14}$$

Let us also remark that condition (3.7) can be written as

$$\phi_{j,\alpha+1}\delta w^+_{j,\alpha+\frac{1}{2}} - \phi_{j,\alpha}\delta w^-_{j,\alpha+\frac{1}{2}} = G_{\alpha+\frac{1}{2}}(\phi_{j,\alpha+1} - \phi_{j,\alpha}). \tag{3.15}$$

### 3.1.2 Momentum conservation jump conditions

From the momentum jump condition (3.4) we have

$$\left[ \left( \sum_{j=0}^{N} \rho_j \phi_j \vec{\boldsymbol{v}}_j;\ \sum_{j=0}^{N} \rho_j \phi_j \vec{\boldsymbol{v}}_j \otimes \vec{\boldsymbol{v}}_j - \boldsymbol{\Sigma} \right) \right]_{|\Gamma_{\alpha+\frac{1}{2}}(t)} \cdot \left( \partial_t z_{\alpha+\frac{1}{2}}, \nabla_x z_{\alpha+\frac{1}{2}}, -1 \right) = 0.$$

Which can also be written as

$$\left[ \boldsymbol{\Sigma} \right]_{|\Gamma_{\alpha+\frac{1}{2}}(t)} \cdot \left( \nabla_x z_{\alpha+\frac{1}{2}}, -1 \right) = \sum_{j=0}^{N} \left[ \left( \rho_j \phi_j \vec{\boldsymbol{v}}_j;\ \rho_j \phi_j \vec{\boldsymbol{v}}_j \otimes \vec{\boldsymbol{v}}_j \right) \right]_{|\Gamma_{\alpha+\frac{1}{2}}(t)} \cdot \left( \partial_t z_{\alpha+\frac{1}{2}}, \nabla_x z_{\alpha+\frac{1}{2}}, -1 \right).$$

Moreover, using (3.7), we have

$$\left[ \left( \rho_j \phi_j \vec{\boldsymbol{v}}_j;\ \rho_j \phi_j \vec{\boldsymbol{v}}_j \otimes \vec{\boldsymbol{v}}_j \right) \right]_{|\Gamma_{\alpha+\frac{1}{2}}(t)} \cdot \left( \partial_t z_{\alpha+\frac{1}{2}}, \nabla_x z_{\alpha+\frac{1}{2}}, -1 \right) = \rho_j\, G_{j,\alpha+\frac{1}{2}}\ [\vec{\boldsymbol{v}}_j]_{|\Gamma_{\alpha+\frac{1}{2}}(t)}.$$

Then, we have that

$$\left[ \boldsymbol{\Sigma} \right]_{|\Gamma_{\alpha+\frac{1}{2}}(t)} \cdot \left( \nabla_x z_{\alpha+\frac{1}{2}}, -1 \right) = \sum_{j=0}^{N} \rho_j\, G_{j,\alpha+\frac{1}{2}}\ [\vec{\boldsymbol{v}}_j]_{|\Gamma_{\alpha+\frac{1}{2}}(t)}.$$

As a consequence, condition (3.4) can be written as

$$\Big[\mathbf{\Sigma}\Big]_{|\Gamma_{\alpha+\frac{1}{2}}(t)}\cdot\vec{\boldsymbol{\eta}}_{\alpha+\frac{1}{2}}=\frac{1}{\sqrt{1+\left|\nabla_x z_{\alpha+\frac{1}{2}}\right|^2}}\sum_{j=0}^{N}\rho_j\,G_{j,\alpha+\frac{1}{2}}\,[\vec{\boldsymbol{v}}_j]_{|\Gamma_{\alpha+\frac{1}{2}}(t)}\,. \tag{3.16}$$

For $\alpha=1,...,N-1$, the total stress writes

$$\mathbf{\Sigma}^{\pm}_{\alpha+\frac{1}{2}}\;=\;-p_{\alpha+\frac{1}{2}}\,\boldsymbol{I}+T^{\pm}_{\alpha+\frac{1}{2}}, \tag{3.17}$$

where $p_{\alpha+\frac{1}{2}}$ is the kinematic pressure and $T^{\pm}_{\alpha+\frac{1}{2}}$ are the limit approximations of $T(\vec{\boldsymbol{v}})$ at $\Gamma_{\alpha+\frac{1}{2}}$. This means that $T^{\pm}_{\alpha+\frac{1}{2}}$ must verify

$$\left(T^{+}_{\alpha+\frac{1}{2}}-T^{-}_{\alpha+\frac{1}{2}}\right)\cdot\vec{\boldsymbol{\eta}}_{\alpha+\frac{1}{2}}=\frac{1}{\sqrt{1+\left|\nabla_x z_{\alpha+\frac{1}{2}}\right|^2}}\sum_{j=0}^{N}\rho_j\,G_{j,\alpha+\frac{1}{2}}\,[\vec{\boldsymbol{v}}_j]_{|\Gamma_{\alpha+\frac{1}{2}}(t)}\,. \tag{3.18}$$

where $G_{j,\alpha+\frac{1}{2}}$ is defined by (3.13).
Moreover, by consistency, $T^{\pm}_{\alpha+\frac{1}{2}}$ should verify

$$\frac{1}{2}\Big(T^{+}_{\alpha+\frac{1}{2}}+T^{-}_{\alpha+\frac{1}{2}}\Big)=\widetilde{T}_{\alpha+\frac{1}{2}}=\left(\begin{array}{c|c}\widetilde{T}_{H,\alpha+\frac{1}{2}} & \widetilde{T}_{xz,\alpha+\frac{1}{2}}\\ \hline \widetilde{T}'_{xz,\alpha+\frac{1}{2}} & \widetilde{T}_{zz,\alpha+\frac{1}{2}}\end{array}\right), \tag{3.19}$$

where $\widetilde{T}_{\alpha+\frac{1}{2}}$ is an approximation of $T(\vec{\boldsymbol{v}})_{|\Gamma_{\alpha+\frac{1}{2}}}$, to be defined.
Concretely, if we set $T(\vec{\boldsymbol{v}})=\mu D(\vec{\boldsymbol{v}})=\mu(\nabla\vec{\boldsymbol{v}}+(\nabla\vec{\boldsymbol{v}})')$, then we define

$$\widetilde{T}_{\alpha+\frac{1}{2}}=\mu\widetilde{D}_{\alpha+\frac{1}{2}}=\mu\left(\begin{array}{cc} D_H\left(\dfrac{\vec{\boldsymbol{u}}^{+}_{H,\alpha+\frac{1}{2}}+\vec{\boldsymbol{u}}^{-}_{H,\alpha+\frac{1}{2}}}{2}\right) & \left(\nabla_x\left(\frac{w^{+}_{\alpha+\frac{1}{2}}+w^{-}_{\alpha+\frac{1}{2}}}{2}\right)\right)'+\vec{\mathbf{Q}}_{H,\alpha+\frac{1}{2}}\\ \nabla_x\left(\frac{w^{+}_{\alpha+\frac{1}{2}}+w^{-}_{\alpha+\frac{1}{2}}}{2}\right)+(\vec{\mathbf{Q}}_{H,\alpha+\frac{1}{2}})' & 2\,Q_{v,\alpha+\frac{1}{2}}\end{array}\right), \tag{3.20}$$

where $\vec{\mathbf{Q}}_{\alpha+\frac{1}{2}}=\vec{\mathbf{Q}}(\widetilde{\vec{\boldsymbol{u}}})$ at $\Gamma_{\alpha+\frac{1}{2}}$ and $\vec{\mathbf{Q}}$ satisfies the equation

$$\vec{\mathbf{Q}}-\partial_z\vec{\boldsymbol{v}}=0,\quad\text{with}\quad\vec{\mathbf{Q}}=(\vec{\mathbf{Q}}_H,Q_v). \tag{3.21}$$

To approximate $\vec{\mathbf{Q}}$, solution of (3.21), we approximate $\vec{\boldsymbol{v}}$ by $\widetilde{\vec{\boldsymbol{u}}}$, that is a linear interpolation in $z$, such that $\widetilde{\vec{\boldsymbol{u}}}_{|z=\frac{1}{2}(z_{\alpha-\frac{1}{2}}+z_{\alpha+\frac{1}{2}})}=\vec{\boldsymbol{u}}_\alpha$.
Finally, we can solve the system defined by (3.18) and the equation resulting to multiply scalarly (3.19) by vector $\vec{\boldsymbol{\eta}}_{\alpha+\frac{1}{2}}$. This way, we obtain the expression of $T^{\pm}_{\alpha+\frac{1}{2}}$, that verifies the jump condition and the consistency condition on the interface. We can solve it easily and we obtain

$$T^{\pm}_{\alpha+\frac{1}{2}}\cdot\vec{\boldsymbol{\eta}}_{\alpha+\frac{1}{2}}=\;\widetilde{T}_{\alpha+\frac{1}{2}}\cdot\vec{\boldsymbol{\eta}}_{\alpha+\frac{1}{2}}\pm\frac{1}{2}\frac{1}{\sqrt{1+\left|\nabla_x z_{\alpha+\frac{1}{2}}\right|^2}}\sum_{j=0}^{N}\rho_j\,G_{j,\alpha+\frac{1}{2}}\,[\vec{\boldsymbol{v}}_j]_{|\Gamma_{\alpha+\frac{1}{2}}(t)}\,. \tag{3.22}$$

## 3.2 Vertical velocity

In this subsection we show how the vertical velocities $w_\alpha$ and $w_{j,\alpha}$ are defined for each layer.

Fist, let us notice that, as $\vec{\boldsymbol{u}}_\alpha$ is a classic solution of the equations in $\Omega_\alpha(t)$, for $z \in ]z_{\alpha-\frac{1}{2}}, z_{\alpha+\frac{1}{2}}[$, the vertical integration of the incompressibility equation leads to the equality

$$w_\alpha(t,x,z) \;=\; w^+_{\alpha-\frac{1}{2}}(t,x) \;-\; (z - z_{\alpha-\frac{1}{2}})\nabla_x \cdot \vec{\boldsymbol{u}}_\alpha(t,x), \quad \text{for } \alpha = 1, ..., N.$$

In addition, from the condition (3.11) at the interfaces, we express the quantities

$$w^+_{\alpha-\frac{1}{2}} \;=\; (\vec{\boldsymbol{u}}_\alpha - \vec{\boldsymbol{u}}_{\alpha-1}) \cdot \nabla_x z_{\alpha-\frac{1}{2}} \;+\; w^-_{\alpha-\frac{1}{2}}. \tag{3.23}$$

where

$$w^-_{\alpha-\frac{1}{2}} = w_{\alpha-1}|_{\Gamma_{\alpha-\frac{1}{2}}(t)} = w^+_{\alpha-\frac{3}{2}} - h_{\alpha-1}\nabla_x \cdot \vec{\boldsymbol{u}}_{H,\alpha-1}. \tag{3.24}$$

Using the horizontal velocities drawn from the model, the averaged vertical velocities in the layers are computed using the following algorithm:

- The quantity $w^+_{\frac{1}{2}}$ is determined, from the given mass transference $G_{\frac{1}{2}}$, through the condition (3.7) at the bottom by
$$w^+_{\frac{1}{2}} = \vec{\boldsymbol{u}}_{H,1} \cdot \nabla_x z_B + \partial_t z_B - G_{\frac{1}{2}}.$$
- Then, for $\alpha = 1, ..., N$ and $z \in ]z_{\alpha-\frac{1}{2}}, z_{\alpha+\frac{1}{2}}[$, we set

$$\begin{aligned} w_\alpha(t,x,z) \;&=\; w^+_{\alpha-\frac{1}{2}}(t,x) \;-\; (z - z_{\alpha-\frac{1}{2}})\nabla_x \cdot \vec{\boldsymbol{u}}_{H,\alpha}(t,x), \\ w^-_{\alpha+\frac{1}{2}} &= w_\alpha|_{\Gamma_{\alpha+\frac{1}{2}}(t)} = w^+_{\alpha-\frac{1}{2}} - h_\alpha \nabla_x \cdot \vec{\boldsymbol{u}}_{H,\alpha}, \\ w^+_{\alpha+\frac{1}{2}} \;&=\; (\vec{\boldsymbol{u}}_{\alpha+1} - \vec{\boldsymbol{u}}_\alpha) \cdot \nabla_x z_{\alpha+\frac{1}{2}} \;+\; w^-_{\alpha+\frac{1}{2}}. \end{aligned} \tag{3.25}$$

### 3.2.1 Vertical velocity of the sediment species

The vertical velocity of the sediment species $j$, for $j = 1, \ldots, N$, inside the layer $\Omega_\alpha$ is defined by

$$w_{j,\alpha}(t,x,z) = w_\alpha(t,x,z) + \delta w_{j,\alpha}(z), \tag{3.26}$$

where $w_\alpha$ is defined by (3.25). Moreover, by assuming a linear profile of $\delta w_{j,\alpha}(z)$, it can be defined in terms of the limits in the layer, that is, in terms of $\delta w^-_{j,\alpha+\frac{1}{2}}$ and $\delta w^+_{j,\alpha-\frac{1}{2}}$. Concretely,

$$\delta w_{j,\alpha}(z) = \delta w^+_{j,\alpha-\frac{1}{2}} + \frac{\delta w^-_{j,\alpha+\frac{1}{2}} - \delta w^+_{j,\alpha-\frac{1}{2}}}{h_\alpha}(z - z_{\alpha-\frac{1}{2}}).$$

Remember that $\delta w^-_{j,\alpha+\frac{1}{2}}$ and $\delta w^+_{j,\alpha+\frac{1}{2}}$ verifies (3.15). Then we have

$$\left\{ \begin{aligned} &\phi_{j,\alpha+1}\delta w^+_{j,\alpha+\frac{1}{2}} - \phi_{j,\alpha}\delta w^-_{j,\alpha+\frac{1}{2}} = (\phi_{j,\alpha+1} - \phi_{j,\alpha})G_{\alpha+\frac{1}{2}}, \\ &\phi_{j,\alpha+1}\delta w^+_{j,\alpha+\frac{1}{2}} + \phi_{j,\alpha}\delta w^-_{j,\alpha+\frac{1}{2}} = 2 < \phi_j \delta w_j >_{\alpha+\frac{1}{2}} . \end{aligned} \right.$$

Thus, we obtain

$$\begin{aligned} \phi_{j,\alpha+1}\delta w^+_{j,\alpha+\frac{1}{2}} \quad &= \quad < \phi_j \delta w_j >_{\alpha+\frac{1}{2}} + \frac{1}{2}(\phi_{j,\alpha+1} - \phi_{j,\alpha})G_{\alpha+\frac{1}{2}}, \\ \phi_{j,\alpha}\,\delta w^-_{j,\alpha+\frac{1}{2}} \quad &= \quad < \phi_j \delta w_j >_{\alpha+\frac{1}{2}} - \frac{1}{2}(\phi_{j,\alpha+1} - \phi_{j,\alpha})G_{\alpha+\frac{1}{2}}. \end{aligned} \tag{3.27}$$

This means that assuming that some approximation of the term $< \phi_j \delta w_j >_{\alpha+\frac{1}{2}}$ is given, then the vertical velocities of each sediment species in the layers can be computed using the following algorithm:

- $w^+_{j,\frac{1}{2}} = w^+_{\frac{1}{2}}$, that is $\delta w^+_{j,\frac{1}{2}} = 0$. It corresponds to consider that the bottom $z = z_{\frac{1}{2}}$ determines the limit of a saturated sediment bottom.

- Then, for $\alpha = 1, ..., N$ and $z \in ]z_{\alpha-\frac{1}{2}}, z_{\alpha+\frac{1}{2}}[$, we set

$$w_{j,\alpha}(t,x,z) \;=\; w_\alpha(t,x,z) + \delta w^+_{j,\alpha-\frac{1}{2}} + \frac{\delta w^-_{j,\alpha+\frac{1}{2}} - \delta w^+_{j,\alpha-\frac{1}{2}}}{h_\alpha}(z - z_{\alpha-\frac{1}{2}}). \tag{3.28}$$

where $w_\alpha(t,x,z)$ is defined by (3.25), and $\delta w^+_{j,\alpha+\frac{1}{2}}$, $\delta w^-_{j,\alpha+\frac{1}{2}}$ by (3.27).

# 4 A particular weak solution with hydrostatic pressure

In this Section we finish the construction of the model under the hypothesis of hydrostatic pressure. This means that

$$p_\alpha(t,x,z) = p_{\alpha+\frac{1}{2}}(t,x) \;+\; \rho(\Phi_\alpha) g (z_{\alpha+\frac{1}{2}} - z), \tag{4.1}$$

with

$$p_{\alpha+\frac{1}{2}}(t,x) = p_S(t,x) + g \sum_{\beta=\alpha+1}^{M} \rho(\Phi_\beta) h_\beta(t,x). \tag{4.2}$$

Here, the component $p_{\alpha+\frac{1}{2}}$ is the kinematic pressure at $\Gamma_{\alpha+\frac{1}{2}}(t)$ and $p_S$ denotes the pressure at the free surface. Then, the unknowns of the systems are the layer depths and the horizontal velocities. As $\vec{\boldsymbol{v}}_{j,\alpha}$ is a weak solution of the equations (2.1) - (2.10) in $\Omega_\alpha(t)$, let us begin by considering the weak formulation of this system in $\Omega_\alpha(t)$ for $\alpha = 1, ..., N$. Assuming $\vec{\boldsymbol{v}}_\alpha \in L^2(0,T; H^1(\Omega_\alpha(t))^3)$, $\partial_t \vec{\boldsymbol{v}}_\alpha \in L^2(0,T; L^2(\Omega_\alpha(t))^3)$ and $p_\alpha \in L^2(0,T; L^2(\Omega_\alpha(t)))$, a weak solution of the original equations in $\Omega_\alpha(t)$ should verify

$$\left\{ \begin{aligned} & \int_{\Omega_\alpha(t)} \left(\partial_t \phi_{j,\alpha} + \nabla \cdot (\phi_{j,\alpha} \vec{\boldsymbol{v}}_{j,\alpha})\right) \varphi \, d\Omega = 0, \\ & \int_{\Omega_\alpha(t)} \sum_{j=0}^{N} \rho_j \partial_t (\phi_{j,\alpha} \vec{\boldsymbol{v}}_{j\alpha}) \cdot \vec{\boldsymbol{\vartheta}} \, d\Omega \quad + \int_{\Omega_\alpha(t)} \sum_{j=0}^{N} \rho_j \nabla \cdot \left(\phi_{j,\alpha} \vec{\boldsymbol{v}}_{j,\alpha} \otimes \vec{\boldsymbol{v}}_{j,\alpha}\right) \cdot \vec{\boldsymbol{\vartheta}} \, d\Omega \\ & \qquad + \int_{\Omega_\alpha(t)} T^E : \nabla \vec{\boldsymbol{\vartheta}} \, d\Omega - \int_{\Omega_\alpha(t)} p \nabla \cdot \vec{\boldsymbol{\vartheta}} \, d\Omega \\ & \qquad + \int_{\Gamma_{\alpha+\frac{1}{2}}(t)} (\boldsymbol{\Sigma}^-_{\alpha+\frac{1}{2}} \vec{\boldsymbol{\vartheta}}) \cdot \vec{\boldsymbol{\eta}}_{\alpha+\frac{1}{2}} d\Gamma \\ & \qquad - \int_{\Gamma_{\alpha-\frac{1}{2}}(t)} (\boldsymbol{\Sigma}^+_{\alpha-\frac{1}{2}} \vec{\boldsymbol{\vartheta}}) \cdot \vec{\boldsymbol{\eta}}_{\alpha-\frac{1}{2}} d\Gamma \\ & \qquad = - \int_{\Omega_\alpha(t)} g \, \rho(\Phi_\alpha) \, \vec{\boldsymbol{k}} \cdot \vec{\boldsymbol{\vartheta}} \, d\Omega. \end{aligned} \right. \tag{4.3}$$

for all $\varphi \in L^2(\Omega_\alpha(t))$ and for all $\vec{\boldsymbol{\vartheta}} \in H^1(\Omega_\alpha(t))^3$ with $\vec{\boldsymbol{\vartheta}}|_{\partial I_F} = 0$.
We consider velocity-pressure pairs with the structure given by (3.5), that satisfy the previous system with particular weak solutions that verify (4.3) for test functions such that $\partial_z \varphi = 0$ and

$$\vec{\boldsymbol{\vartheta}}(t,x,z) = \left(\vec{\boldsymbol{\vartheta}}_H(t,x),\; (z - z_B) \, V(t,x)\right)', \tag{4.4}$$

where $\vartheta$ and $V(t,x)$ are smooth functions that do no depend on $z$.
Following a similar approach as in [11], after some easy calculations we get

□ *Mass conservation law*

$$\partial_t (\phi_{j,\alpha} h_\alpha) + \nabla_x \cdot (\phi_{j,\alpha} h_\alpha \vec{\boldsymbol{u}}_\alpha) = G_{j,\alpha+\frac{1}{2}} - G_{j,\alpha-\frac{1}{2}}, \quad j = 0, 1, \ldots, N, \quad \alpha = 0, ..., M, \tag{4.5}$$

where $G_{j,\alpha+\frac{1}{2}}$ is defined by (3.13).

Remark that taking into account $\displaystyle\sum_{j=0}^{N}\phi_{j,\alpha}=1$, we get that

$$\partial_t h_\alpha + \nabla_x\cdot(h_\alpha \vec{\boldsymbol{u}}_\alpha) = G_{\alpha+\frac{1}{2}} - G_{\alpha-\frac{1}{2}},\quad \alpha=0,...,M. \tag{4.6}$$

Moreover, combining (4.5) with (4.6) and using (3.12) we have

$$h_\alpha\partial_t\phi_{j,\alpha} + h_\alpha\vec{\boldsymbol{u}}_\alpha\nabla_x\phi_{j,\alpha} = -\phi_{j,\alpha}(\delta w^-_{j,\alpha+\frac{1}{2}} + \delta w^+_{j,\alpha-\frac{1}{2}}),\quad \alpha=0,...,M. \tag{4.7}$$

□ *Momentum conservation.*

$$\begin{aligned}
&\partial_t(\rho(\Phi_\alpha)h_\alpha\vec{\boldsymbol{u}}_\alpha) + \nabla_x\cdot\Big(\rho(\Phi_\alpha)h_\alpha\vec{\boldsymbol{u}}_\alpha\otimes\vec{\boldsymbol{u}}_\alpha\Big) + \int_{z_{\alpha-\frac{1}{2}}}^{z_{\alpha+\frac{1}{2}}}\nabla_x p_\alpha dz - \nabla_x\cdot(h_\alpha T_H)\\
&+(\widetilde{T}_{H,\alpha+\frac{1}{2}}(\nabla_x z_{\alpha+\frac{1}{2}})' - \widetilde{T}_{xz,\alpha+\frac{1}{2}}) - (\widetilde{T}_{H,\alpha-\frac{1}{2}}(\nabla_x z_{\alpha-\frac{1}{2}})' - \widetilde{T}^+_{xz,\alpha-\frac{1}{2}})\\
&= \frac{\vec{\boldsymbol{u}}_{\alpha+1}+\vec{\boldsymbol{u}}_\alpha}{2}\sum_{j=0}^{N}\rho_j G_{j,\alpha+\frac{1}{2}} - \frac{\vec{\boldsymbol{u}}_\alpha+\vec{\boldsymbol{u}}_{\alpha-1}}{2}\sum_{j=0}^{N}\rho_j G_{j,\alpha-\frac{1}{2}}
\end{aligned} \tag{4.8}$$

Let us introduce the following notation,

- $\displaystyle \overline{p}_\alpha = p_S + g\sum_{\beta=\alpha+1}^{M}\rho(\Phi_\beta)h_\beta + g\rho(\Phi_\alpha)\frac{h_\alpha}{2}$,
- $\displaystyle \bar{z}_\alpha = z_b + \sum_{\beta=1}^{\alpha-1}h_b + \frac{h_\alpha}{2}$.

Then, we obtain the following system for $\alpha=1,...,M$,

$$\left\{\begin{aligned}
&\partial_t(\phi_{j,\alpha}h_\alpha) + \nabla_x\cdot(\phi_{j,\alpha}h_\alpha\vec{\boldsymbol{u}}_\alpha) \;=\; <\phi_j(G-\delta w_j)>_{\alpha+\frac{1}{2}} - <\phi_j(G-\delta w_j)>_{\alpha-\frac{1}{2}} \text{ for } j=0,\dots,N.\\
&\partial_t(\rho(\Phi_\alpha)h_\alpha\vec{\boldsymbol{u}}_\alpha) + \nabla_x\cdot\Big(\rho(\Phi_\alpha)h_\alpha\vec{\boldsymbol{u}}_\alpha\otimes\vec{\boldsymbol{u}}_\alpha\Big) + h_\alpha\Bigg(\nabla_x\overline{p}_\alpha + g\rho(\Phi_\alpha)\nabla_x\bar{z}_\alpha\Bigg) - \nabla_x\cdot(h_\alpha T_H)\\
&\quad+(\widetilde{T}_{H,\alpha+\frac{1}{2}}(\nabla_x z_{\alpha+\frac{1}{2}})' - \widetilde{T}_{xz,\alpha+\frac{1}{2}}) - (\widetilde{T}_{H,\alpha-\frac{1}{2}}(\nabla_x z_{\alpha-\frac{1}{2}})' - \widetilde{T}_{xz,\alpha-\frac{1}{2}})\\
&\quad= \frac{\vec{\boldsymbol{u}}_{\alpha+1}+\vec{\boldsymbol{u}}_\alpha}{2}\sum_{j=0}^{N}\rho_j <\phi_j(G-\delta w_j)>_{\alpha+\frac{1}{2}} - \frac{\vec{\boldsymbol{u}}_\alpha+\vec{\boldsymbol{u}}_{\alpha-1}}{2}\sum_{j=0}^{N}\rho_j <\phi_j(G-\delta w_j)>_{\alpha-\frac{1}{2}}.
\end{aligned}\right. \tag{4.9}$$

where $\widetilde{T}_{H,\alpha+\frac{1}{2}}$ and $\widetilde{T}_{xz,\alpha-\frac{1}{2}}$ are defined by (3.19),

$$<\phi_j(G-\delta w_j)>_{\alpha+\frac{1}{2}} = \frac{\phi_{j,\alpha+1}+\phi_{j,\alpha}}{2}(G_{\alpha+\frac{1}{2}} - \delta w_{j,\alpha+\frac{1}{2}}) \tag{4.10}$$

and

$$\frac{\vec{\boldsymbol{u}}_{\alpha+1}+\vec{\boldsymbol{u}}_\alpha}{2}\sum_{j=0}^{N}\rho_j <\phi_j(G-\delta w_j)>_{\alpha+\frac{1}{2}} = \frac{\vec{\boldsymbol{u}}_{\alpha+1}+\vec{\boldsymbol{u}}_\alpha}{2}\Bigg(\sum_{j=0}^{N}\rho_j\frac{\phi_{j,\alpha}+\phi_{j,\alpha+1}}{2}(G_{\alpha+\frac{1}{2}} - \delta w_{j,\alpha+\frac{1}{2}})\Bigg). \tag{4.11}$$

In Appendix A, we propose another definition of terms (4.10) and (4.11), that could be seen as an upwind approximation of the ones given previously.

## 5 Closure and reformulation of the model

For the sake of simplicity, we consider here only a one-dimensional horizontal space and, in the sequel, we shall denote the horizontal velocities $\vec{\boldsymbol{u}}_\alpha$ merely by $u_\alpha$.

**Assumption 1.** *We consider layers having thickness proportional to the total height. That is for* $\alpha = 1, \dots, M,\ h_\alpha = l_\alpha\, h$ *with* $l_\alpha$ *a positive constant. Hence we have*

$$\sum_{\alpha=1}^{M} l_\alpha = 1. \tag{5.1}$$

From Assumption 1, summing the equations (4.5) up to $\alpha = 1, ..., M$, yields

$$G_{\alpha+\frac{1}{2}} - G_{\frac{1}{2}} = \sum_{\beta=1}^{\alpha} \left(\partial_t h_\beta + \partial_x(h_\beta u_\beta)\right) \tag{5.2}$$

and for the particular value $\alpha = M$, we get the global continuity equation

$$\partial_t h + \partial_x \left( h \sum_{\beta=1}^{M} l_\beta u_\beta \right) = G_{M+1/2} - G_{\frac{1}{2}}. \tag{5.3}$$

Now, from (5.2), assuming $G_{M+\frac{1}{2}} = 0$, and using the global continuity equation we get

$$\begin{aligned} G_{\alpha+\frac{1}{2}} &= G_{\frac{1}{2}} + \sum_{\beta=1}^{\alpha} l_\beta \big(\partial_t h + \partial_x(h u_\beta)\big) \\ &= G_{\frac{1}{2}} + \sum_{\beta=1}^{\alpha} l_\beta \left( \partial_x \left(h u_\beta\right) - \sum_{\gamma=1}^{M} \partial_x \left(l_\gamma h u_\gamma\right) - G_{\frac{1}{2}} \right). \end{aligned}$$

Therefore we can set

$$G_{\alpha+\frac{1}{2}} = (1 - L_\alpha)\ G_{\frac{1}{2}} + \sum_{\gamma=1}^{M} \xi_{\alpha,\gamma} \partial_x(h u_\gamma), \quad \alpha = 1, \dots, M, \tag{5.4}$$

where for $\alpha,\ \gamma \in \{1, \dots, M\}$, we define $L_\alpha := l_1 + \cdots + l_\alpha$ and

$$\xi_{\alpha,\gamma} := \sum_{\beta=1}^{\alpha} (\delta_{\beta\gamma} - l_\beta) l_\gamma = \begin{cases} \big(1 - (l_1 + \cdots + l_\alpha)\big) l_\gamma & \text{if } \gamma \le \alpha, \\ \\ -(l_1 + \cdots + l_\alpha) l_\gamma & \text{otherwise,} \end{cases}$$

$\delta_{\beta\gamma}$ being the standard Kronecker symbol. Thus, we explicitly obtain the mass transference across interfaces in terms of the velocities at each layer.

*Remark* 5.1. *In light of* (5.1) *we have* $\xi_{M,\gamma} = 0$ *for all* $\gamma = 1, \dots, M$. *In addition, setting* $\xi_{0,\gamma} = 0$ *for all* $\gamma = 1, \dots, M$, *we notice that* $\xi_{\alpha,\gamma} = \xi_{\alpha-1,\gamma} + (\delta_{\alpha\gamma} - l_\alpha) l_\gamma$ *for all* $\alpha, \gamma = 1, \dots, M$.

Let us introduce the notation

$$r_{j,\alpha} = \phi_{j,\alpha} h, \quad q_\alpha = \rho(\Phi_\alpha) h u_\alpha, \quad \text{for } \alpha = 1, \dots, M, \quad j = 1, \dots, N, \tag{5.5}$$

and

$$m_\alpha \equiv m_\alpha(h, r_{1,\alpha}, \dots, r_{N,\alpha}) = \rho(\Phi_\alpha) h = \rho_0 \left( h + \sum_{j=1}^{N} \left( \frac{\rho_j}{\rho_0} - 1 \right) r_{j,\alpha} \right). \tag{5.6}$$

Then system (4.9) reduces to the equations for variables $h, r_{j,\alpha}, q_\alpha$ given by

$$\partial_t h + \partial_x \left( \sum_{\beta=1}^{M} \left( h l_\beta \frac{q_\beta}{m_\beta} \right) \right) = G_{M+1/2} - G_{1/2} \tag{5.7}$$

$$\begin{aligned} \partial_t (r_{j,\alpha}) + \partial_x \left( \frac{r_{j,\alpha} q_\alpha}{m_\alpha} \right) &= \frac{1}{l_\alpha} \left\{ \frac{r_{j,\alpha+1} + r_{j,\alpha}}{2h} G_{\alpha+\frac{1}{2}} - \frac{r_{j,\alpha} + r_{j,\alpha-1}}{2h} G_{\alpha-\frac{1}{2}} \right\} \\ - \frac{1}{l_\alpha} \left\{ < \phi_j \delta w_j >_{\alpha+\frac{1}{2}} - < \phi_j \delta w_j >_{\alpha-\frac{1}{2}} \right\}, & \qquad \text{for } j = 1, \ldots, N, \end{aligned} \tag{5.8}$$

and

$$\begin{aligned} &\partial_t q_\alpha + \partial_x \left( \frac{q_\alpha^2}{m_\alpha} + h \left( p_S + \frac{g}{2} l_\alpha m_\alpha + g \sum_{\beta=\alpha+1}^{M} l_\beta m_\beta \right) \right) \\ &= \left( p_S + g \sum_{\beta=\alpha+1}^{M} l_\beta m_\beta \right) \partial_x h - g m_\alpha \partial_x z_b - g m_\alpha L_{\alpha-1} \partial_x h \\ &+ \frac{1}{l_\alpha} \left\{ \left( \frac{1}{2} \frac{q_{\alpha+1}}{m_{\alpha+1}} + \frac{q_\alpha}{m_\alpha} \right) \frac{m_{\alpha+1} + m_\alpha}{2h} G_{\alpha+1/2} - \frac{1}{2} \left( \frac{q_\alpha}{m_\alpha} + \frac{q_{\alpha-1}}{m_{\alpha-1}} \right) \frac{m_\alpha + m_{\alpha-1}}{2h} G_{\alpha-1/2} \right\} \\ &- \frac{1}{l_\alpha} \left\{ \frac{1}{2} \left( \frac{q_{\alpha+1}}{m_{a+1}} + \frac{q_\alpha}{m_a} \right) < \sum_{j=0}^{N} \rho_j \phi_j \delta w_j >_{\alpha+\frac{1}{2}} + \left( \frac{q_\alpha}{m_a} + \frac{q_{\alpha-1}}{m_{\alpha-1}} \right) < \sum_{j=0}^{N} \rho_j \phi_j \delta w_j >_{\alpha-\frac{1}{2}} \right\} \\ &- \partial_x (h (T_{xx}^E)_\alpha) \\ &+ \frac{1}{l_\alpha} < (T_{xx}^E, T_{zx}^E) >_{\alpha+\frac{1}{2}} \cdot (\nabla_x z_{\alpha+\frac{1}{2}}, -1)^t \\ &- \frac{1}{l_\alpha} < (T_{xx}^E, T_{zx}^E) >_{\alpha-\frac{1}{2}} \cdot (\nabla_x z_{\alpha-\frac{1}{2}}, -1)^t \end{aligned} \tag{5.9}$$

with $G_{\alpha+1/2}$ given by (5.4).
The full system given by (5.4), (5.7), (5.8), (5.9) could be written under the structure of a hyperbolic system with conservative flux, non-conservative products and source terms:

$$\partial_t \boldsymbol{w} + \partial_x \boldsymbol{F}(\boldsymbol{w}) = \boldsymbol{B}(\boldsymbol{w}) \partial_x \boldsymbol{w} + \boldsymbol{S}(\boldsymbol{w}) \partial_x z_b + \boldsymbol{E}(\boldsymbol{w}) + \boldsymbol{\Psi}(\boldsymbol{w}), \tag{5.10}$$

where

$$\boldsymbol{w} = (h, r_{1,1}, \ldots, r_{1,M}, r_{2,1}, \ldots, r_{N,M}, q_1, \ldots, q_M), \tag{5.11}$$

and the definitions of $\boldsymbol{F}(\boldsymbol{w})$, $\boldsymbol{B}(\boldsymbol{w})$, $\boldsymbol{S}(\boldsymbol{w})$, $\boldsymbol{E}(\boldsymbol{w})$, and $\boldsymbol{\Psi}(\boldsymbol{w})$ are described in Appendix B.
The system (5.10) can be reformulated as

$$\partial_t \boldsymbol{w} \ + \ \boldsymbol{A}(\boldsymbol{w}) \partial_x \boldsymbol{w} \ = \ \boldsymbol{S}(\boldsymbol{w}) \partial_x z_b + \boldsymbol{E}(\boldsymbol{w}) + \boldsymbol{\Psi}(\boldsymbol{w}), \tag{5.12}$$

where $\boldsymbol{A}(\boldsymbol{w}) = \boldsymbol{J}(\boldsymbol{w}) - \boldsymbol{B}(\boldsymbol{w})$ with $\boldsymbol{J}(\boldsymbol{w}) = \dfrac{\partial \boldsymbol{F}(\boldsymbol{w})}{\partial \boldsymbol{w}}$ the Jacobian matrix of $\boldsymbol{F}$ which is described in Appendix B.

## 5.1 Compact formulation

Remark that the size of system (5.4), (5.7), (5.8), (5.9) is $(N+2)M$ which could be large, especially if we consider several sediment species. In order to reduce the computational cost of the numerical resolution of the system, we shall consider first a reduced system which, for one-dimensional horizontal space, has size $2M+1$.

Let us consider the variables

$$\widetilde{\boldsymbol{w}} = (h, m_1, \dots, m_M, q_1, q_2, ..., q_M)^{'} \in \mathbb{R}^{2M+1}. \tag{5.13}$$

From equation (5.8) and (5.6 ) one gets

$$\begin{aligned} \partial_t m_\alpha + \partial_x q_\alpha = & \frac{1}{l_\alpha}\Big\{ \Big( \frac{m_{\alpha+1}+m_\alpha}{2h} G_{\alpha+\frac{1}{2}} - \frac{m_\alpha + m_{\alpha-1}}{2h} G_{\alpha-\frac{1}{2}} \Big) \Big\} \\ & - \frac{1}{l_\alpha}\Big\{ < \sum_{j=0}^{N} \rho_j \phi_j \delta w_j >_{\alpha+\frac{1}{2}} + < \sum_{j=0}^{N} \rho_j \phi_j \delta w_j >_{\alpha-\frac{1}{2}} \Big\}. \end{aligned} \tag{5.14}$$

Once $\widetilde{\boldsymbol{w}}$ is obtained , we can compute $\boldsymbol{w}$ by the procedure described in Section 6
System given by (5.4), (5.7), (5.9), (5.14) can be written under the structure of an hyperbolic system with conservative flux, non-conservative products and source terms, more precisely,

$$\partial_t \widetilde{\boldsymbol{w}} + \partial_x \widetilde{\boldsymbol{F}}(\widetilde{\boldsymbol{w}}) = \widetilde{\boldsymbol{B}}(\widetilde{\boldsymbol{w}})\partial_x \widetilde{\boldsymbol{w}} + \widetilde{\boldsymbol{S}}(\widetilde{\boldsymbol{w}})\partial_x z_b + \widetilde{\boldsymbol{E}}(\widetilde{\boldsymbol{w}}) + \widetilde{\boldsymbol{\Psi}}(\widetilde{\boldsymbol{w}}), \tag{5.15}$$

where $\widetilde{\boldsymbol{w}} = (h, m_1, \dots, m_M, q_1, q_2, ..., q_M)^t \in \mathbb{R}^{2M+1}$ is the unknown vector, $\widetilde{\boldsymbol{F}} : \mathbb{R}^{2M+1} \to \mathbb{R}^{2M+1}$ is a regular vector function, $\widetilde{\boldsymbol{B}} : \mathbb{R}^{2M+1} \to \mathcal{M}_{2M+1}(\mathbb{R})$ is a matrix function, where $\mathcal{M}_n(\mathbb{R})$ is the space of real $n \times n$ matrices ($n \in \mathbb{N}^*$), $\widetilde{\boldsymbol{S}}, \widetilde{\boldsymbol{E}}, \widetilde{\boldsymbol{\Psi}} : \mathbb{R}^{2M+1} \to \mathbb{R}^{2M+1}$ are vectorial functions, and $z_b : \mathbb{R} \to \mathbb{R}$ is a real scalar function. The form (5.15) constitutes a classic simplified model type for multiphase or multilayer flows in the literature. In Appendix C we exhibit the algebraic expressions of the terms $\widetilde{\boldsymbol{F}}(\widetilde{\boldsymbol{w}}) = (\widetilde{\boldsymbol{F}}_\alpha(\widetilde{\boldsymbol{w}}))_{\alpha=0,1,\dots,2M}$, $\widetilde{\boldsymbol{S}}(\widetilde{\boldsymbol{w}})$, $\widetilde{\boldsymbol{E}}(\widetilde{\boldsymbol{w}}) = (\widetilde{\boldsymbol{E}}_\alpha(\widetilde{\boldsymbol{w}}))_{\alpha=0,1,\dots,2M}$, $\widetilde{\boldsymbol{\Psi}}(\widetilde{\boldsymbol{w}}) = (\widetilde{\boldsymbol{\Psi}}_\alpha(\widetilde{\boldsymbol{w}}))_{\alpha=0,1,\dots,2M}$ and $\widetilde{\boldsymbol{B}}(\widetilde{\boldsymbol{w}}) = (\widetilde{\boldsymbol{B}}_{\alpha,\beta}(\widetilde{\boldsymbol{w}}))_{\alpha,\beta=0,1,\dots,2M}$ involved in (5.15).
System (5.15) can also be reformulated as

$$\partial_t \widetilde{\boldsymbol{w}} \ + \ \widetilde{\boldsymbol{A}}(\widetilde{\boldsymbol{w}})\partial_x \widetilde{\boldsymbol{w}} \ = \ \widetilde{\boldsymbol{S}}(\widetilde{\boldsymbol{w}})\partial_x z_b + \widetilde{\boldsymbol{E}}(\widetilde{\boldsymbol{w}}) + \widetilde{\boldsymbol{\Psi}}(\widetilde{\boldsymbol{w}}), \tag{5.16}$$

where $\widetilde{\boldsymbol{A}}(\widetilde{\boldsymbol{w}}) = \widetilde{\boldsymbol{J}}(\widetilde{\boldsymbol{w}}) - \widetilde{\boldsymbol{B}}(\widetilde{\boldsymbol{w}})$ with $\widetilde{\boldsymbol{J}}(\widetilde{\boldsymbol{w}}) = \dfrac{\partial \widetilde{\boldsymbol{F}}(\widetilde{\boldsymbol{w}})}{\partial \widetilde{\boldsymbol{w}}}$ the Jacobian matrix of $\widetilde{\boldsymbol{F}}$. The matrix $\widetilde{\boldsymbol{J}}(\widetilde{\boldsymbol{w}}) = (\widetilde{\boldsymbol{J}}_{\alpha,\beta}(\widetilde{\boldsymbol{w}}))_{\alpha,\beta=0,1,\dots,2M} \in \mathcal{M}_{(2N+1)}(\mathbb{R})$ is described in Appendix C.

# 6 Numerical approximation

## 6.1 Definition of the scheme

The numerical approximation of the model is based on a standard finite volume method combined with a three-step splitting procedure. The procedure will be detailed afterwards, but the main idea is the following. As usual in finite volume schemes, we subdivide the horizontal spatial domain into standard computational cells $I_i = [x_{i-1/2}, x_{i+1/2}]$, and assume an approximation at time $t_n$ in each cell $\boldsymbol{w}_i^n$ and the corresponding values $\widetilde{\boldsymbol{w}}_i^n$.

- In the first step, we rule out the contribution of source terms $\widetilde{\boldsymbol{E}}(\widetilde{\boldsymbol{w}})$ and $\widetilde{\boldsymbol{\Psi}}(\widetilde{\boldsymbol{w}})$ in (5.15) and perform a path-conservative scheme to obtain the approximations of this system at time $t_{n+1}$ $\widetilde{\boldsymbol{w}}_i^{n+1/3}$. Then, the values values $\boldsymbol{w}_i^{n+1/3}$ are recovered from $\widetilde{\boldsymbol{w}}_i^{n+1/3}$ by an upwind procedure.
- In the second step, we include the effects of the source term $\boldsymbol{E}(\boldsymbol{w})$ by an implicit Euler scheme, obtaining the approximations $\boldsymbol{w}_i^{n+2/3}$
- In the third step, we include the effects of the source term $\boldsymbol{\Psi}(\boldsymbol{w})$ by a semi-implicit approach, obtaining the approximations $\boldsymbol{w}_i^{n+1}$ at time $t_{n+1}$. Finally, $\widetilde{\boldsymbol{w}}_i^{n+1}$ are updated from $\boldsymbol{w}_i^{n+1}$ using (5.6).

We proceed now to describe each step more precisely.

### 6.1.1 First step: path-conservative scheme

In the first step, we rule out the contributions of the source terms $\widetilde{\boldsymbol{E}}(\widetilde{\boldsymbol{w}})$ and $\widetilde{\boldsymbol{\Psi}}(\widetilde{\boldsymbol{w}})$ in (5.15) and then apply a finite volume scheme. The resulting system can be written as

$$\partial_t \widetilde{\boldsymbol{W}} + \widetilde{\mathcal{A}}(\widetilde{\boldsymbol{W}}) \cdot \partial_x \widetilde{\boldsymbol{W}} = 0, \tag{6.1}$$

where $\widetilde{\boldsymbol{W}}$ is the concatenated vector $\widetilde{\boldsymbol{W}} := (\widetilde{\boldsymbol{w}}, z_b)^{\mathrm{t}} \in \widetilde{\Omega}$ for some open convex domain $\widetilde{\Omega} \subset \mathbb{R}^{2M+1}$ Solutions of (6.1) may develop discontinuities and, due to the non-divergence form of the equations, the notion of weak solution in the sense of distributions cannot be used. The theory introduced by Dal Maso, LeFloch, and Murat [10] is followed here to define weak solutions. This theory allows one to define the non-conservative product $\mathcal{A}(\widetilde{\boldsymbol{W}}) \cdot \partial_x \widetilde{\boldsymbol{W}}$ as a bounded measure provided a family of Lipschitz continuous paths $\psi : [0,1] \times \widetilde{\Omega} \times \widetilde{\Omega} \to \widetilde{\Omega}$ is prescribed, which must satisfy certain natural regularity conditions. We will consider here the family of straight segments. Then, a path-conservative numerical scheme in the sense defined by Parés in [24] can be used to compute the approximations $\widetilde{\boldsymbol{w}}_i^{n+1/3}$ at time $t^{n+1}$ of the considered system. The scheme may be written in the general form

$$\widetilde{\boldsymbol{w}}_i^{n+1/3} = \widetilde{\boldsymbol{w}}_i^n - \frac{\Delta t}{\Delta x}\left( \widetilde{\mathcal{F}}_{i+1/2}^n - \widetilde{\mathcal{F}}_{i-1/2}^n - \frac{1}{2}\big(\widetilde{\mathcal{B}}_{i+1/2}^n + \widetilde{\mathcal{B}}_{i-1/2}^n + \widetilde{\mathcal{S}}_{i+1/2}^n + \widetilde{\mathcal{S}}_{i-1/2}^n\big)\right), \tag{6.2}$$

where the expressions $\mathcal{F}_{i+1/2}^n$ and $\mathcal{B}_{i+1/2}^n$ are defined as follows:

$$\begin{aligned}\widetilde{\mathcal{F}}_{i+1/2}^n := \; & \frac{1}{2}\big(\widetilde{\boldsymbol{F}}(\widetilde{\boldsymbol{w}}_i^n) + \widetilde{\boldsymbol{F}}(\widetilde{\boldsymbol{w}}_{i+1}^n)\big) \\ & - \frac{1}{2}\boldsymbol{Q}_{i+1/2}^n \left( \widetilde{\boldsymbol{w}}_{i+1}^n - \widetilde{\boldsymbol{w}}_i^n - \widetilde{\Lambda}_{i+1/2}^n \widetilde{\boldsymbol{S}}_{i+1/2}^n \big((z_b)_{i+1}^n - (z_b)_i^n\big)\right),\end{aligned} \tag{6.3}$$

$$\widetilde{\mathcal{B}}_{i+1/2}^n = \widetilde{\boldsymbol{B}}_{i+1/2}^n \big(\widetilde{\boldsymbol{w}}_{i+1}^n - \widetilde{\boldsymbol{w}}_i^n\big), \tag{6.4}$$

$$\widetilde{\mathcal{S}}_{i+1/2}^n = \widetilde{\boldsymbol{S}}_{i+1/2}^n \big((z_b)_{i+1}^n - (z_b)_i^n\big), \tag{6.5}$$

with

$$\widetilde{\mathcal{B}}_{i+1/2} = \int_0^1 \widetilde{\boldsymbol{B}}(\psi(s, \widetilde{\boldsymbol{w}}_i, \widetilde{\boldsymbol{w}}_{i+1}))ds, \tag{6.6}$$

$$\widetilde{\mathcal{S}}_{i+1/2} = \int_0^1 \widetilde{\boldsymbol{S}}(\psi(s, \widetilde{\boldsymbol{w}}_i, \widetilde{\boldsymbol{w}}_{i+1}))ds. \tag{6.7}$$

$$\tag{6.8}$$

The matrix $\widetilde{\Lambda}(\widetilde{\boldsymbol{w}})$ represents an approximation of the inverse of $\widetilde{\boldsymbol{A}}(\widetilde{\boldsymbol{w}}) = \widetilde{\boldsymbol{J}}(\widetilde{\boldsymbol{w}}) - \widetilde{\boldsymbol{B}}(\widetilde{\boldsymbol{w}})$ given by (C.2) - (C.7) In addition, $\widetilde{\boldsymbol{Q}}_{i+1/2}^n$ is the numerical viscosity matrix whose definition identifies the particular finite volume method used. For example, the Roe method is defined by $\widetilde{\boldsymbol{Q}}_{i+1/2} = |\widetilde{\boldsymbol{A}}_{i+1/2}|$, where $\widetilde{\boldsymbol{A}}_{i+1/2}$ is the Roe matrix defined in the sense of Toumi (see [25, 32]). An interesting alternative to Roe method for system with a great number of unknowns are PVM ("polynomial viscosity matrix") methods (see [7]). At this step, we obtain from (6.2) the intermediate solution

$$\widetilde{\boldsymbol{w}}_i^{n+1/3} = \big(h_i^{n+1/3}, (m_1)_i^{n+1/3}, \ldots, (m_M)_i^{n+1/3}, (q_1)_i^{n+1/3}, \ldots, (q_M)_i^{n+1/3}\big)^{\mathrm{t}}.$$

In this paper we will use a HLL type PVM scheme. We refer the reader to [7] for the details.
Once the approximations $\widetilde{\boldsymbol{w}}_i^{n+1/3}$ are computed, $\boldsymbol{w}_i^{n+1/3}$ are recovered from $\widetilde{\boldsymbol{w}}_i^{n+1/3}$, but to do that, the values $(r_{j,\alpha})_i^{n+1/3}$ for $j = 1, \ldots, N$ and $\alpha = 1, \ldots, M$ should be computed. This is done using the following upwind scheme:

$$(r_{j,\alpha})_i^{n+1/3} = (r_{j,\alpha})_i^n - \frac{\Delta t}{\Delta x}\left( (\mathcal{F}^{r_{j,\alpha}})_{i+1/2}^n - (\mathcal{F}^{r_{j,\alpha}})_{i-1/2}^n - \frac{1}{2}\Pi_{r_{j,\alpha}}\left(\mathcal{B}_{i+1/2}^n + \mathcal{B}_{i-1/2}^n\right)\right), \tag{6.9}$$

where

$$\mathcal{B}_{i+1/2}^n = \boldsymbol{B}_{i+1/2}^n\big(\boldsymbol{w}_{i+1}^n - \boldsymbol{w}_i^n\big), \tag{6.10}$$

$$(\mathcal{F}^{r_{j,\alpha}})_{i+1/2}^n = \begin{cases} \dfrac{(r_{j,\alpha})_i^n}{(m_\alpha)_i^n}\mathcal{F}_{i+1/2}^n, & \text{if } \Pi_{m_\alpha}\mathcal{F}_{i+1/2}^n \geq 0 \\ \\ \dfrac{(r_{j,\alpha})_{i+1}^n}{(m_\alpha)_{i+1}^n}\mathcal{F}_{i+1/2}^n, & \text{otherwise} \end{cases} \tag{6.11}$$

and $\Pi_{r_{j,\alpha}}$, $\Pi_{m_\alpha}$, denotes respectively the canonical projection on the variables $r_{j,\alpha}$ and $m_\alpha$.

*Remark 6.1. The definition of $(r_{j,\alpha})_i^{n+1/3}$ is consistent with that of $(m_\alpha)_i^{n+1/3}$ in the sense that extending (6.9) for $j = 0$, then we have*

$$(\rho(\Phi_\alpha)h)_i^{n+1/3} = \sum_{j=0}^{N}\rho_j(r_{j,\alpha})_i^{n+1/3} = (m_\alpha)_i^{n+1/3}.$$

### 6.1.2 Second step: viscosity effects

In this second step we take into account the contribution of the source term $\boldsymbol{E}(\boldsymbol{w})$ in (5.10). As those terms are related to friction at the interfaces and mass flux exchange at the bottom, we propose to use an implicit update. Moreover, if we assume a non-penetrable bottom layer, the term $G_{\frac{1}{2}}$ vanishes and only friction terms are retained. Finally, if friction terms are neglected, then $\boldsymbol{E}(\boldsymbol{w})$ vanishes and this step is no longer needed.

$$\boldsymbol{w}_i^{n+2/3} = \boldsymbol{w}_i^{n+1/3} + \Delta t\,\boldsymbol{E}\big(\boldsymbol{w}_i^{n+2/3}\big). \tag{6.12}$$

### 6.1.3 Third step: deposition effects

Finally, we have to apply the source term $\boldsymbol{\Psi}$ which stands for the transfer between the layers due to deposition effects. To do so, following Section 3.2.1, we need to describe the terms $< \phi_j \delta w_j >_{\alpha+\frac{1}{2}}$. As it was mention in Section 2.1, we may assume that the hindered settling velocity of the sediment species is given by

$$\delta w_j = -\chi(\phi) w_{s_j}, \text{ for } j = 1, \ldots, N, \tag{6.13}$$

where $w_{s_j}$ is the terminal velocity of the species $j$ and $\chi(\phi)$ is the hindered settling factor given by (2.14).
It is clear that the deposition at the interface $\alpha + \frac{1}{2}$ depends on the sediment present on the upper layer and on whether the lower layer $\alpha$ is saturated or not. We thus propose to define

$$< \phi_j \delta w_j >_{\alpha+\frac{1}{2}} = -\phi_{j,\alpha+1}\chi(\Phi_\alpha) w_{s_j}, \text{ for } j = 1, \ldots, N. \tag{6.14}$$

Based on this definition, we set

$$< \phi_j \delta w_j >_{\alpha+\frac{1}{2}} = \frac{w_{s_j}}{h^{n+2/3}}(r_{j,\alpha+1})_i^{n+1}\chi((\phi_\alpha)_i^{n+2/3}), \text{ for } j = 1, \ldots, N \tag{6.15}$$

and perform an update of $r_{j,\alpha}$ by

$$l_\alpha(r_{j,\alpha})_i^{n+1} = l_\alpha(r_{j,\alpha})_i^{n+2/3} + \frac{w_{s_j}\Delta t}{h^{n+2/3}}\left((r_{j,\alpha+1})_i^{n+1}\chi((\phi_\alpha)_i^{n+2/3}) - (r_{j,\alpha})_i^{n+1}\chi((\phi_{\alpha-1})_i^{n+2/3})\right), \tag{6.16}$$

where we are considering $r_{j,M+1}=0$ and $\phi_{-1}=0$.
This can be solved explicitly by

$$(r_{j,\alpha})_i^{n+1}=\frac{1}{l_\alpha+(k_j)_i^{n+2/3}\chi((\phi_{\alpha-1})_i^{n+2/3})}\left(l_\alpha(r_{j,\alpha})_i^{n+2/3}+(k_j)_i^{n+2/3}\chi((\phi_\alpha)_i^{n+2/3})(r_{j,\alpha+1})_i^{n+1}\right) \tag{6.17}$$

with $(k_j)_i^{n+2/3}=\dfrac{w_{s_j}\Delta t}{h^{n+2/3}}$, $r_{j,M+1}=0$ and $\phi_{-1}=0$.
Now, the variables $m_\alpha$ and $q_\alpha$ have to be updated by

$$\begin{aligned}l_\alpha(m_\alpha)_i^{n+1}&=l_\alpha(m_\alpha)_i^{n+2/3}\\&+\sum_{j=0}^N\rho_j\frac{w_{s_j}\Delta t}{h^{n+2/3}}\left((r_{j,\alpha+1})_i^{n+1}\chi((\phi_\alpha)_i^{n+2/3})-(r_{j,\alpha})_i^{n+1}\chi((\phi_{\alpha-1})_i^{n+2/3})\right),\end{aligned} \tag{6.18}$$

$$\begin{aligned}l_\alpha(q_\alpha)_i^{n+1}&=l_\alpha(q_\alpha)_i^{n+2/3}\\&+\sum_{j=0}^N\rho_j\frac{w_{s_j}\Delta t}{h^{n+2/3}}\left[\frac{1}{2}\left(\frac{(q_{\alpha+1})_i^{n+2/3}}{(m_{\alpha+1})_i^{n+2/3}}+\frac{(q_\alpha)_i^{n+2/3}}{(m_\alpha)_i^{n+2/3}}\right)(r_{j,\alpha+1})_i^{n+1}\chi((\phi_\alpha)_i^{n+2/3})\right.\\&\left.-\frac{1}{2}\left(\frac{(q_\alpha)_i^{n+2/3}}{(m_\alpha)_i^{n+2/3}}+\frac{(q_{\alpha-1})_i^{n+2/3}}{(m_{\alpha-1})_i^{n+2/3}}\right)(r_{j,\alpha})_i^{n+1}\chi((\phi_{\alpha-1})_i^{n+2/3})\right],\end{aligned} \tag{6.19}$$

*Remark* 6.2. *Definition (6.17) satisfies the following properties:*

- $(r_{j,\alpha})_i^{n+1}$ *are non-negative provided that the approximations* $(r_{j,\alpha})_i^{n+2/3}$ *are non-negative.*
- *Sediment mass is preserved*

$$\sum_{\alpha=1}^M l_\alpha(r_{j,\alpha})_i^{n+1}=\sum_{\alpha=1}^M l_\alpha(r_{j,\alpha})_i^{n+2/3}.$$

- *Following (2.5), we can easily generalize (6.16) for freshwater volume fractions by*

$$l_\alpha(r_{0,\alpha})_i^{n+1}=l_\alpha(r_{0,\alpha})_i^{n+2/3}-\sum_{j=1}^N\frac{w_{s_j}\Delta t}{h^{n+2/3}}\left((r_{j,\alpha+1})_i^{n+1}\chi((\phi_\alpha)_i^{n+2/3})-(r_{j,\alpha})_i^{n+1}\chi((\phi_{\alpha-1})_i^{n+2/3})\right),$$

*which grants that*

$$\sum_{\alpha=1}^M l_\alpha(r_{0,\alpha})_i^{n+1}=\sum_{\alpha=1}^M l_\alpha(r_{0,\alpha})_i^{n+2/3},\quad \text{and}\quad \sum_{j=0}^N(r_{j,\alpha})_i^{n+1}=\sum_{j=0}^N(r_{j,\alpha})_i^{n+2/3}.$$

- *If* $(\phi_\alpha)_i^{n+2/3}=\phi_{max}$ *then*

$$(r_{j,\alpha})_i^{n+1}=\frac{l_\alpha}{l_\alpha+(k_j)_i^{n+2/3}\chi((\phi_{\alpha-1})_i^{n+2/3})}(r_{j,\alpha})_i^{n+2/3}$$

*and we get* $(\phi_\alpha)_i^{n+2/3}\le\phi_{max}$. *This means that solid concentration does not increase in the cells which are saturated.*

# 7 Numerical simulations

The objective of this section is to show the potential of this model by showing two different tests: the first one corresponds to the simulation of a hyperpycnal plume and the second one represents a hypopycnal plume.

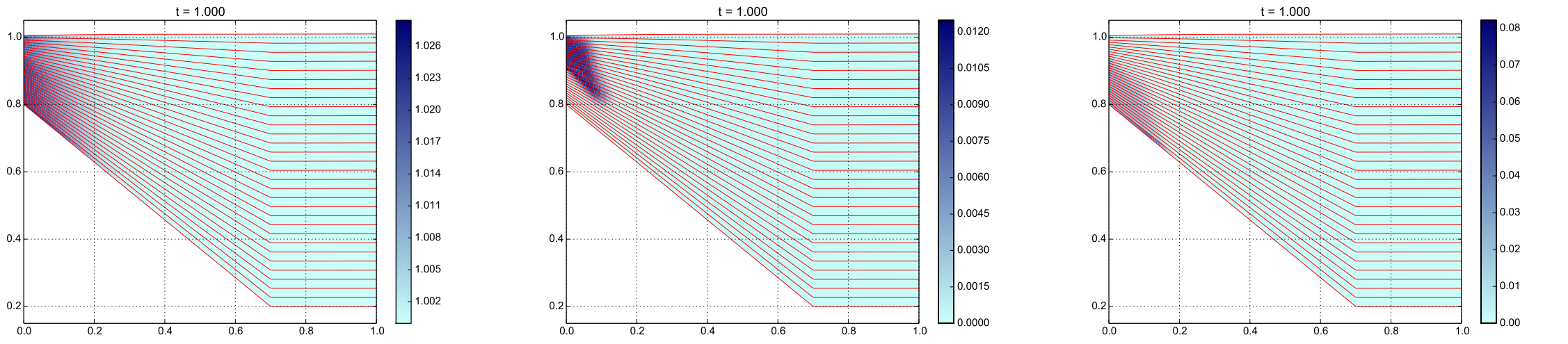

(a) Relative density $\rho(\Phi_\alpha)/\rho_0$ (b) First species concentration $\phi_{1,\alpha}$ (c) Second species concentration $\phi_{2,\alpha}$

Figure 2: Sediment distribution by layers at time $t = 1$

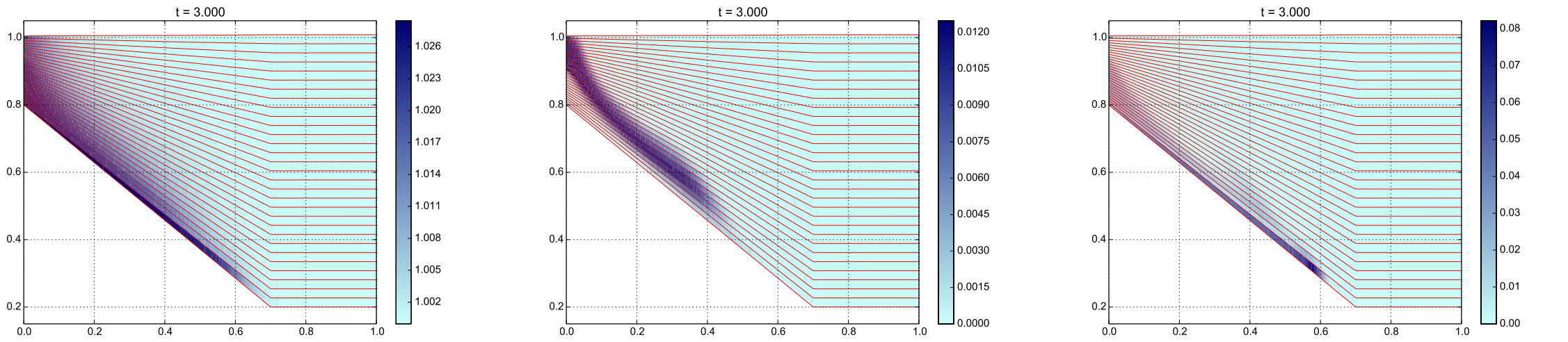

(a) Relative density $\rho(\Phi_\alpha)/\rho_0$ (b) First species concentration $\phi_{1,\alpha}$ (c) Second species concentration $\phi_{2,\alpha}$

Figure 3: Sediment distribution by layers at time $t = 3$

## 7.1 Simulation of a hyperpycnal plume

In this first test we consider a bottom given by

$$z_B(x) = \begin{cases} 0.2 - \dfrac{6}{7}(x - 0.7), & \text{if } x \leq 0.7, \\ 0.2, & \text{otherwise,} \end{cases}$$

in the domain $[0, 1]$ and we set initially clear water satisfying a lake-at-rest steady state, that is, $h(t = 0, x) + z_B(x) = 0$ and $u_\alpha(t = 0, x) = 0$ for each layer $\alpha = 1, \ldots, M$. We have used here $M = 30$ layers and 200 points on the domain. We shall consider two sediment species of density $1150kg/m^3$ and $1250kg/m^3$ respectively. As boundary conditions we set open conditions on the right hand side and we impose on the left hand side

$$u_\alpha(x, t = 0) = 0.2, \qquad \text{for } \alpha = 1, \ldots, M, \tag{7.1}$$

and

$$\phi_{1,\alpha} = 0.01, \text{ for } \alpha \geq M/3 \quad \text{and} \quad \phi_{2,\alpha} = 0.02, \text{ for } \alpha \leq 2M/3. \tag{7.2}$$

The settling velocity is set to 0.015 for the first species and 0.025 for the second one.
Figures 2, 3, and 4 show the evolution of the plume that plunges into the ambient water generating a hyperpycnal plume. The first plot in each figure represents the relative density $\rho_\alpha(\Phi_\alpha)/\rho_0$, while the second and third plots in each figure represents the volumetric concentration of each sediment species by layer. We observe that the vertical distribution of sediment is obtained with detail. The first specie, which is less dense, remains on top of the second one, more dense, as expected. The same behaviour can be seen in Figures 5 and 6 where the vertical concentration of each sediment species is shown at two points: $x = 0.05$, near the left boundary, and $x = 0.3$, in the mid-region of the slope. We remark how particles tend to go down due to deposition effects. Again observe the great detail on the vertical distribution obtained. Moreover, the model allows us to recover the vertical

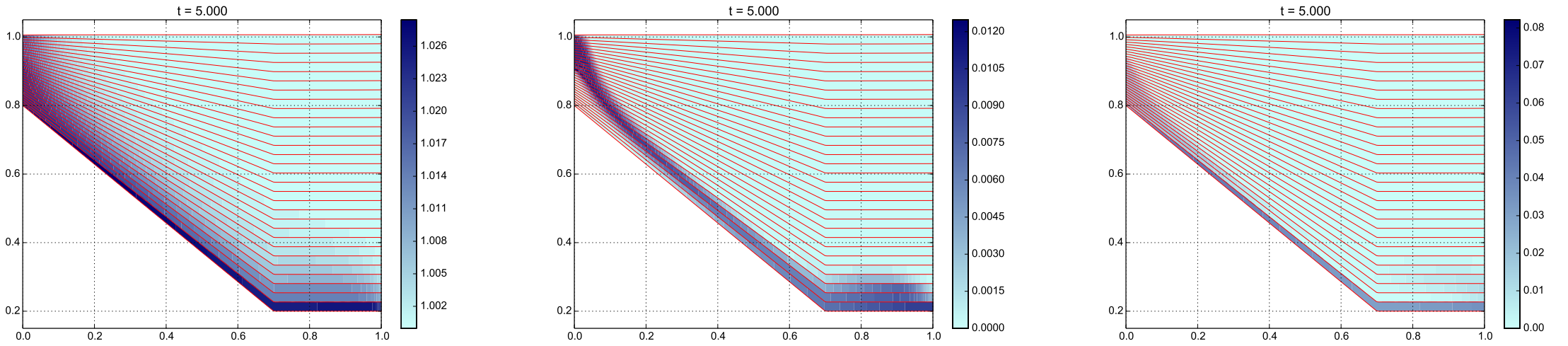

(a) Relative density $\rho(\Phi_\alpha)/\rho_0$ (b) First species concentration $\phi_{1,\alpha}$ (c) Second species concentration $\phi_{2,\alpha}$

Figure 4: Sediment distribution by layers at time $t = 5$

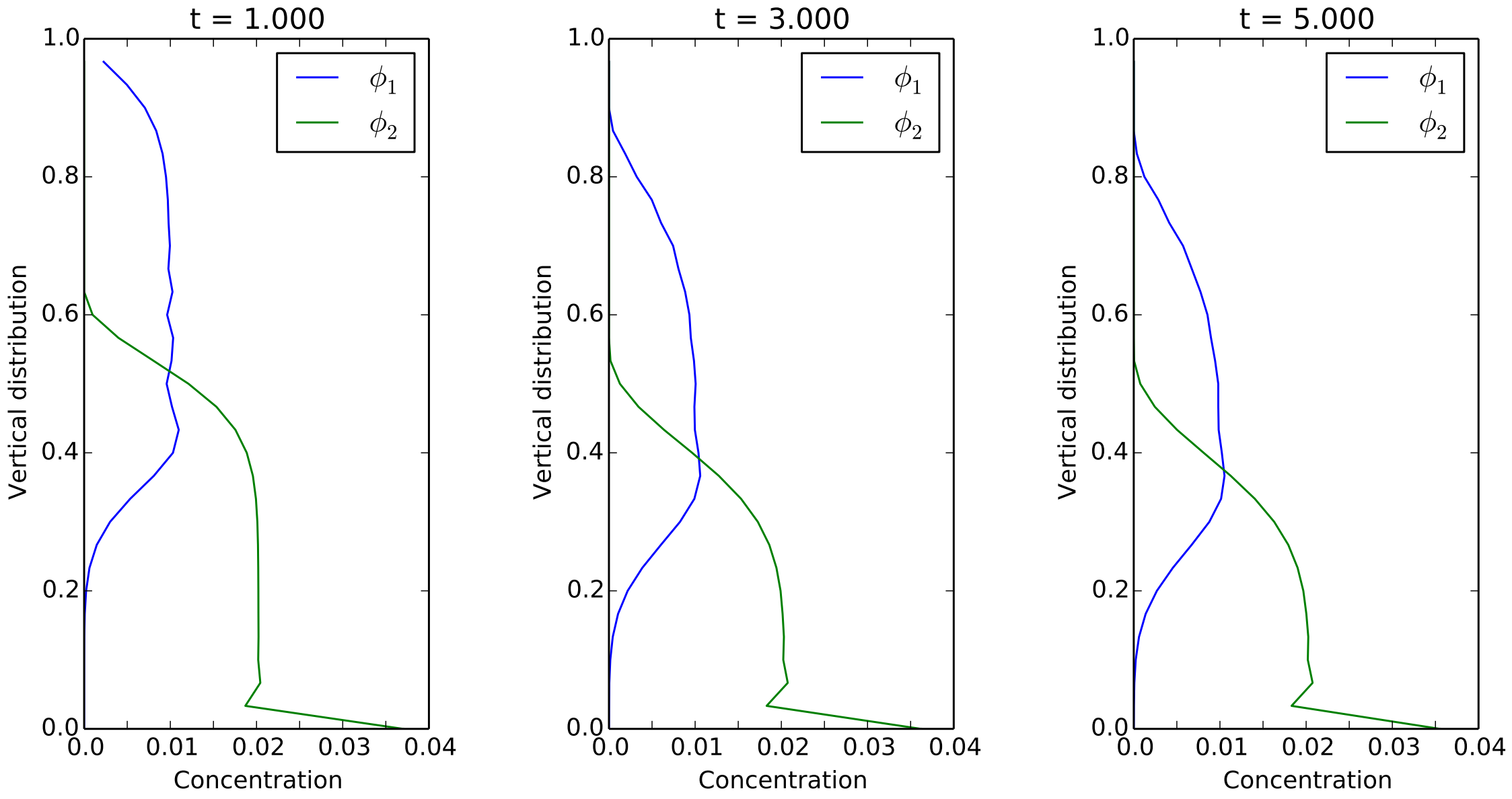

(a) (b) (c)

Figure 5: Vertical concentration distribution at $x = 0.05$. On the y-axis 0 corresponds to bottom level and 1 to the surface.

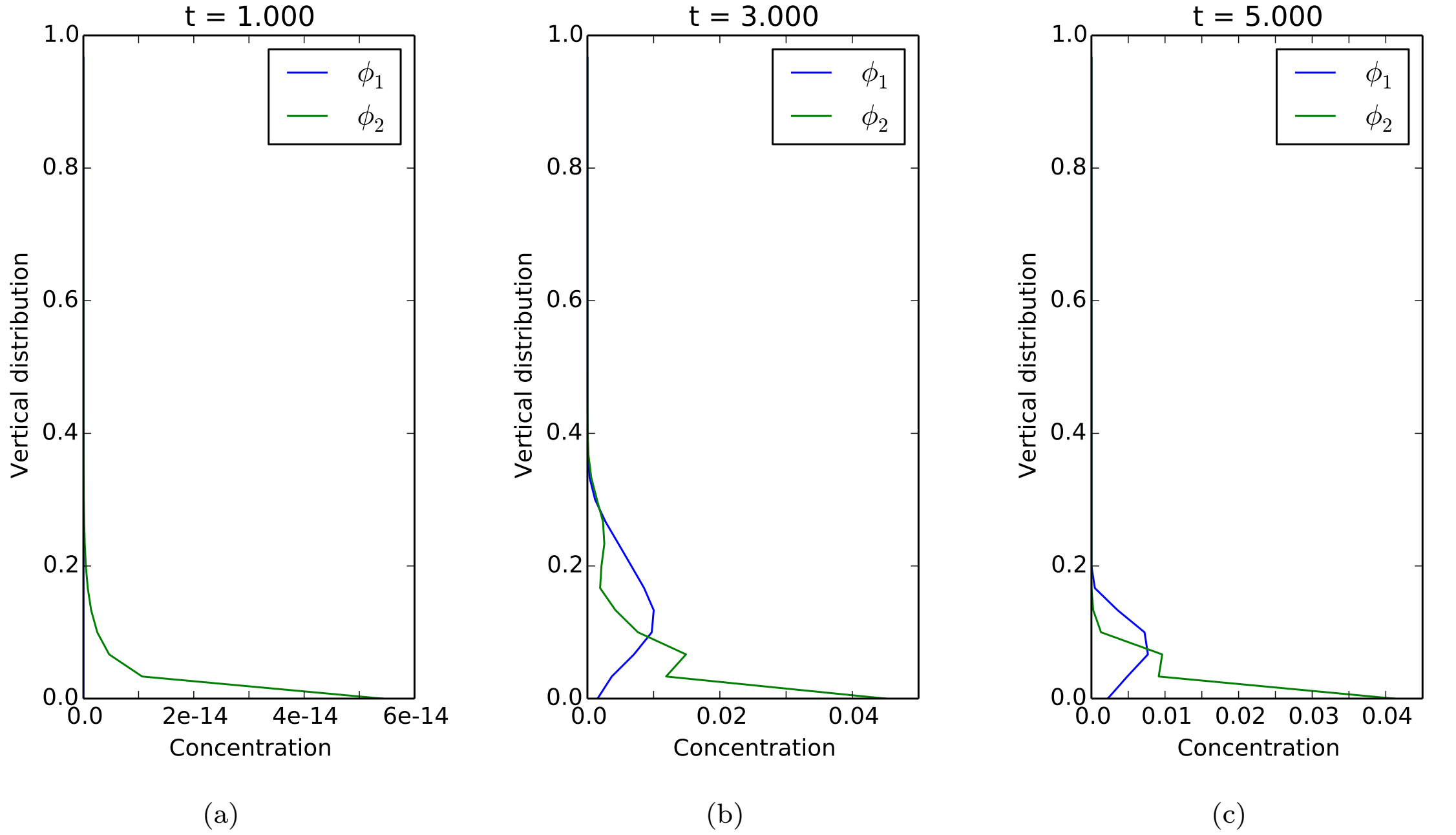

Figure 6: Vertical concentration distribution at $x = 0.3$. On the y-axis 0 corresponds to bottom level and 1 to the surface.

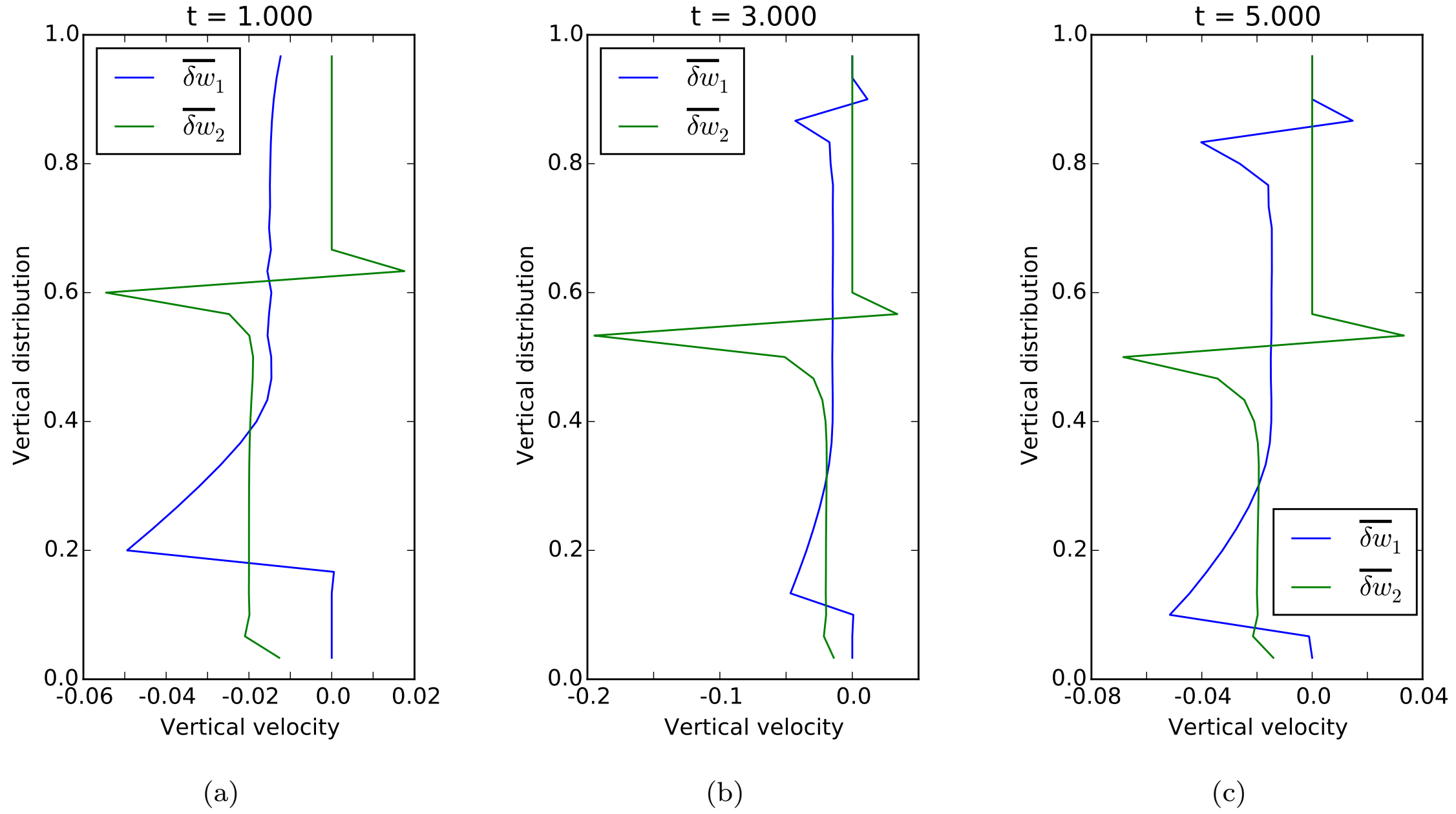

Figure 7: Average vertical velocity $\overline{\delta w}_{j,\alpha+1/2}$ at $x = 0.05$. On the y-axis 0 corresponds to bottom level and 1 to the surface.

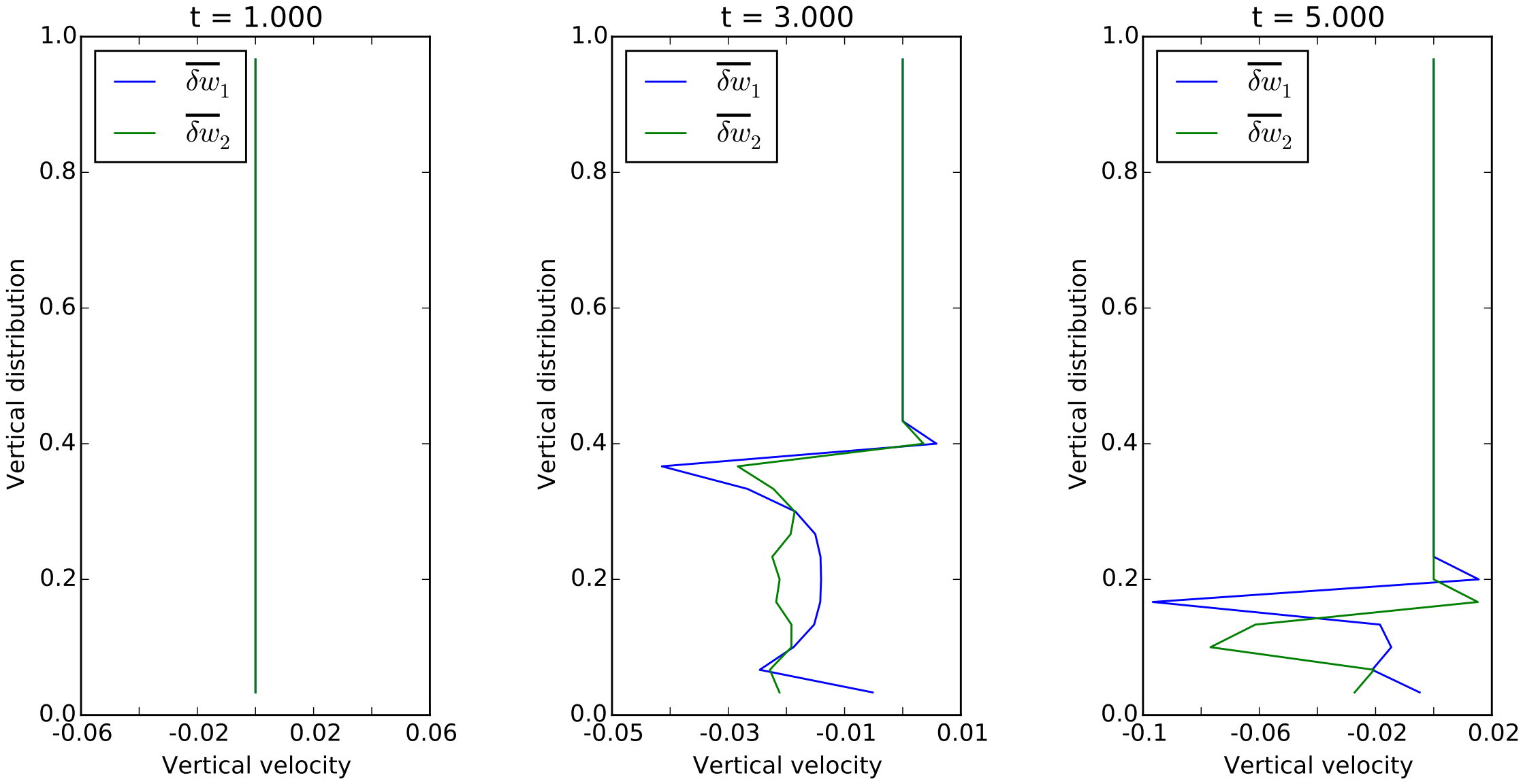

(a) (b) (c)

Figure 8: Average vertical velocity $\overline{\delta w}_{j,\alpha+1/2}$ at $x = 0.3$. On the y-axis 0 corresponds to bottom level and 1 to the surface.

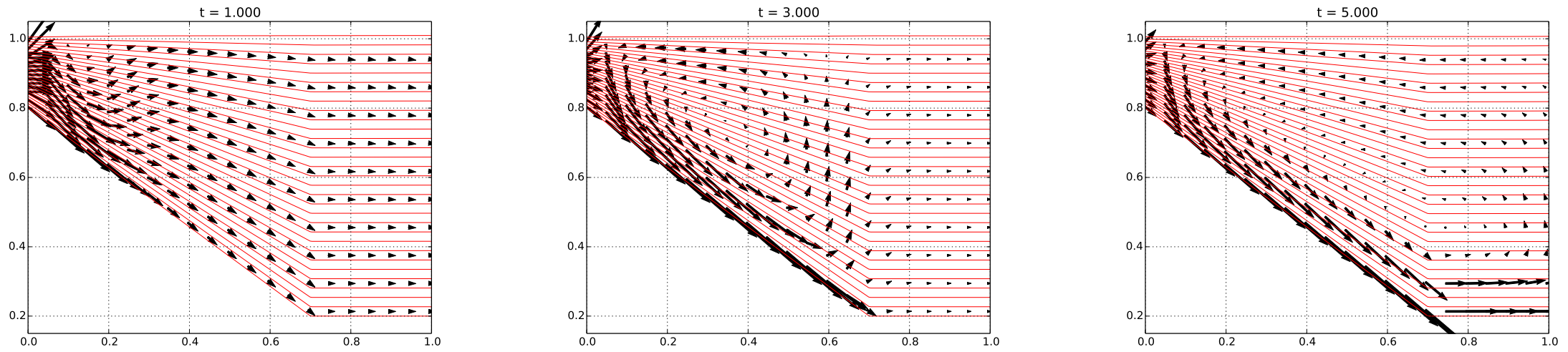

(a) (b) (c)

Figure 9: Velocity vector by layers

velocities of each sediment species. In Figures 7 and 8 we show the average velocities at each interface $\overline{\delta w}_{j,\alpha+\frac{1}{2}} = 0.5(\delta w^{+}_{j,\alpha+\frac{1}{2}} + \delta w^{-}_{j,\alpha+\frac{1}{2}})$ given by (3.27). It is also interesting the formation at the head of the plume in Figure 4 due to the change of slope in the bottom.
In Figures 9 we show the velocity field for different times. We can see that the hyperpycnal plume forms a recirculation of the receiving ambient water.
All the effects described previously are recovered thanks to this multi-layer model. We remark again that this could not be possible with a one layer model as the one introduced in [21].

## 7.2 Simulation of a hypopycnal plume

The objective of this test is to show the versatility of the model. It allows to study many physical situations even the case of an hypopycnal plume. Let us assume a river with freshwater density $\rho_w$, carrying sediment in suspension, comes into the ocean with a given density $\rho_o$, greater than the freshwater density $\rho_w$. When the sediment of the mixture is less than the sediment of the ocean, the river will form a plume that floats and go up to the surface of the ocean. The model derived here assumes that water has density $\rho_0$ and we have $N$ sediment species in suspension of density $\rho_j$, $j = 1, \ldots, N$. So a priori this model would not be suitable to simulate hypopycnal plumes because the water density is always the same. Nevertheless, we may assume that $\rho_0 \equiv \rho_o$ corresponds to the ambient water and we may assimilate one of the species to the freshwater of the river, that is, $\rho_1 = \rho_w$ which will give the desired results.
For instance, consider the following topography

$$z_B(x) = \begin{cases} 0.2 - \dfrac{3}{7}(x - 0.7), & \text{if } x \leq 0.7, \\ 0.2, & \text{otherwise,} \end{cases}$$

in the domain $[0, 1]$. Assume that ambient water has density higher than freshwater density $\rho_0 = 1020 kg/m^3$. We set initially the ambient water satisfying a lake-at-rest steady state, that is, $h(t = 0, x) + z_B(x) = 0$ and $u_\alpha(t = 0, x) = 0$ for each layer $\alpha = 1, \ldots, M$. We have used here $M = 30$ layers and 200 points on the domain.
Now, consider two different species: the first one, assimilated to freshwater of the river, with density $\rho_1 = 1000 kg/m^3$ and the second one, assimilated to the sediment in suspension, with density $\rho_2 = 1150 kg/m^3$. As boundary conditions we set open conditions on the right hand side and we impose on the left hand side

$$u_\alpha(x, t = 0) = 0.01, \qquad \text{for } \alpha = 1, \ldots, M, \tag{7.3}$$

and

$$\phi_{1,\alpha} = 0.95, \text{ for } \alpha \geq M/3 \quad \text{and} \quad \phi_{2,\alpha} = 0.05, \text{ for } \alpha \leq 2M/3. \tag{7.4}$$

The settling velocity is set to 0.0005 for the sediment species.
Figures 10, 11, and 12 show the evolution of the plume. In the first plot of each of the Figures we remark that density of the mixture $\rho(\Phi_\alpha)$ is smaller than the one of ambient water $\rho_0$. This will originate the hypopycnal plume. Again we observe that the vertical distribution of sediment is obtained with detail. In Figures 13, 14, and 15 we show the vertical distribution of freshwater $\phi_1$ and sediment $\phi_2$ at three points: $x = 0.1$, $x = 0.35$, and $x = 0.8$. In Figures 16, 17, and 18 we show, again at those points, the average velocities at each interface $\overline{\delta w}_{j,\alpha+\frac{1}{2}} = 0.5(\delta w^{+}_{j,\alpha+\frac{1}{2}} + \delta w^{-}_{j,\alpha+\frac{1}{2}})$ given by (3.27).
Moreover, the velocity field in Figure 19 shows some interesting profiles. While the river goes up and to the right, it interacts with the ambient water forming some kind of turbulence. As a consequence the ambient water will flow to the left under the plume.
We remark that the model reproduces the desired results and that the numerical scheme can handle such situations.

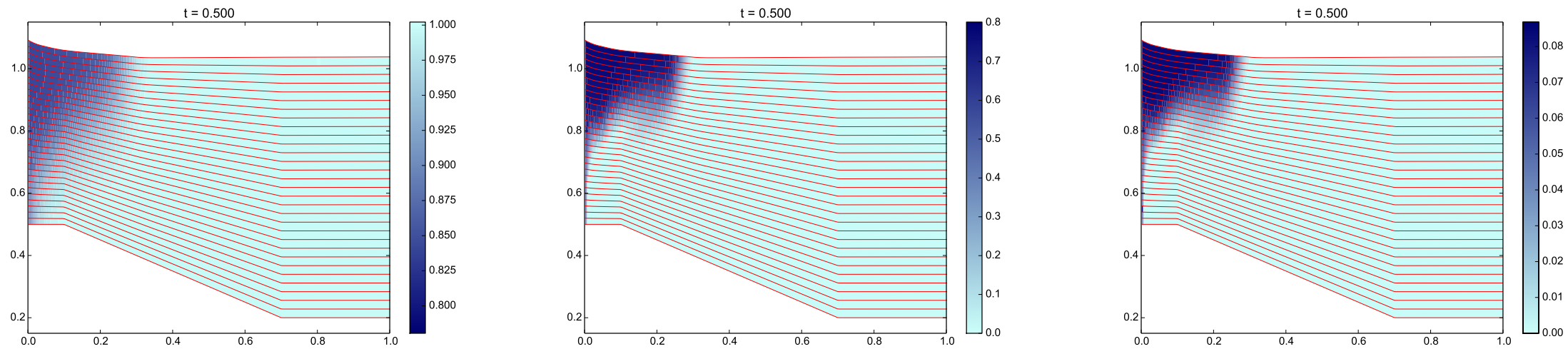

(a) Relative density $\rho(\Phi_\alpha)/\rho_0$ (b) First species concentration $\phi_{1,\alpha}$ (c) Second species concentration $\phi_{2,\alpha}$

Figure 10: Sediment distribution by layers at time $t = 0.5$

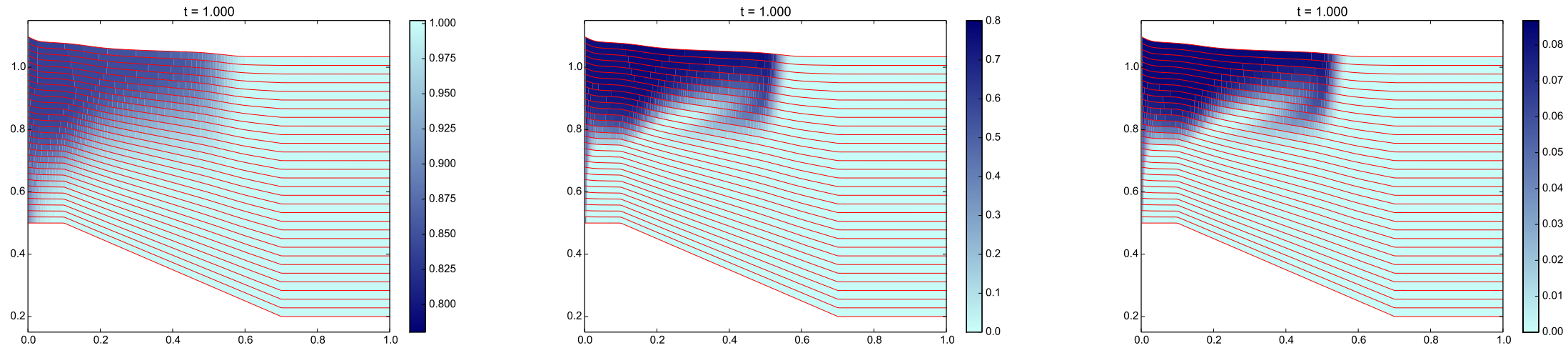

(a) Relative density $\rho(\Phi_\alpha)/\rho_0$ (b) First species concentration $\phi_{1,\alpha}$ (c) Second species concentration $\phi_{2,\alpha}$

Figure 11: Sediment distribution by layers at time $t = 1$

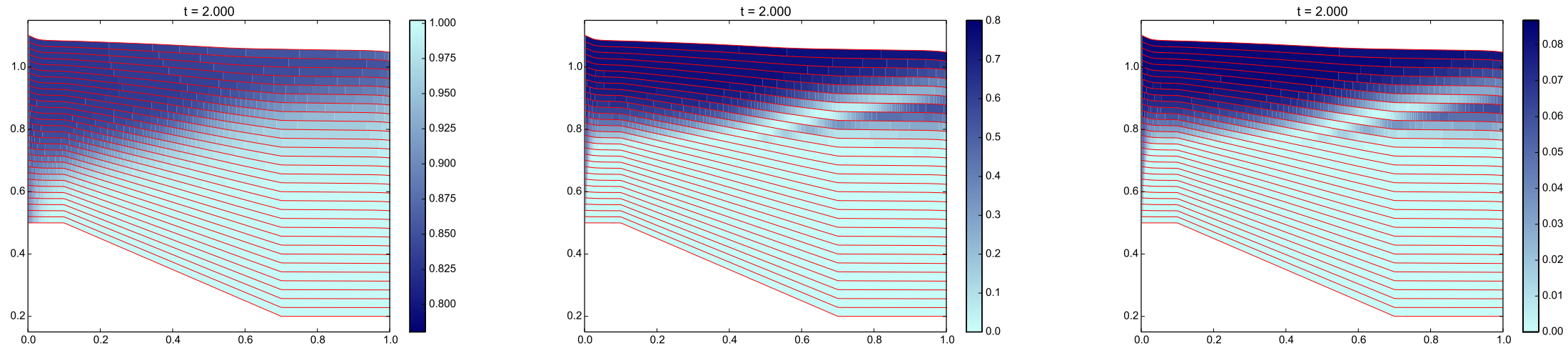

(a) Relative density $\rho(\Phi_\alpha)/\rho_0$ (b) First species concentration $\phi_{1,\alpha}$ (c) Second species concentration $\phi_{2,\alpha}$

Figure 12: Sediment distribution by layers at time $t = 2$

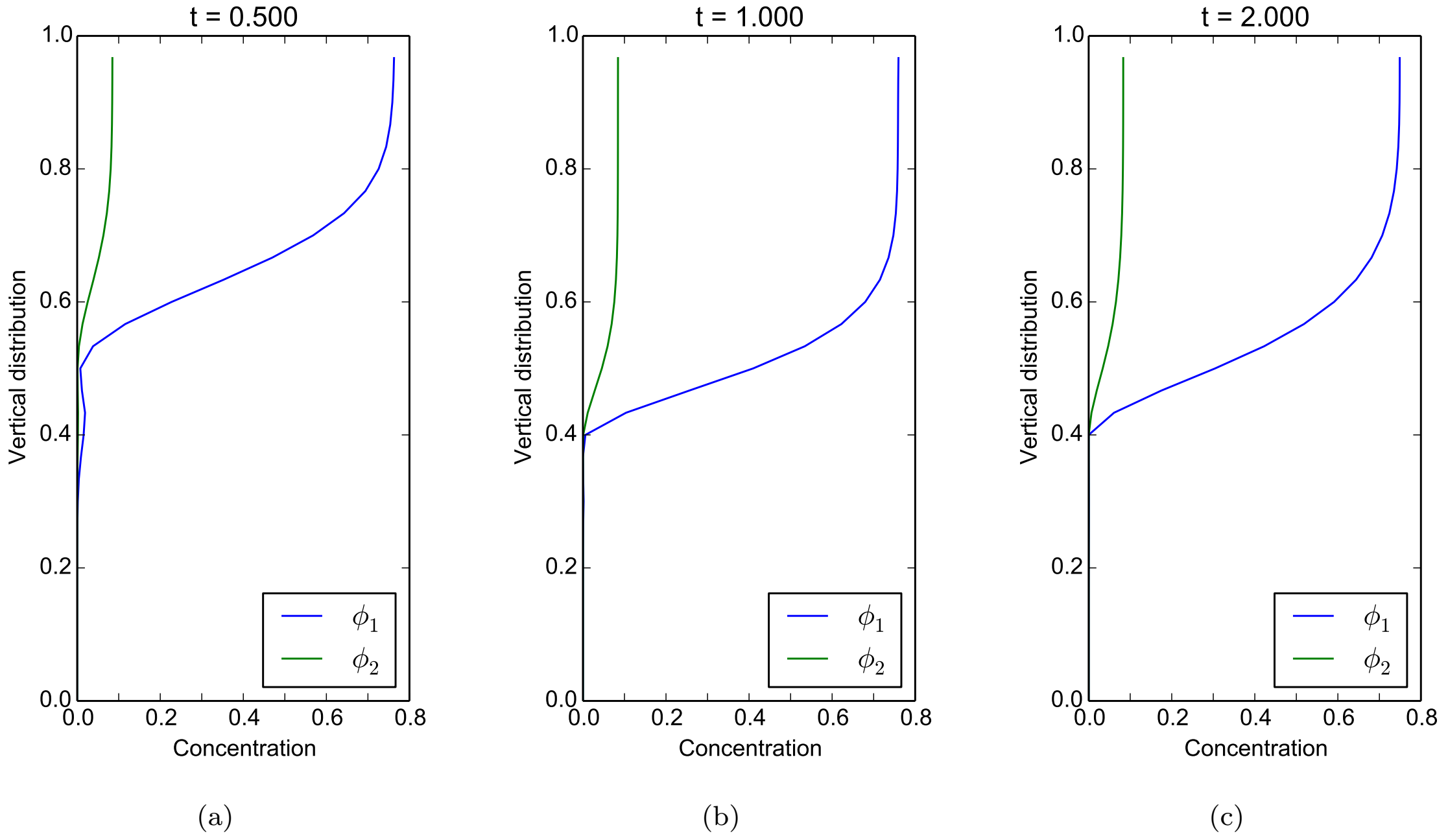

Figure 13: Vertical concentration distribution at $x = 0.1$. On the y-axis 0 corresponds to bottom level and 1 to the surface.

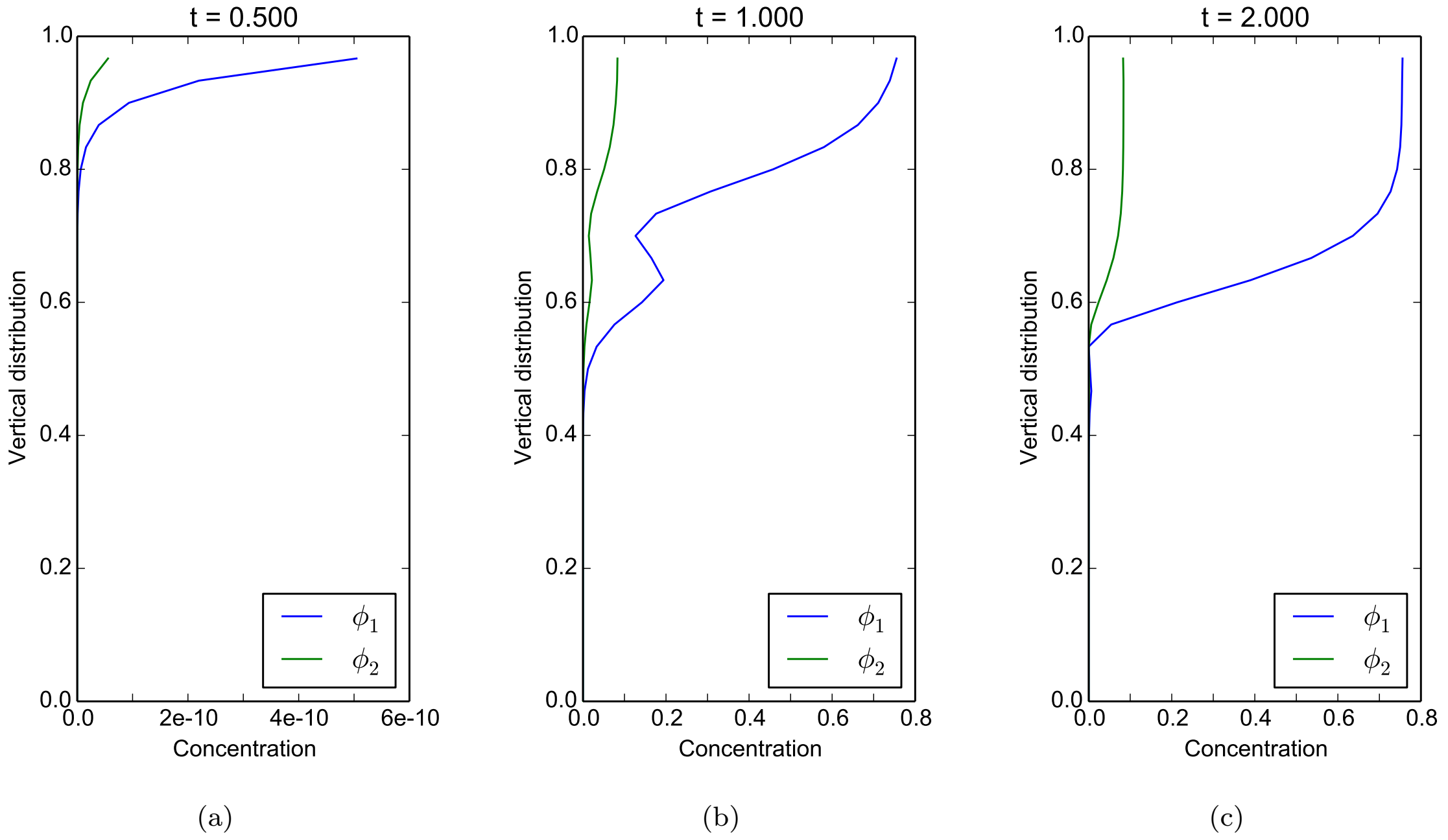

Figure 14: Vertical concentration distribution at $x = 0.35$. On the y-axis 0 corresponds to bottom level and 1 to the surface.

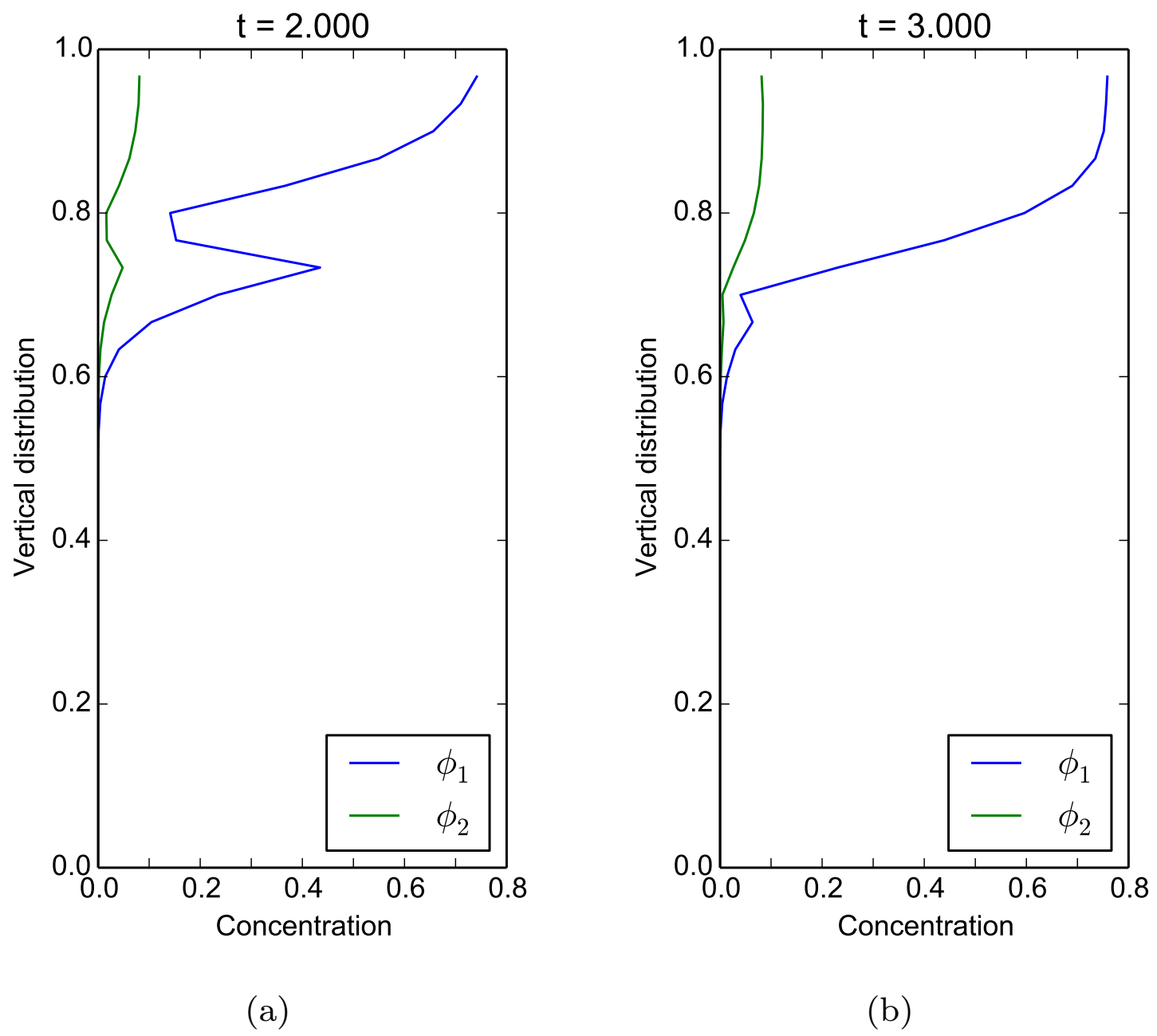

Figure 15: Vertical concentration distribution at $x = 0.8$. On the y-axis 0 corresponds to bottom level and 1 to the surface.

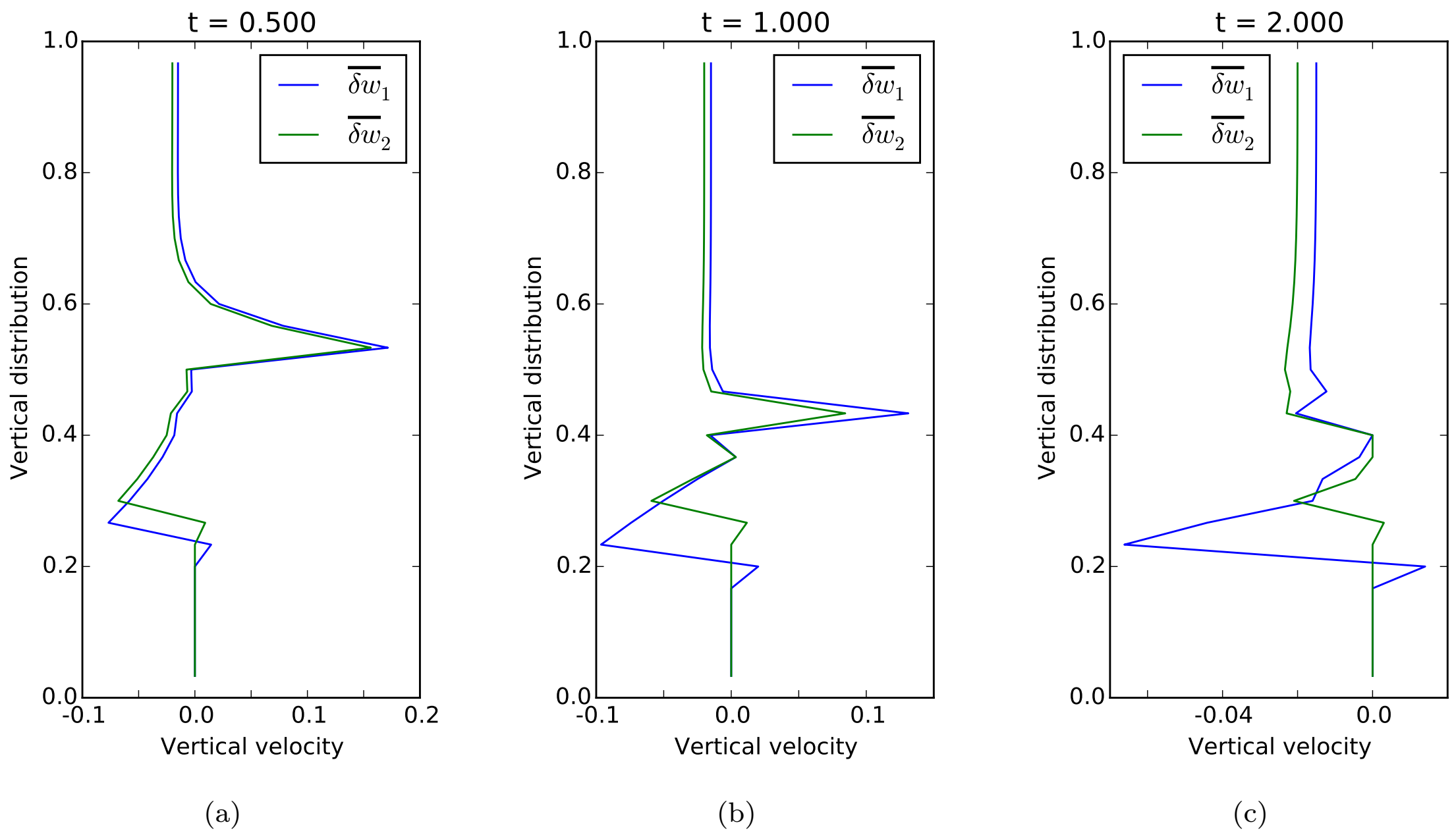

Figure 16: Average vertical velocity $\overline{\delta w}_{j,\alpha+1/2}$ at $x = 0.1$. On the y-axis 0 corresponds to bottom level and 1 to the surface.

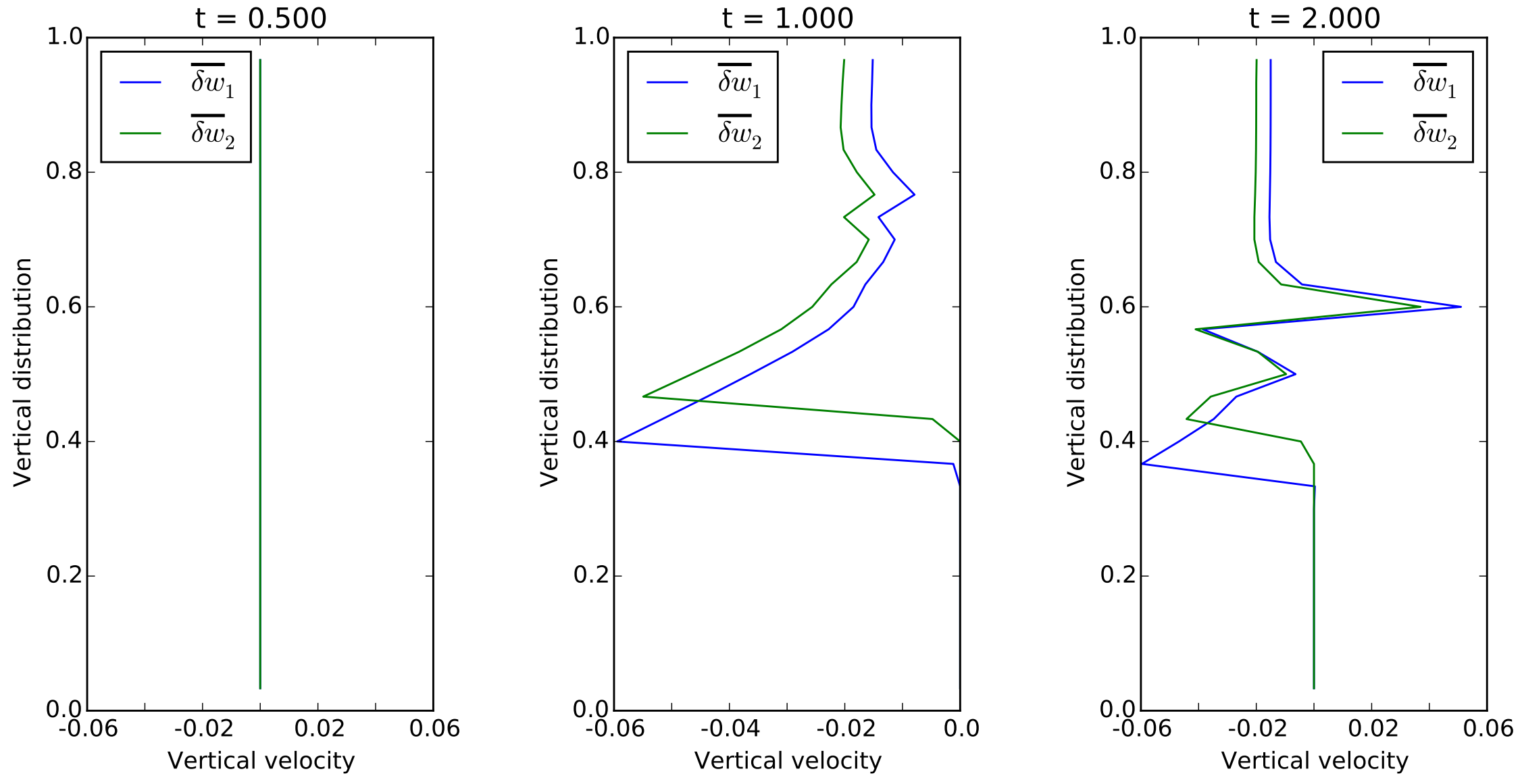

(a) (b) (c)

Figure 17: Average vertical velocity $\overline{\delta w}_{j,\alpha+1/2}$ at $x = 0.35$. On the y-axis 0 corresponds to bottom level and 1 to the surface.

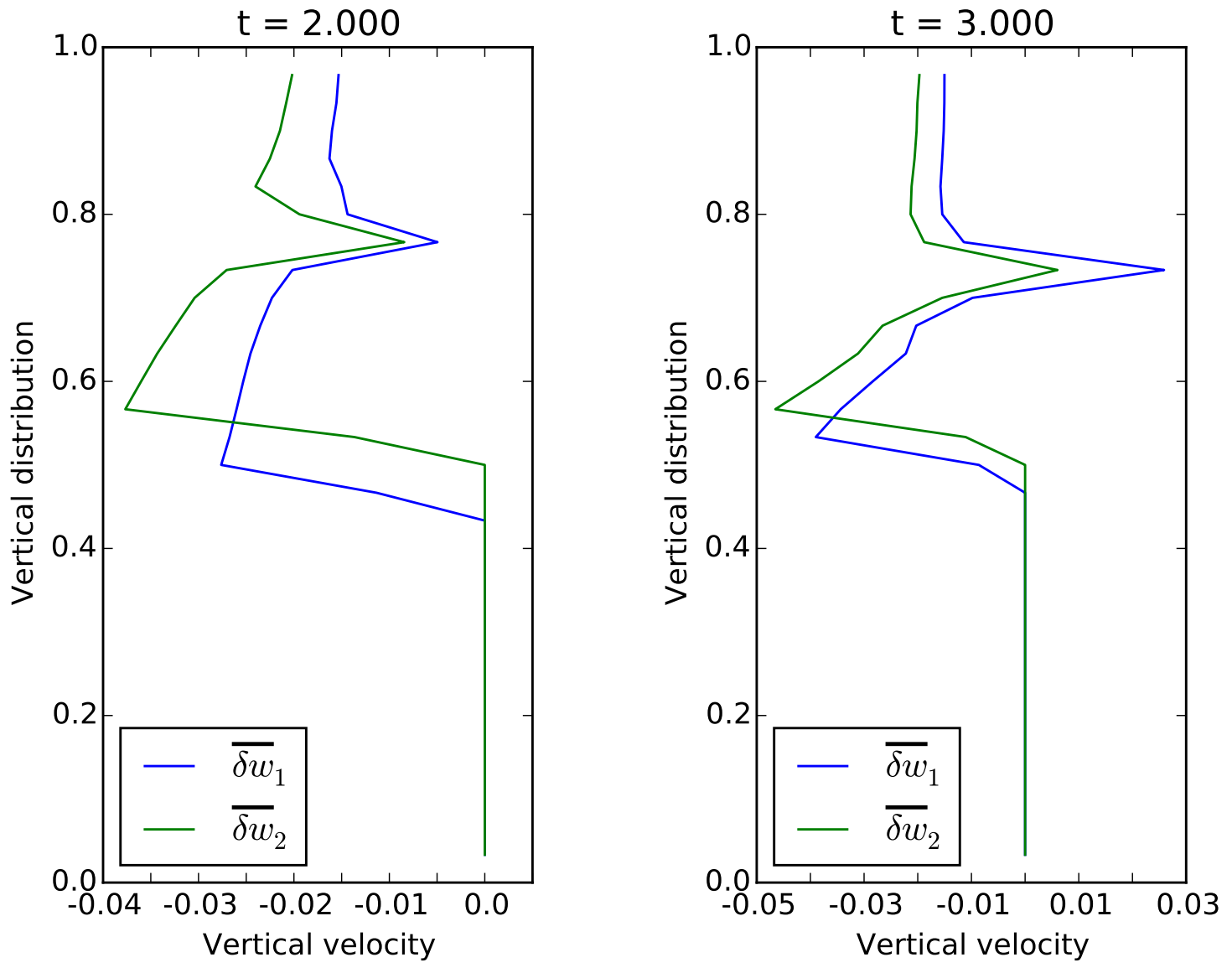

(a) (b)

Figure 18: Average vertical velocity $\overline{\delta w}_{j,\alpha+1/2}$ at $x = 0.8$. On the y-axis 0 corresponds to bottom level and 1 to the surface.

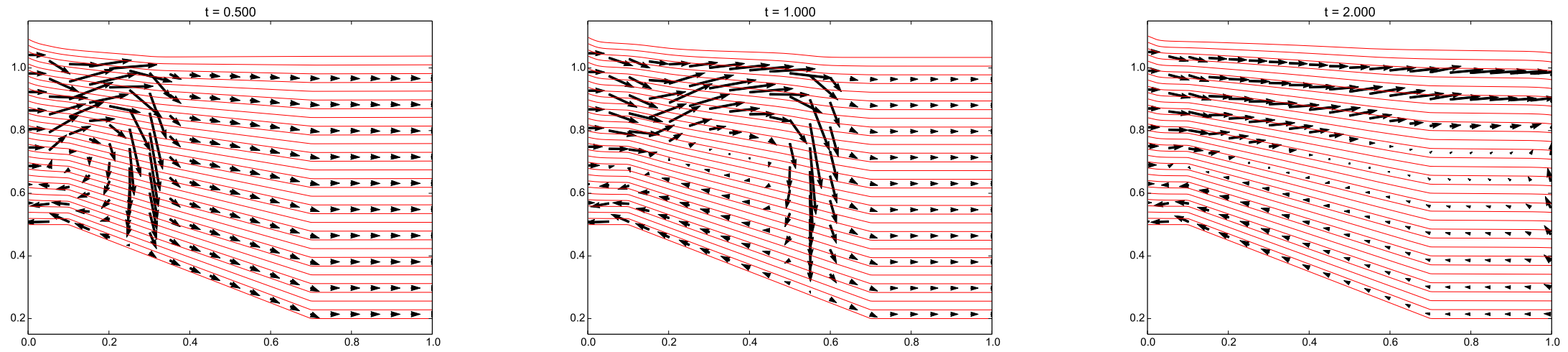

(a) (b) (c)

Figure 19: Velocity vector by layers

# 8 Conclusion

In this work, we have proposed a multi-layer shallow-water type model for the simulation of particle driven gravity currents.The model allows to simulate hyperpycnal and hypopycnal plumes. This technique allows to describe the vertical distribution of sediment overcoming the lack of information given by some more simple models. The numerical approximation of the model is based on a path-conservative finite volume method combined with a three-step splitting procedure. Two numerical experiments consisting on the simulation of a hyperpycnal and hypopycnal plume on a simplify geometry have been performed. The results are promising and this model could help to better understanding these phenomena.

# Acknowledgement

This research has been partially supported by the Spanish Government Research projects MTM2012-38383-C02-01 and MTM2012-38383-C02-02 , and Andalusian Government Research projects P11-FQM-8179 and P11-RNM-7069.

# A Upwind approximation of the transfer terms

Remember that the transfer terms (4.10) and (4.11) present in (4.9) are defined in a 'centered' way at each interface. Nevertheless, transfer terms also admit another definition that could be seen as an 'upwind' approximation see for example [3] and [11]. Thus, in the case of the multilayer shallow water system without sediments, the momentum transfer term (4.11) reduces to

$$\frac{\vec{\boldsymbol{u}}_{\alpha+1}+\vec{\boldsymbol{u}}_{\alpha}}{2}G_{\alpha+1/2}. \tag{A.1}$$

According to [3] and [11], (A.1) could be replaced by

$$u^{upwind}G_{\alpha+1/2}$$

where

$$u^{upwind}=\begin{cases} u_\alpha & \text{if } G_{\alpha+1/2}<0,\\ u_{\alpha+1} & \text{if } G_{\alpha+1/2}>0.\end{cases}$$

In [11], it has been proved that this definition is equivalent to introducing a numerical vertical diffusion term in the model, proportional to $(l_\alpha + l_{\alpha+1})/2$, what improves its stability and tends to zero when the number of layers tends to infinity.

Nevertheless, here it is not so simple, because $\rho$ is not constant. In what follows we propose an upwind definition of the transfer terms (4.10) and (4.11) in terms of the sign of the mass transfer for each species

$$\rho_j(G_{\alpha+\frac{1}{2}}-\delta w_{j,\alpha+\frac{1}{2}}).$$

Let us first consider an upwind definition of the term (4.10). If we denote by

$$(A)_{\alpha+\frac{1}{2}} = \frac{\phi_{j,\alpha+1}+\phi_{j,\alpha}}{2}(G_{\alpha+\frac{1}{2}} - \delta w_{j,\alpha+\frac{1}{2}}),$$

then we could define an upwind approximation of $(A)_{\alpha+\frac{1}{2}}$ as follows:

$$(A)^{up}_{\alpha+\frac{1}{2}} = \left(\frac{\phi_{j,\alpha+1}+\phi_{j,\alpha}}{2} + \frac{1}{2}\mathrm{sgn}(G_{\alpha+\frac{1}{2}} - \delta w_{j,\alpha+\frac{1}{2}})(\phi_{j,\alpha+1}-\phi_{j,\alpha})\right)(G_{\alpha+\frac{1}{2}} - \delta w_{j,\alpha+\frac{1}{2}}).$$

Note that

$$(A)^{up}_{\alpha+\frac{1}{2}} = (A)_{\alpha+\frac{1}{2}} + \frac{1}{2}|G_{\alpha+\frac{1}{2}} - \delta w_{j,\alpha+\frac{1}{2}}|(\phi_{j,\alpha+1}-\phi_{j,\alpha}).$$

If we replace the term (4.10) by $(A)^{up}_{\alpha+\frac{1}{2}}$ in the $N+1$ continuity equations of (4.9), we obtain the following

$$\partial_t(\phi_{j,\alpha}h_\alpha) + \nabla_x\cdot(\phi_{j,\alpha}h_\alpha\vec{\boldsymbol{u}}_\alpha) +$$

$$-\frac{1}{2}|G_{\alpha+\frac{1}{2}} - \delta w_{j,\alpha+\frac{1}{2}}|(\phi_{j,\alpha+1}-\phi_{j,\alpha}) + \frac{1}{2}|G_{\alpha-\frac{1}{2}} - \delta w_{j,\alpha-\frac{1}{2}}|(\phi_{j,\alpha}-\phi_{j,\alpha-1}) =$$

$$\frac{\phi_{j,\alpha+1}+\phi_{j,\alpha}}{2}(G_{\alpha+\frac{1}{2}} - \delta w_{j,\alpha+\frac{1}{2}}) - \frac{\phi_{j,\alpha}+\phi_{j,\alpha-1}}{2}(G_{\alpha-\frac{1}{2}} - \delta w_{j,\alpha-\frac{1}{2}}).$$

for $j=0,\ldots,N$. Finally, note that these equations could be seen as an approximation of (4.9) plus the vertical diffusion term:

$$-\partial_z\left(\frac{\Delta\tilde{z}}{2}|G-\delta w_j|\partial_z\phi_j\right),$$

being $G-\delta w_j$ a function such that $G-\delta w_j|_{z_{\alpha+\frac{1}{2}}} = G_{\alpha+\frac{1}{2}} - \delta w_{j,\alpha+\frac{1}{2}}$ and $\Delta\tilde{z}|_{z_{\alpha+\frac{1}{2}}} = (h_\alpha + h_{\alpha+1})/2$. We analogously proceed with term (4.11). If we denote by

$$(B)_{\alpha+\frac{1}{2}} = \frac{\vec{\boldsymbol{u}}_{\alpha+1}+\vec{\boldsymbol{u}}_\alpha}{2}\left(\sum_{j=0}^{N}\rho_j\frac{\phi_{j,\alpha}+\phi_{j,\alpha+1}}{2}(G_{\alpha+\frac{1}{2}} - \delta w_{j,\alpha+\frac{1}{2}})\right),$$

then we could define an upwind approximation of $(B)_{\alpha+\frac{1}{2}}$ as follows:

$$(B)^{up}_{\alpha+\frac{1}{2}} = \sum_{j=0}^{N}\rho_j\Bigg\{\left(\frac{\vec{\boldsymbol{u}}_{\alpha+1}+\vec{\boldsymbol{u}}_\alpha}{2} + \frac{1}{2}\mathrm{sgn}(G_{\alpha+\frac{1}{2}} - \delta w_{j,\alpha+\frac{1}{2}})(\vec{\boldsymbol{u}}_{\alpha+1}-\vec{\boldsymbol{u}}_\alpha)\right)\frac{\phi_{j,\alpha}+\phi_{j,\alpha+1}}{2}$$
$$+\frac{1}{2}\mathrm{sgn}(G_{\alpha+\frac{1}{2}} - \delta w_{j,\alpha+\frac{1}{2}})(\phi_{j,\alpha+1}-\phi_{j,\alpha})\frac{\vec{\boldsymbol{u}}_{\alpha+1}+\vec{\boldsymbol{u}}_\alpha}{2}\Bigg\}(G_{\alpha+\frac{1}{2}} - \delta w_{j,\alpha+\frac{1}{2}}).$$

The first term inside the brackets corresponds to an upwind approximation of the velocity times the centered approximation of sediment concentration. The second one corresponds to a centered approximation of the velocity times the diffusive term appearing in the concentration equation previously defined. Note that the following equality holds:

$$(B)^{up}_{\alpha+\frac{1}{2}} = (B)_{\alpha+\frac{1}{2}} + \frac{1}{2}\sum_{j=0}^{N}\rho_j|G_{\alpha+\frac{1}{2}} - \delta w_{j,\alpha+\frac{1}{2}}|(\phi_{j,\alpha+1}u_{\alpha+1} - \phi_{j,\alpha}u_\alpha).$$

Finally, we could conclude that $(B)^{up}_{\alpha+\frac{1}{2}}$ corresponds to approximating the original momentum equations plus a vertical diffusion term:

$$-\partial_z\left(\frac{\Delta\tilde{z}}{2}|G-\delta w_j|\partial_z(\phi_j u)\right).$$

# B Description of the terms appearing in the full formulation

We describe now the terms appearing in (5.10):

- $\boldsymbol{F}(w) = (\boldsymbol{F}_0(\boldsymbol{w})|\boldsymbol{F}_{(j-1)\cdot M+\alpha-1}(\boldsymbol{w})|\boldsymbol{F}_{N\cdot M+\alpha}(\boldsymbol{w})) \in \mathbb{R}^{(N+1)\cdot M+1}$

$$\begin{aligned}
\boldsymbol{F}_0(\boldsymbol{w}) &= h\sum_{\beta=1}^{M} l_\beta \frac{q_\beta}{m_\beta}, \\
\boldsymbol{F}_{(j-1)\cdot M+\alpha-1}(\boldsymbol{w}) &= r_{j,\alpha}\frac{q_\alpha}{m_\alpha}, \text{ for } j=1,\ldots,N, \alpha=1,\ldots,M \\
\boldsymbol{F}_{N\cdot M+\alpha}(\boldsymbol{w}) &= \frac{q_\alpha^2}{m_a} + h\left(p_S + \frac{g}{2}l_\alpha m_\alpha + g\sum_{\beta=\alpha+1}^{M} l_\beta m_\beta\right), \text{ for } \alpha=1,\ldots,M.
\end{aligned} \tag{B.1}$$

- $\boldsymbol{B}(\boldsymbol{w})$ has the following block structure

$$\boldsymbol{B}(\boldsymbol{w}) = \left(\begin{array}{c|c|c} 0 & 0 & 0 \\ \hline \boldsymbol{B}_{\mu,0}(\boldsymbol{w}) & \boldsymbol{B}_{\mu,\nu}(\boldsymbol{w}) & \boldsymbol{B}_{\mu,N\cdot M+\beta}(\boldsymbol{w}) \\ \hline \boldsymbol{B}_{N\cdot M+\alpha,0}(\boldsymbol{w}) & \boldsymbol{B}_{N\cdot M+\alpha,\nu}(\boldsymbol{w}) & \boldsymbol{B}_{N\cdot M+\alpha,N\cdot M+\beta}(\boldsymbol{w}) \end{array}\right) \in \mathcal{M}_{(N+1)\cdot M+1}(\mathbb{R}), \tag{B.2}$$

where for every $\alpha \in \{1,\ldots,M\}$, $\beta \in \{1,\ldots,M\}$, and $j \in \{1,\ldots,N\}$, and the corresponding values $\mu = (j-1)M+\alpha-1$, $\nu = (j-1)M+\beta-1$ we have

$$
\begin{aligned}
\boldsymbol{B}_{0,0}(\boldsymbol{w}) &= \boldsymbol{B}_{0,\nu}(\boldsymbol{w}) = \boldsymbol{B}_{0,N\cdot M+\beta}(\boldsymbol{w}) = 0,\\
\boldsymbol{B}_{\mu,0}(\boldsymbol{w}) &= \frac{1}{l_\alpha}\sum_{\gamma=1}^{M}\left[\left(\frac{r_{j,\alpha+1}+r_{j,\alpha}}{2h}\xi_{\alpha,\gamma} - \frac{r_{j,\alpha}+r_{j,\alpha-1}}{2h}\xi_{\alpha-1,\gamma}\right)\left(\frac{q_\gamma}{m_\gamma} - \frac{hq_\gamma}{m_\gamma^2}\rho_0\right)\right]\\
\boldsymbol{B}_{\mu,\nu}(\boldsymbol{w}) &= \frac{1}{l_\alpha}\left(\frac{r_{j,\alpha+1}+r_{j,\alpha}}{2h}\xi_{\alpha,\beta} - \frac{r_{j,\alpha}+r_{j,\alpha-1}}{2h}\xi_{\alpha-1,\beta}\right)\frac{-hq_\beta}{m_\beta^2}(\rho_j-\rho_0)\\
\boldsymbol{B}_{\mu,N\cdot M+\beta}(\boldsymbol{w}) &= \frac{1}{l_\alpha}\left(\frac{r_{j,\alpha+1}+r_{j,\alpha}}{2h}\xi_{\alpha,\beta} - \frac{r_{j,\alpha}+r_{j,\alpha-1}}{2h}\xi_{\alpha-1,\beta}\right)\frac{h}{m_\beta}\\
\boldsymbol{B}_{N\cdot M+\alpha,0}(\widetilde{\boldsymbol{w}}) &= \left(p_S + g\sum_{\beta=\alpha+1}^{M} l_\beta m_\beta\right) - g m_\alpha L_{\alpha-1}\\
&\quad + \frac{1}{l_\alpha}\left\{\sum_{\gamma=1}^{M}\left[\left(\frac{1}{2}\left(\frac{q_{\alpha+1}}{m_{a+1}} + \frac{q_\alpha}{m_a}\right)\frac{m_{\alpha+1}+m_\alpha}{2h}\xi_{\alpha,\gamma}\right.\right.\right.\\
&\quad\left.\left.\left. -\frac{1}{2}\left(\frac{q_\alpha}{m_a} + \frac{q_{\alpha-1}}{m_{\alpha-1}}\right)\frac{m_\alpha+m_{\alpha-1}}{2h}\xi_{\alpha-1,\gamma}\right)\left(\frac{q_\gamma}{m_\gamma} - \frac{hq_\gamma}{m_\gamma^2}\rho_0\right)\right]\right\}\\
\boldsymbol{B}_{N\cdot M+\alpha,\nu}(\boldsymbol{w}) &= \frac{1}{l_\alpha}\left(\frac{1}{2}\left(\frac{q_{\alpha+1}}{m_{a+1}} + \frac{q_\alpha}{m_a}\right)\frac{m_{\alpha+1}+m_\alpha}{2h}\xi_{\alpha,\beta}\right.\\
&\quad\left. -\frac{1}{2}\left(\frac{q_\alpha}{m_a} + \frac{q_{\alpha-1}}{m_{\alpha-1}}\right)\frac{m_\alpha+m_{\alpha-1}}{2h}\xi_{\alpha-1,\beta}\right)\frac{-hq_\beta}{m_\beta^2}(\rho_j-\rho_0)\\
\boldsymbol{B}_{N\cdot M+\alpha,N\cdot M+\beta}(\boldsymbol{w}) &= \frac{1}{l_\alpha}\left(\frac{1}{2}\left(\frac{q_{\alpha+1}}{m_{a+1}} + \frac{q_\alpha}{m_a}\right)\frac{m_{\alpha+1}+m_\alpha}{2h}\xi_{\alpha,\beta}\right.\\
&\quad\left. -\frac{1}{2}\left(\frac{q_\alpha}{m_a} + \frac{q_{\alpha-1}}{m_{\alpha-1}}\right)\frac{m_\alpha+m_{\alpha-1}}{2h}\xi_{\alpha-1,\beta}\right)\frac{h}{m_\beta}
\end{aligned}
\tag{B.3}
$$

- $\boldsymbol{E}(\boldsymbol{w}) = (\boldsymbol{E}_0(\boldsymbol{w})|(\mathcal{E}(\boldsymbol{w}))^t|\boldsymbol{E}_{N\cdot M+\alpha}(\boldsymbol{w})) \in \mathbb{R}^{(N+1)M+1}$, $\alpha \in \{1,\ldots,M\}$, with

$$\boldsymbol{E}_0(\boldsymbol{w}) = -G_{\frac{1}{2}}, \tag{B.4}$$

$$
\begin{aligned}
\boldsymbol{E}_{N\cdot M+\alpha}(\boldsymbol{w}) &= \frac{1}{l_\alpha}\left(\frac{1}{2}\left(\frac{q_{\alpha+1}}{m_{\alpha+1}} + \frac{q_\alpha}{m_\alpha}\right)\frac{m_{\alpha+1}+m_\alpha}{2h}(1-L_\alpha)\right.\\
&\quad\left. -\frac{1}{2}\left(\frac{q_\alpha}{m_\alpha} + \frac{q_{\alpha-1}}{m_{\alpha-1}}\right)\frac{m_\alpha+m_{\alpha-1}}{2h}(1-L_{\alpha-1})\right)G_{\frac{1}{2}}\\
&\quad - \partial_x(h(T_{xx}^E)_\alpha)\\
&\quad + \frac{1}{l_\alpha} < (T_{xx}^E, T_{zx}^E) >_{\alpha+\frac{1}{2}} \cdot(\nabla_x z_{\alpha+\frac{1}{2}}, -1)^t\\
&\quad - \frac{1}{l_\alpha} < (T_{xx}^E, T_{zx}^E) >_{\alpha-\frac{1}{2}} \cdot(\nabla_x z_{\alpha-\frac{1}{2}}, -1)^t
\end{aligned}
\tag{B.5}
$$

$$\mathcal{E}(\boldsymbol{w}) = (\mathcal{E}_\mu(\boldsymbol{w}))_{\mu=1,\ldots,N\cdot M}$$

and for $j = 1,\ldots,N, \alpha = 1,\ldots,M$ we set $\mu = (j-1)\cdot M + \alpha - 1$ and

$$\mathcal{E}_\mu(\boldsymbol{w}) = \frac{1}{l_\alpha}\left(\frac{r_{j,\alpha+1}+r_{j,\alpha}}{2h}(1-L_\alpha) - \frac{r_{j,\alpha}+r_{j,\alpha-1}}{2h}(1-L_{\alpha-1})\right)G_{\frac{1}{2}}, \tag{B.6}$$

- $\boldsymbol{\Psi}(\boldsymbol{w}) = (0|(\ominus(\boldsymbol{w}))^t|\boldsymbol{\Psi}_{N\cdot M+\alpha}(\boldsymbol{w})) \in \mathbb{R}^{(N+1)M+1}$,, $\alpha \in \{1,\cdots,M\}$ with

$$\ominus(\boldsymbol{w}) = (\ominus_\mu(\boldsymbol{w}))_{\mu=1,\ldots,N\cdot M}$$

and for $j = 1, \ldots, N, \alpha = 1, \ldots, M$ we set $\mu = (j-1) \cdot M + \alpha - 1$ and

$$\ominus_\mu(\boldsymbol{w}) = -\frac{1}{l_\alpha}\Bigg\{ < \phi_j \delta w_j >_{\alpha+\frac{1}{2}} + < \phi_j \delta w_j >_{\alpha-\frac{1}{2}} \Bigg\}. \tag{B.7}$$

$$\begin{aligned} \boldsymbol{\Psi}_{N\cdot M+\alpha}(\boldsymbol{w}) = -\frac{1}{l_\alpha}\Bigg\{ \frac{1}{2}\left(\frac{q_{\alpha+1}}{m_{a+1}} + \frac{q_\alpha}{m_a}\right) < \sum_{j=0}^{N} \rho_j \phi_j \delta w_j >_{\alpha+\frac{1}{2}} \\ + \left(\frac{q_\alpha}{m_a} + \frac{q_{\alpha-1}}{m_{\alpha-1}}\right) < \sum_{j=0}^{N} \rho_j \phi_j \delta w_j >_{\alpha-\frac{1}{2}} \Bigg\}. \end{aligned} \tag{B.8}$$

When the system is written in the form (5.12)

- $\boldsymbol{J}(\boldsymbol{w})$ has the following block structure:

$$\boldsymbol{J}(\boldsymbol{w}) = \left(\begin{array}{c|c|c} \boldsymbol{J}_{0,0}(\boldsymbol{w}) & \boldsymbol{J}_{0,\nu}(\boldsymbol{w}) & \boldsymbol{J}_{0,N\cdot M+\beta}(\boldsymbol{w}) \\ \hline \boldsymbol{J}_{\mu,0}(\boldsymbol{w}) & \boldsymbol{J}_{\mu,\nu}(\boldsymbol{w}) & \boldsymbol{J}_{\mu,N\cdot M+\beta}(\boldsymbol{w}) \\ \hline \boldsymbol{J}_{N\cdot M+\alpha,0}(\boldsymbol{w}) & \boldsymbol{J}_{N\cdot M+\alpha,\nu}(\boldsymbol{w}) & \boldsymbol{J}_{N\cdot M+\alpha,N\cdot M+\beta}(\boldsymbol{w}) \end{array}\right) \in \mathcal{M}_{(N+1)\cdot M+1}(\mathbb{R}), \tag{B.9}$$

where for every $\alpha \in \{1, \ldots, M\}$, $\beta \in \{1, \ldots, M\}$, and $j \in \{1, \ldots, N\}$, and corresponding $\mu = (j-1)M + \alpha - 1$, $\nu = (j-1)M + \beta - 1$, the components of $\boldsymbol{J}(\boldsymbol{w})$ are written as follows:

$$\begin{aligned} \boldsymbol{J}_{0,0}(\boldsymbol{w}) &= \sum_{\beta=1}^{M} l_\beta \frac{q_\beta}{m_\beta} \\ \boldsymbol{J}_{0,\nu}(\boldsymbol{w}) &= -l_\beta \frac{h q_\beta}{m_\beta^2}(\rho_j - \rho_0) \\ \boldsymbol{J}_{0,N\cdot M+\beta}(\boldsymbol{w}) &= l_\beta \frac{h}{m_\beta} \\ \boldsymbol{J}_{\mu,0}(\boldsymbol{w}) &= 0 \\ \boldsymbol{J}_{\mu,\nu}(\boldsymbol{w}) &= \left(\frac{q_\alpha}{m_\alpha} - r_{j,\alpha}\frac{q_\alpha}{m_\alpha^2}\right)\delta_{\alpha,\beta} \\ \boldsymbol{J}_{\mu,N\cdot M+\beta}(\boldsymbol{w}) &= \frac{r_{j,\alpha}}{m_\alpha}\delta_{\alpha,\beta} \\ \boldsymbol{J}_{N\cdot M+\alpha,0}(\boldsymbol{w}) &= \widetilde{\boldsymbol{J}}_{M+\alpha,0}(\widetilde{\boldsymbol{w}}) + \rho_0 \sum_{\gamma=1}^{M} \widetilde{\boldsymbol{J}}_{M+\alpha,\gamma}(\widetilde{\boldsymbol{w}}), \\ \boldsymbol{J}_{N\cdot M+\alpha,\nu}(\boldsymbol{w}) &= (\rho_j - \rho_0)\widetilde{\boldsymbol{J}}_{M+\alpha,\beta}(\widetilde{\boldsymbol{w}}), \\ \boldsymbol{J}_{N\cdot M+\alpha,N\cdot M+\beta}(\boldsymbol{w}) &= \widetilde{\boldsymbol{J}}_{M+\alpha,\beta}(\widetilde{\boldsymbol{w}}), \end{aligned} \tag{B.10}$$

where

$$\begin{aligned} \widetilde{\boldsymbol{J}}_{M+\alpha,0}(\widetilde{\boldsymbol{w}}) &= p_S + g \sum_{\gamma=\alpha+1}^{M} l_\gamma m_\gamma + \frac{g}{2} l_a m_\alpha, \\ \widetilde{\boldsymbol{J}}_{M+\alpha,\beta}(\widetilde{\boldsymbol{w}}) &= -\frac{q_\alpha^2}{m_\alpha^2}\delta_{\alpha,\beta} + h\left(\sum_{\gamma=\alpha+1}^{M} l_\gamma \delta_{\gamma,\beta} + \frac{g}{2} l_\alpha \delta_{\alpha,\beta}\right) \\ \widetilde{\boldsymbol{J}}_{M+\alpha,M+\beta}(\widetilde{\boldsymbol{w}}) &= 2\frac{q_\alpha}{m_\alpha}\delta_{\alpha,\beta}. \end{aligned} \tag{B.11}$$

## C Description of the terms appearing in the compact formulation

We give here a detailed description of the different terms appearing in (5.15):

- $\widetilde{\boldsymbol{F}}(\widetilde{\boldsymbol{w}}) = (\boldsymbol{F}_0(\boldsymbol{w})|\widetilde{\boldsymbol{F}}_\alpha(\widetilde{\boldsymbol{w}})|\widetilde{\boldsymbol{F}}_{M+\alpha}(\widetilde{\boldsymbol{w}}))^t \in \mathbb{R}^{2M+1}$, where the blocks are defined as follows

$$\begin{aligned} \widetilde{\boldsymbol{F}}_\alpha(\widetilde{\boldsymbol{w}}) &= q_\alpha, \text{for } \alpha = 1, \dots, M, \\ \widetilde{\boldsymbol{F}}_{M+\alpha}(\widetilde{\boldsymbol{w}}) &= \boldsymbol{F}_{N\cdot M+\alpha}(\boldsymbol{w}), \text{for } \alpha = 1, \dots, M. \end{aligned} \tag{C.1}$$

- $\widetilde{\boldsymbol{B}}(\widetilde{\boldsymbol{w}})$ is given by the following block structure

$$\widetilde{\boldsymbol{B}}(\widetilde{\boldsymbol{w}}) = \left(\begin{array}{c|c|c} 0 & 0 & 0 \\ \hline \widetilde{\boldsymbol{B}}_{\alpha,0}(\widetilde{\boldsymbol{w}}) & \widetilde{\boldsymbol{B}}_{\alpha,\beta}(\widetilde{\boldsymbol{w}}) & \widetilde{\boldsymbol{B}}_{\alpha,M+\beta}(\widetilde{\boldsymbol{w}}) \\ \hline \widetilde{\boldsymbol{B}}_{M+\alpha,0}(\widetilde{\boldsymbol{w}}) & \widetilde{\boldsymbol{B}}_{M+\alpha,\beta}(\widetilde{\boldsymbol{w}}) & \widetilde{\boldsymbol{B}}_{M+\alpha,M+\beta}(\widetilde{\boldsymbol{w}}) \end{array}\right) \in \mathcal{M}_{2M+1}(\mathbb{R}) \tag{C.2}$$

  where for each $\alpha \in \{1, \dots, M\}$, and $\beta \in \{1, \dots, M\}$

$$\widetilde{\boldsymbol{B}}_{0,0}(\widetilde{\boldsymbol{w}}) = \widetilde{\boldsymbol{B}}_{0,\beta}(\widetilde{\boldsymbol{w}}) = \widetilde{\boldsymbol{B}}_{0,M+\beta}(\widetilde{\boldsymbol{w}}) = 0,$$

$$\begin{aligned} \widetilde{\boldsymbol{B}}_{\alpha,0}(\widetilde{\boldsymbol{w}}) &= \frac{1}{l_\alpha} \sum_{\gamma=1}^{M} \left[ \left( \frac{m_{\alpha+1} + m_\alpha}{2h} \xi_{\alpha,\gamma} - \frac{m_\alpha + m_{\alpha-1}}{2h} \xi_{\alpha-1,\gamma} \right) \frac{q_\gamma}{m_\gamma} \right] \\ \widetilde{\boldsymbol{B}}_{\alpha,\beta}(\widetilde{\boldsymbol{w}}) &= \frac{1}{l_\alpha} \left( \frac{m_{\alpha+1} + m_\alpha}{2h} \xi_{\alpha,\beta} - \frac{m_\alpha + m_{\alpha-1}}{2h} \xi_{\alpha-1,\beta} \right) \frac{-hq_\beta}{m_\beta^2} \\ \widetilde{\boldsymbol{B}}_{\alpha,M+\beta}(\widetilde{\boldsymbol{w}}) &= \frac{1}{l_\alpha} \left( \frac{m_{\alpha+1} + m_\alpha}{2h} \xi_{\alpha,\beta} - \frac{m_\alpha + m_{\alpha-1}}{2h} \xi_{\alpha-1,\beta} \right) \frac{h}{m_\beta} \end{aligned}$$

$$\begin{aligned} \widetilde{\boldsymbol{B}}_{M+\alpha,0}(\widetilde{\boldsymbol{w}}) &= \left( p_S + g \sum_{\beta=\alpha+1}^{M} l_\beta m_\beta \right) - g m_\alpha L_{\alpha-1} \\ &\quad + \frac{1}{l_\alpha} \left\{ \sum_{\gamma=1}^{M} \left[ \left( \frac{1}{2} \left( \frac{q_{\alpha+1}}{m_{a+1}} + \frac{q_\alpha}{m_a} \right) \frac{m_{\alpha+1} + m_\alpha}{2h} \xi_{\alpha,\gamma} \right.\right.\right. \\ &\quad \left.\left.\left. - \frac{1}{2} \left( \frac{q_\alpha}{m_a} + \frac{q_{\alpha-1}}{m_{\alpha-1}} \right) \frac{m_\alpha + m_{\alpha-1}}{2h} \xi_{\alpha-1,\gamma} \right) \frac{q_\gamma}{m_\gamma} \right] \right\} \\ \widetilde{\boldsymbol{B}}_{M+\alpha,\beta}(\widetilde{\boldsymbol{w}}) &= \frac{1}{l_\alpha} \left( \frac{1}{2} \left( \frac{q_{\alpha+1}}{m_{a+1}} + \frac{q_\alpha}{m_a} \right) \frac{m_{\alpha+1} + m_\alpha}{2h} \xi_{\alpha,\beta} \right. \\ &\quad \left. - \frac{1}{2} \left( \frac{q_\alpha}{m_a} + \frac{q_{\alpha-1}}{m_{\alpha-1}} \right) \frac{m_\alpha + m_{\alpha-1}}{2h} \xi_{\alpha-1,\beta} \right) \frac{-hq_{\mathcal{B}}}{m_\beta^2} \\ \widetilde{\boldsymbol{B}}_{M+\alpha,M+\beta}(\widetilde{\boldsymbol{w}}) &= \frac{1}{l_\alpha} \left( \frac{1}{2} \left( \frac{q_{\alpha+1}}{m_{a+1}} + \frac{q_\alpha}{m_a} \right) \frac{m_{\alpha+1} + m_\alpha}{2h} \xi_{\alpha,\beta} \right. \\ &\quad \left. - \frac{1}{2} \left( \frac{q_\alpha}{m_a} + \frac{q_{\alpha-1}}{m_{\alpha-1}} \right) \frac{m_\alpha + m_{\alpha-1}}{2h} \xi_{\alpha-1,\beta} \right) \frac{h}{m_\beta} \end{aligned} \tag{C.3}$$

- $\widetilde{\boldsymbol{S}}(\widetilde{\boldsymbol{w}}) = (\widetilde{\boldsymbol{S}}_0 \,|\, \widetilde{\boldsymbol{S}}_\alpha \,|\, \widetilde{\boldsymbol{S}}_{M+\alpha}(\widetilde{\boldsymbol{w}}))^t = (0 \,|\, 0 \,|\, \widetilde{\boldsymbol{S}}_{M+\alpha}(\widetilde{\boldsymbol{w}}))^t \in \mathbb{R}^{2M+1}$, where for each $\alpha \in \{1, \dots, M\}$,

the block structure is given by:

$$\begin{aligned} \widetilde{\boldsymbol{S}}_0(\boldsymbol{w}) = \widetilde{\boldsymbol{S}}_\alpha(\boldsymbol{w}) = 0, \text{ for } \alpha = 0, 1, \dots, M, \\ \widetilde{\boldsymbol{S}}_{M+\alpha}(\boldsymbol{w}) = g m_\alpha, \text{ for } \alpha = 1, \dots, M, \end{aligned} \tag{C.4}$$

- $\widetilde{\boldsymbol{E}}(\widetilde{\boldsymbol{w}}) = (\boldsymbol{E}_0(\boldsymbol{w}) \,|\, \widetilde{\boldsymbol{E}}_\alpha(\widetilde{\boldsymbol{w}}) \,|\, \widetilde{\boldsymbol{E}}_{M+\alpha}(\widetilde{\boldsymbol{w}}))^t \in \mathbb{R}^{2M+1}$, where for each $\alpha \in \{1, \dots, M\}$, the block structure is given by:

$$\begin{aligned} \widetilde{\boldsymbol{E}}_\alpha(\widetilde{\boldsymbol{w}}) &= \frac{1}{l_\alpha} \left( \frac{m_{\alpha+1} + m_\alpha}{2h} (1 - L_\alpha) - \frac{m_\alpha + m_{\alpha-1}}{2h} (1 - L_{\alpha-1}) \right) G_{\frac{1}{2}} \\ \widetilde{\boldsymbol{E}}_{M+\alpha}(\widetilde{\boldsymbol{w}}) &= \boldsymbol{E}_{N\cdot M+\alpha}(\boldsymbol{w}) \end{aligned} \tag{C.5}$$

- $\widetilde{\boldsymbol{\Psi}}(\widetilde{\boldsymbol{w}}) = (0 \,|\, \widetilde{\boldsymbol{\Psi}}_\alpha(\widetilde{\boldsymbol{w}}) \,|\, \widetilde{\boldsymbol{\Psi}}_{M+\alpha}(\widetilde{\boldsymbol{w}}))^t \in \mathbb{R}^{2M+1}$, where for each $\alpha \in \{1, \dots, M\}$, the block structure is given by:

$$\begin{aligned} \widetilde{\boldsymbol{\Psi}}_\alpha(\widetilde{\boldsymbol{w}}) &= -\frac{1}{l_\alpha} \left\{ < \sum_{j=0}^{N} \rho_j \phi_j \delta w_j >_{\alpha+\frac{1}{2}} + < \sum_{j=0}^{N} \rho_j \phi_j \delta w_j >_{\alpha-\frac{1}{2}} \right\} \\ \widetilde{\boldsymbol{\Psi}}_{M+\alpha}(\widetilde{\boldsymbol{w}}) &= \boldsymbol{\Psi}_{N\cdot M+\alpha}(\boldsymbol{w}). \end{aligned} \tag{C.6}$$

When (5.15) can also be reformulated as in (5.16), then we have:

- $\widetilde{\boldsymbol{J}}(\widetilde{\boldsymbol{w}})$ is given by the following block structure

$$\widetilde{\boldsymbol{J}}(\widetilde{\boldsymbol{w}}) = \left( \begin{array}{c|c|c} \widetilde{\boldsymbol{J}}_{0,0}(\widetilde{\boldsymbol{w}}) & \widetilde{\boldsymbol{J}}_{0,\beta}(\widetilde{\boldsymbol{w}}) & \widetilde{\boldsymbol{J}}_{0,M+\beta}(\widetilde{\boldsymbol{w}}) \\ \hline \widetilde{\boldsymbol{J}}_{\alpha,0}(\widetilde{\boldsymbol{w}}) & \widetilde{\boldsymbol{J}}_{\alpha,\beta}(\widetilde{\boldsymbol{w}}) & \widetilde{\boldsymbol{J}}_{\alpha,M+\beta}(\widetilde{\boldsymbol{w}}) \\ \hline \widetilde{\boldsymbol{J}}_{M+\alpha,0}(\widetilde{\boldsymbol{w}}) & \widetilde{\boldsymbol{J}}_{M+\alpha,\beta}(\widetilde{\boldsymbol{w}}) & \widetilde{\boldsymbol{J}}_{M+\alpha,M+\beta}(\widetilde{\boldsymbol{w}}) \end{array} \right) \in \mathcal{M}_{2M+1}(\mathbb{R}) \tag{C.7}$$

where for each $\alpha \in \{1, \dots, M\}$, and $\beta \in \{1, \dots, M\}$

$$\begin{aligned} \widetilde{\boldsymbol{J}}_{0,0}(\widetilde{\boldsymbol{w}}) &= \sum_{\beta=1}^{M} l_\beta \frac{q_\beta}{m_\beta} \\ \widetilde{\boldsymbol{J}}_{0,\beta}(\widetilde{\boldsymbol{w}}) &= -l_\beta \frac{h q_\beta}{m_\beta^2} \\ \widetilde{\boldsymbol{J}}_{0,M+\beta}(\widetilde{\boldsymbol{w}}) &= l_\beta \frac{h}{m_\beta} \\ \\ \widetilde{\boldsymbol{J}}_{\alpha,0}(\widetilde{\boldsymbol{w}}) &= \widetilde{\boldsymbol{J}}_{\alpha,\beta}(\widetilde{\boldsymbol{w}}) = 0 \\ \widetilde{\boldsymbol{J}}_{\alpha,M+\beta}(\widetilde{\boldsymbol{w}}) &= \delta_{\alpha,\beta} \end{aligned} \tag{C.8}$$

and $\widetilde{\boldsymbol{J}}_{M+\alpha,0}(\widetilde{\boldsymbol{w}})$, $\widetilde{\boldsymbol{J}}_{M+\alpha,\beta}(\widetilde{\boldsymbol{w}})$, and $\widetilde{\boldsymbol{J}}_{M+\alpha,M+\beta}(\widetilde{\boldsymbol{w}})$ are defined in (B.11).

# D A particular weak solution with hydrostatic pressure: Deduction of equations

We detail here the calculations needed to obtain the system detailed in Section 4.

□ *Mass conservation.*

We choose a scalar test function $\varphi = \varphi(t,x)$ independent of $z$. Then, in general for a weak solution $\vec{\boldsymbol{v}}_j$ the mass conservation equation yields for all $\alpha = 0, ..., M$, $j = 0, 1, \ldots, N$

$$
\begin{aligned}
0 = & \int_{\Omega_\alpha(t)} (\partial_t \phi_j + \nabla \cdot (\phi_j \vec{\boldsymbol{v}}_j))\, \varphi \, d\Omega \\
= & \int_{I_F(t)} \varphi(t,x) \left\{ \int_{z_{\alpha-\frac{1}{2}}}^{z_{\alpha+\frac{1}{2}}} \Big( \partial_t \phi_j + \nabla_x \cdot (\phi_j \vec{\boldsymbol{u}}) + \partial_z(\phi_j w_j) \Big) dz \right\} dx \\
= & \int_{I_F(t)} \varphi(t,x) \Bigg\{ \partial_t \int_{z_{\alpha-\frac{1}{2}}}^{z_{\alpha+\frac{1}{2}}} \phi_j dz + \nabla_x \cdot \int_{z_{\alpha-\frac{1}{2}}}^{z_{\alpha+\frac{1}{2}}} (\phi_j \vec{\boldsymbol{u}})\, dz \\
& - \phi_{j,\alpha} \partial_t z_{\alpha+\frac{1}{2}} - \phi_{j,\alpha} \vec{\boldsymbol{u}}^-_{\alpha+\frac{1}{2}} \cdot \nabla_x z_{\alpha+\frac{1}{2}} + \phi_{j,\alpha} w^-_{j,\alpha+\frac{1}{2}} \\
& + \phi_{j,\alpha} \partial_t z_{\alpha-\frac{1}{2}} + \phi_{j,\alpha} \vec{\boldsymbol{u}}^+_{\alpha-\frac{1}{2}} \cdot \nabla_x z_{\alpha-\frac{1}{2}} - \phi_{j,\alpha} w^+_{j,\alpha-\frac{1}{2}} \Bigg\} dx.
\end{aligned}
$$

Moreover, noticing that $\partial_t h_\alpha = \partial_t z_{\alpha+\frac{1}{2}} - \partial_t z_{\alpha-\frac{1}{2}}$, we obtain the equation

$$
\begin{aligned}
0 = & \int_{I_F(t)} \varphi(t,x) \Bigg\{ \partial_t(\phi_{j,\alpha} h_\alpha) \; + \; \nabla_x \cdot (\phi_{j,\alpha} h_\alpha \, \vec{\boldsymbol{u}}_\alpha \,) \\
& - \phi_{j,\alpha} \partial_t z_{\alpha+\frac{1}{2}} - \phi_{j,\alpha} \vec{\boldsymbol{u}}^-_{\alpha+\frac{1}{2}} \cdot \nabla_x z_{\alpha+\frac{1}{2}} + \phi_{j,\alpha} w^-_{j,\alpha+\frac{1}{2}} \\
& + \phi_{j,\alpha} \partial_t z_{\alpha-\frac{1}{2}} + \phi_{j,\alpha} \vec{\boldsymbol{u}}^+_{\alpha-\frac{1}{2}} \cdot \nabla_x z_{\alpha-\frac{1}{2}} - \phi_{j,\alpha} w^+_{j,\alpha-\frac{1}{2}} \Bigg\} dx,
\end{aligned}
\tag{D.1}
$$

for all $\phi(t,.) \in L^2(I_F(t))$.
Applying the equation (D.1) to $\vec{\boldsymbol{u}}$, and taking into account (3.7) and (3.8), we obtain the mass conservation laws (4.5)

□ *Momentum conservation.*
We consider tests functions $\vec{\boldsymbol{\vartheta}} \in H^1(\Omega_\alpha)$ verifying (4.4). We can develop the weak formulation (4.3) taking into account the structure of $\vec{\boldsymbol{v}}$, performing an integration with respect to the variable $z$ and identifying each of the two components of the vector test functions. However, the hydrostatic pressure framework allows to drop the equations that correspond to the vertical component. That is equivalent to identifying the weak formulation for test functions in the form $(\vec{\boldsymbol{\vartheta}}_H, 0)'$, with $\vec{\boldsymbol{\vartheta}}_H = \vec{\boldsymbol{\vartheta}}_H(t,x)$ independent of $z$, with $\vec{\boldsymbol{v}}_H|_{\partial I_F} = 0$. Then, from (4.3) and using these test functions, for the horizontal momentum conservation equation we obtain,

$$
\begin{aligned}
& \int_{\Omega_\alpha(t)} \sum_{j=0}^{N} \rho_j \partial_t(\phi_{j,\alpha} \vec{\boldsymbol{u}}_\alpha) \cdot \vec{\boldsymbol{\vartheta}}_H \, d\Omega + \int_{\Omega_\alpha(t)} \sum_{j=0}^{N} \rho_j \nabla_x \cdot \Big( \phi_{j,\alpha} \vec{\boldsymbol{u}}_\alpha \otimes \vec{\boldsymbol{u}}_\alpha \Big) \cdot \vec{\boldsymbol{\vartheta}}_H \, d\Omega \\
& + \int_{\Omega_\alpha(t)} \sum_{j=0}^{N} \rho_j \partial_z(\phi_{j,\alpha} w_{j,\alpha} \vec{\boldsymbol{u}}_\alpha) \cdot \vec{\boldsymbol{\vartheta}}_H \, d\Omega + \int_{\Omega_\alpha(t)} T_{H,\alpha} : \nabla_x \vec{\boldsymbol{\vartheta}} \, d\Omega - \int_{\Omega_\alpha(t)} p_\alpha \nabla_x \cdot \vec{\boldsymbol{\vartheta}}_H \, d\Omega \\
& + \int_{\Gamma_{\alpha+\frac{1}{2}}(t)} (\boldsymbol{\Sigma}^-_{\alpha+\frac{1}{2}} (\vec{\boldsymbol{\vartheta}}_H, 0)') \cdot \vec{\boldsymbol{\eta}}_{\alpha+\frac{1}{2}} d\Gamma - \int_{\Gamma_{\alpha-\frac{1}{2}}(t)} (\boldsymbol{\Sigma}^+_{\alpha-\frac{1}{2}} (\vec{\boldsymbol{\vartheta}}_H, 0)') \cdot \vec{\boldsymbol{\eta}}_{\alpha-\frac{1}{2}} d\Gamma = 0,
\end{aligned}
\tag{D.2}
$$

for all $\alpha = 1, ..., M$, where

$$T_{H,\alpha} = T_H(\vec{\boldsymbol{v}}_\alpha).$$

Taking into account that $\Omega_\alpha(t) = I_F \times [z_{\alpha-\frac{1}{2}}(t), z_{\alpha+\frac{1}{2}}(t)]$ and the hypothesis on the independence in $z$ of $\vec{\boldsymbol{u}}_\alpha$ and $\vec{\boldsymbol{\vartheta}}_H$, we develop in what follows each one of the components of previous equation:

- $\displaystyle\int_{\Omega_\alpha(t)} \sum_{j=0}^{N} \rho_j \partial_t(\phi_{j,\alpha}\vec{\boldsymbol{u}}_\alpha)\cdot\vec{\boldsymbol{\vartheta}}_H \, d\Omega = \int_{I_F}\left(\int_{z_{\alpha-\frac{1}{2}}}^{z_{\alpha+\frac{1}{2}}} \partial_t(\rho(\Phi_\alpha)\vec{\boldsymbol{u}}_\alpha)\cdot\vec{\boldsymbol{\vartheta}}_H dx\right)dz$
$\displaystyle= \int_{I_F} h_\alpha \partial_t(\rho(\Phi_\alpha)\vec{\boldsymbol{u}}_\alpha)\cdot\vec{\boldsymbol{\vartheta}}_H dx.$
- $\displaystyle\int_{\Omega_\alpha(t)} \sum_{j=0}^{N} \rho_j \nabla_x\cdot\Big(\phi_{j,\alpha}\vec{\boldsymbol{u}}_\alpha\otimes\vec{\boldsymbol{u}}_\alpha\Big)\cdot\vec{\boldsymbol{\vartheta}}_H \, d\Omega = \int_{I_F} h_\alpha\nabla_x\cdot\Big(\rho(\Phi_\alpha)\vec{\boldsymbol{u}}_\alpha\otimes\vec{\boldsymbol{u}}_\alpha\Big)\cdot\vec{\boldsymbol{\vartheta}}_H \, dx.$
- $\displaystyle\int_{\Omega_\alpha(t)} \sum_{j=0}^{N} \rho_j \partial_z(\phi_{j,\alpha}w_{j,\alpha}\vec{\boldsymbol{u}}_\alpha)\cdot\vec{\boldsymbol{\vartheta}}_H \, d\Omega = \sum_{j=0}^{N}\int_{I_F}\int_{z_{\alpha-\frac{1}{2}}}^{z_{\alpha+\frac{1}{2}}} \rho_j\partial_z(\phi_{j,\alpha}w_{j,\alpha}\vec{\boldsymbol{u}}_\alpha)\cdot\vec{\boldsymbol{\vartheta}}_H dzdx$
$\displaystyle= \sum_{j=0}^{N}\int_{I_F}\rho_j\phi_{j,\alpha}(w^-_{j,\alpha+\frac{1}{2}} - w^+_{j,\alpha-\frac{1}{2}})\vec{\boldsymbol{u}}_\alpha\cdot\vec{\boldsymbol{\vartheta}}_H dx$
- $\displaystyle\int_{\Omega_\alpha(t)} T_{H,\alpha} : \nabla_x\vec{\boldsymbol{\vartheta}}_H \, d\Omega = \int_{I_F}\left(\int_{z_{\alpha-\frac{1}{2}}}^{z_{\alpha+\frac{1}{2}}} T_{H,\alpha}:\nabla_x\vec{\boldsymbol{\vartheta}}_H dz\right)dx = \int_{I_F} h_\alpha T_{H,\alpha}:\nabla_x\vec{\boldsymbol{\vartheta}}_H dx$
$\displaystyle= -\int_{I_F}\nabla_x\cdot(h_\alpha T_{H,\alpha})\cdot\vec{\boldsymbol{\vartheta}}_H dx.$
- $\displaystyle\int_{\Omega_\alpha(t)} p_\alpha\nabla_x\cdot\vec{\boldsymbol{\vartheta}}_H \, d\Omega = \int_{I_F}\left(\int_{z_{\alpha-\frac{1}{2}}}^{z_{\alpha+\frac{1}{2}}} p_\alpha dz\right)\nabla\cdot\vec{\boldsymbol{\vartheta}}_H dx = -\int_{I_F}\vec{\boldsymbol{\vartheta}}_H \cdot\nabla_x\left(\int_{z_{\alpha-\frac{1}{2}}}^{z_{\alpha+\frac{1}{2}}} p_\alpha dz\right)dx$
$\displaystyle= -\int_{I_F}\vec{\boldsymbol{\vartheta}}_H \cdot\left(\int_{z_{\alpha-\frac{1}{2}}}^{z_{\alpha+\frac{1}{2}}}\nabla_x p_\alpha dz + p_{\alpha+\frac{1}{2}}\nabla_x z_{\alpha+\frac{1}{2}} - p_{\alpha-\frac{1}{2}}\nabla_x z_{\alpha-\frac{1}{2}}\right)dx$
$\displaystyle= -\int_{I_F}\vec{\boldsymbol{\vartheta}}_H\cdot\left(\int_{z_{\alpha-\frac{1}{2}}}^{z_{\alpha+\frac{1}{2}}}\nabla_x p_\alpha dz\right)dx - \int_{I_F} p_{\alpha+\frac{1}{2}}(\vec{\boldsymbol{\vartheta}}_H,0)'\cdot\eta_{\alpha+\frac{1}{2}}\sqrt{1+\left|\nabla_x z_{\alpha+\frac{1}{2}}\right|^2}dx$
$\displaystyle+ \int_{I_F} p_{\alpha-\frac{1}{2}}(\vec{\boldsymbol{\vartheta}}_H,0)'\cdot\eta_{\alpha-\frac{1}{2}}\sqrt{1+\left|\nabla_x z_{\alpha-\frac{1}{2}}\right|^2}dx.$
- $\displaystyle\int_{\Gamma_{\alpha+\frac{1}{2}}(t)}(\boldsymbol{\Sigma}^-_{\alpha+\frac{1}{2}}(\vec{\boldsymbol{\vartheta}}_H,0)')\cdot\vec{\boldsymbol{\eta}}_{\alpha+\frac{1}{2}}d\Gamma = \int_{I_F}(\boldsymbol{\Sigma}^-_{\alpha+\frac{1}{2}}(\vec{\boldsymbol{\vartheta}}_H,0)')\cdot\vec{\boldsymbol{\eta}}_{\alpha+\frac{1}{2}}\sqrt{1+\left|\nabla_x z_{\alpha+\frac{1}{2}}\right|^2}dx.$

Moreover, as $\boldsymbol{\Sigma}^-_{\alpha+\frac{1}{2}} + p_{\alpha+\frac{1}{2}}\boldsymbol{I} = T^-_{\alpha+\frac{1}{2}}$, we can do the following simplification,

$$\int_{I_F}(\boldsymbol{\Sigma}^-_{\alpha+\frac{1}{2}}(\vec{\boldsymbol{\vartheta}}_H,0)')\cdot\vec{\boldsymbol{\eta}}_{\alpha+\frac{1}{2}}\sqrt{1+\left|\nabla_x z_{\alpha+\frac{1}{2}}\right|^2}dx + \int_{I_F} p_{\alpha+\frac{1}{2}}(\vec{\boldsymbol{\vartheta}}_H,0)'\cdot\eta_{\alpha+\frac{1}{2}}\sqrt{1+\left|\nabla_x z_{\alpha+\frac{1}{2}}\right|^2}dx$$
$$= \int_{I_F}(T^-_{\alpha+\frac{1}{2}}(\vec{\boldsymbol{\vartheta}}_H,0)')\cdot(\nabla_x z_{\alpha+\frac{1}{2}},-1)'dx = \int_{I_F}(T^-_{H,\alpha+\frac{1}{2}}(\nabla_x z_{\alpha+\frac{1}{2}})' - T^-_{xz,\alpha+\frac{1}{2}})\cdot\vec{\boldsymbol{\vartheta}}_H dx.$$

Where we have used that as $T^-_{\alpha+\frac{1}{2}}$ is a symmetric matrix. Then

$$(T^-_{\alpha+\frac{1}{2}}(\vec{\boldsymbol{\vartheta}}_H,0)')\cdot(\nabla_x z_{\alpha+\frac{1}{2}},-1)' = (T^-_{\alpha+\frac{1}{2}}(\nabla_x z_{\alpha+\frac{1}{2}},-1)')\cdot(\vec{\boldsymbol{\vartheta}}_H,0)'.$$

Then, the weak formulation (D.2), corresponding to the horizontal momentum equation for this set of tests functions, can be written as follows:

$$\int_{I_F} \vec{\boldsymbol{\vartheta}}_H \cdot \Bigg( \begin{array}{l} h_\alpha \partial_t(\rho(\Phi_\alpha)\vec{\boldsymbol{u}}_\alpha) + h_\alpha \nabla_x \cdot \Big(\rho(\Phi_\alpha)\vec{\boldsymbol{u}}_\alpha \otimes \vec{\boldsymbol{u}}_\alpha\Big) + \sum_{j=0}^{N} \rho_j \phi_{j,\alpha}(w^-_{j,\alpha+\frac{1}{2}} - w^+_{j,\alpha-\frac{1}{2}})\vec{\boldsymbol{u}}_\alpha \\ + \int_{z_{\alpha-\frac{1}{2}}}^{z_{\alpha+\frac{1}{2}}} \nabla_x p_\alpha dz - \nabla_x \cdot (h_\alpha T_H^E) \\ +(T^-_{H,\alpha+\frac{1}{2}}(\nabla_x z_{\alpha+\frac{1}{2}})' - T^-_{xz,\alpha+\frac{1}{2}}) - (T^+_{H,\alpha-\frac{1}{2}}(\nabla_x z_{\alpha-\frac{1}{2}})' - T^+_{xz,\alpha-\frac{1}{2}}) \end{array} \Bigg) dx = 0, \quad \forall \vec{\boldsymbol{\vartheta}}_H. \tag{D.3}$$

Taking into account (3.22) we deduce,

$$\begin{array}{l} h_\alpha \partial_t(\rho(\Phi_\alpha)\vec{\boldsymbol{u}}_\alpha) + h_\alpha \nabla_x \cdot \Big(\rho(\Phi_\alpha)\vec{\boldsymbol{u}}_\alpha \otimes \vec{\boldsymbol{u}}_\alpha\Big) + \sum_{j=0}^{N} \rho_j \phi_{j,\alpha}(w^-_{j,\alpha+\frac{1}{2}} - w^+_{j,\alpha-\frac{1}{2}})\vec{\boldsymbol{u}}_\alpha \\ + \int_{z_{\alpha-\frac{1}{2}}}^{z_{\alpha+\frac{1}{2}}} \nabla_x p_\alpha dz - \nabla_x \cdot (h_\alpha T_H^E) \\ +(\widetilde{T}_{H,\alpha+\frac{1}{2}}(\nabla_x z_{\alpha+\frac{1}{2}})' - \widetilde{T}_{xz,\alpha+\frac{1}{2}}) - (\widetilde{T}_{H,\alpha-\frac{1}{2}}(\nabla_x z_{\alpha-\frac{1}{2}})' - \widetilde{T}_{xz,\alpha-\frac{1}{2}}) \\ = \dfrac{\vec{\boldsymbol{u}}_{\alpha+1} - \vec{\boldsymbol{u}}_\alpha}{2} \displaystyle\sum_{j=0}^{N} \rho_j G_{j,\alpha+\frac{1}{2}} - \dfrac{\vec{\boldsymbol{u}}_\alpha - \vec{\boldsymbol{u}}_{\alpha-1}}{2} \sum_{j=0}^{N} \rho_j G_{j,\alpha-\frac{1}{2}} \end{array} \tag{D.4}$$

Using (4.7) we may write (D.4) in the form

$$\begin{array}{l} h_\alpha \rho(\Phi_\alpha) \partial_t \vec{\boldsymbol{u}}_\alpha + h_\alpha \rho(\Phi_\alpha) \nabla_x \cdot \Big(\vec{\boldsymbol{u}}_\alpha \otimes \vec{\boldsymbol{u}}_\alpha\Big) + \displaystyle\int_{z_{\alpha-\frac{1}{2}}}^{z_{\alpha+\frac{1}{2}}} \nabla_x p_\alpha dz - \nabla_x \cdot (h_\alpha T_H^E) \\ +(\widetilde{T}_{H,\alpha+\frac{1}{2}}(\nabla_x z_{\alpha+\frac{1}{2}})' - \widetilde{T}_{xz,\alpha+\frac{1}{2}}) - (\widetilde{T}_{H,\alpha-\frac{1}{2}}(\nabla_x z_{\alpha-\frac{1}{2}})' - \widetilde{T}_{xz,\alpha-\frac{1}{2}}) \\ = \dfrac{\vec{\boldsymbol{u}}_{\alpha+1} - \vec{\boldsymbol{u}}_\alpha}{2} \displaystyle\sum_{j=0}^{N} \rho_j G_{j,\alpha+\frac{1}{2}} - \dfrac{\vec{\boldsymbol{u}}_\alpha - \vec{\boldsymbol{u}}_{\alpha-1}}{2} \sum_{j=0}^{N} \rho_j G_{j,\alpha-\frac{1}{2}} \\ -\vec{\boldsymbol{u}}_\alpha \displaystyle\sum_{j=0}^{N} \rho_j \phi_{j,\alpha}(w^-_{\alpha+\frac{1}{2}} - w^+_{\alpha-\frac{1}{2}}) \end{array} \tag{D.5}$$

Finally, we can obtain the momentum equations by combining previous equation with (4.5). If we multiply (4.5) by $\rho_j$, for $j = 0, \ldots, N$ and by summing up these equations we obtain

$$\partial_t(\rho(\Phi_\alpha)h_\alpha) + \nabla_x \cdot (\rho(\Phi_\alpha)h_\alpha \vec{\boldsymbol{u}}_\alpha) = \sum_{j=0}^{N} \rho_j G_{j,\alpha+\frac{1}{2}} - \sum_{j=0}^{N} \rho_j G_{j,\alpha-\frac{1}{2}}. \tag{D.6}$$

Finally, if we sum up (D.5) with the result of multiplying (D.6) by $\vec{\boldsymbol{u}}_\alpha$, and taking into account (3.14), we obtain the evolution equation for the momentum at each layer:

$$\partial_t(\rho(\Phi_\alpha)h_\alpha\vec{\boldsymbol{u}}_\alpha) + \nabla_x \cdot \Big(\rho(\Phi_\alpha)h_\alpha\vec{\boldsymbol{u}}_\alpha \otimes \vec{\boldsymbol{u}}_\alpha\Big) + h_\alpha\rho(\Phi_\alpha)\Big(\nabla_x \cdot \vec{\boldsymbol{u}}_\alpha\Big)\vec{\boldsymbol{u}}_\alpha + \int_{z_{\alpha-\frac{1}{2}}}^{z_{\alpha+\frac{1}{2}}} \nabla_x p_\alpha dz - \nabla_x \cdot (h_\alpha T_H)$$

$$+(\widetilde{T}_{H,\alpha+\frac{1}{2}}(\nabla_x z_{\alpha+\frac{1}{2}})' - \widetilde{T}_{xz,\alpha+\frac{1}{2}}) - (\widetilde{T}_{H,\alpha-\frac{1}{2}}(\nabla_x z_{\alpha-\frac{1}{2}})' - \widetilde{T}^+_{xz,\alpha-\frac{1}{2}})$$

$$= \frac{\vec{\boldsymbol{u}}_{\alpha+1} + \vec{\boldsymbol{u}}_\alpha}{2} \sum_{j=0}^{N} \rho_j G_{j,\alpha+\frac{1}{2}} - \frac{\vec{\boldsymbol{u}}_\alpha + \vec{\boldsymbol{u}}_{\alpha-1}}{2} \sum_{j=0}^{N} \rho_j G_{j,\alpha-\frac{1}{2}}$$

$$-\vec{\boldsymbol{u}}_\alpha \sum_{j=0}^{N} \rho_j \phi_{j,\alpha}(w^-_{\alpha+\frac{1}{2}} - w^+_{\alpha-\frac{1}{2}}). \tag{D.7}$$

Remark that $\nabla \cdot v = 0$ which means:

$$h_\alpha \nabla_x \cdot \vec{\boldsymbol{u}}_\alpha + (w^-_{\alpha+\frac{1}{2}} - w^+_{\alpha-\frac{1}{2}}) = 0, \tag{D.8}$$

and (4.8) follows.
Now, by (4.1) and (4.2), we obtain

$$\int_{z_{\alpha-\frac{1}{2}}}^{z_{\alpha+\frac{1}{2}}} \nabla_x p_\alpha dz = h_\alpha \Bigg( \nabla_x \Big( p_S + g \sum_{\beta=\alpha+1}^{M} \rho(\Phi_\beta) h_\beta + g\rho(\Phi_\alpha)\frac{h_\alpha}{2}\Big) + g\rho(\Phi_\alpha)\nabla_x \Big( z_b + \sum_{\beta=1}^{\alpha-1} h_b + \frac{h_\alpha}{2}\Big) \Bigg). \tag{D.9}$$

Then, (D.9) can be rewritten as

$$\int_{z_{\alpha-\frac{1}{2}}}^{z_{\alpha+\frac{1}{2}}} \nabla_x p_\alpha dz = h_\alpha \Big( \nabla_x \overline{p}_\alpha + g\rho(\Phi_\alpha)\nabla_x \bar{z}_\alpha \Big).$$

and (4.9) follows.